\documentclass[aoas]{imsart}

\RequirePackage{amsthm,amsmath,amsfonts,amssymb}
\RequirePackage[authoryear]{natbib}

\startlocaldefs

\makeatletter % Avoid "Submitted to ..." on title page for arxiv.
\def\journal@name{}
\makeatother

\numberwithin{equation}{section}
\usepackage{etoc}
\usepackage{bibunits} % Separate main-paper and supplement bibliographies

\usepackage[normalem]{ulem} % to cross out

\usepackage{url}
\usepackage{mathrsfs}
\usepackage{dsfont}
\usepackage{paralist}
\usepackage{caption}
\usepackage{subcaption}
\usepackage{booktabs}
\usepackage{multirow}
\usepackage[dvipsnames]{xcolor}
\usepackage{hyperref}

\hypersetup{
colorlinks=true,
urlcolor=blue,
}

\newcommand{\R}{\mathbb{R}}
\newcommand{\E}{\mathds{E}}

\newcommand{\N}{\mathbb{N}}
\newcommand{\eps}{\varepsilon}
\newcommand{\bm}[1]{\boldsymbol{#1}}

\newcommand{\Exp}{\operatorname{\mathds E}}
\newcommand{\Var}{\operatorname{Var}}
\newcommand{\Cov}{\operatorname{Cov}}

\newcommand{\GEV}{\operatorname{GEV}}

\newcommand{\argmax}{\operatornamewithlimits{\arg\max}}

\newcommand{\PR}{\mathrm{PR}}

\usepackage{pgfplots}
\pgfplotsset{compat=1.18} 

\newcommand{\obs}{{\mathrm{obs}}}

\newcommand{\sobs}{{\mathrm{sobs}}}
\newcommand{\smod}{{\mathrm{smod}}}
\newcommand{\sfull}{{\mathrm{sfull}}}

\newcommand{\model}{\mathrm{mod}}

\newcommand{\fact}{\mathrm{fact}}
\newcommand{\cntr}{\mathrm{cntr}}

\usepackage{siunitx}
\usepackage{algorithm}
\usepackage{algpseudocode}
\theoremstyle{definition}
\newtheorem{example}{Example}

\endlocaldefs

\begin{document}
\etocdepthtag.toc{main}
% Main-paper citation unit: only citations before \putbib are collected here.
\begin{bibunit}[imsart-nameyear]
\begin{frontmatter}
\title{Evidence Synthesis in Probabilistic Extreme Event Attribution: From Attribution Measures to Model Parameters}
\runtitle{Evidence Synthesis in Extreme Event Attribution}

\begin{aug}
%%%%%%%%%%%%%%%%%%%%%%%%%%%%%%%%%%%%%%%%%%%%%%%
%% Only one address is permitted per author. %%
%% Only division, organization and e-mail is %%
%% included in the address.                  %%
%% Additional information such as            %%
%% identifying the corresponding author must %%
%% be included in in the Acknowledgments     %%
%% section if necessary.                     %%
\author[A]{\fnms{Erik}~\snm{Haufs}\ead[label=e1]{erik.haufs@rub.de}\orcid{0009-0008-8194-7445}}
\author[A]{\fnms{Axel}~\snm{Bücher}\ead[label=e2]{axel.buecher@rub.de}\orcid{0000-0002-1947-1617}}
\author[B]{\fnms{Jonas}~\snm{Schröter}\ead[label=e3]{jonas.schroeter@dwd.de}\orcid{0009-0008-4488-3104}}
%%%%%%%%%%%%%%%%%%%%%%%%%%%%%%%%%%%%%%%%%%%%%%
%% Addresses                                %%
%%%%%%%%%%%%%%%%%%%%%%%%%%%%%%%%%%%%%%%%%%%%%%
\address[A]{Ruhr-Universität Bochum, Fakultät für Mathematik\printead[presep={,\ }]{e1,e2}}

\address[B]{Deutscher Wetterdienst (DWD), Regionales Klimabüro Potsdam\printead[presep={,\ }]{e3}}
\end{aug}

\begin{abstract}
Probabilistic extreme event attribution aims to quantify how anthropogenic climate change has altered the likelihood or intensity of a class of extreme events. Existing studies commonly combine evidence from observational products and climate-model ensembles by first estimating attribution measures, such as probability ratios or intensity changes, for each data source and then synthesizing the resulting estimates. We critically assess this approach, identify potential shortcomings of a respective benchmark procedure from the literature, and propose both targeted modifications and a new parameter-level synthesis method. The latter combines estimates of the underlying nonstationary distributional regression parameters, thereby enabling inference across multiple event thresholds and counterfactual climate conditions. In controlled simulation studies, the proposed modifications substantially improve upon the benchmark procedure, while parameter-level synthesis provides competitive overall performance. The practical usefulness is illustrated through a case study of the heavy precipitation associated with Storm Boris in September 2024.
\end{abstract}


\end{frontmatter}

\section{Introduction}
\label{sec:introduction}

Extreme event attribution seeks to assess whether, and to what extent, anthropogenic climate change has influenced an observed extreme event. One line of research addresses this question from a primarily physical perspective by investigating how the event would have unfolded under different background climate conditions, conditional on selected dynamical features of the observed event \citep[the storyline approach;][]{Shepherd2016,Shepherd2019}. 
A complementary line of research, originating with \cite{Allen2003}, takes a more explicitly statistical perspective and asks how climate change has altered the probability or intensity of a specified class of extreme events. This framework, commonly referred to as \emph{probabilistic extreme event attribution}, is the focus of the present paper.

Within this framework, attribution is based on comparing the distribution of an event-related quantity under the factual climate with its distribution under a counterfactual climate without, or with reduced, anthropogenic influence. For a prescribed event threshold, one may compare the corresponding exceedance probabilities under the two climates; alternatively, for a fixed exceedance probability, one may compare the associated event intensities or return levels. Attribution is then summarized by suitable contrasts of these climate-specific quantities, such as probability or risk ratios and changes in event intensity or return level. Estimation is commonly based on nonstationary distributional regression models whose parameters depend on covariates such as the smoothed global mean surface temperature (GMST) anomaly, which serves as a proxy for anthropogenic climate change \citep{philip2020, jvo21}. Estimates of the resulting attribution measures, together with uncertainty assessments such as confidence intervals, have become standard outputs of operational attribution studies \citep[among others]{VanOldenborgh2017, Vautard2019, tradowsky2023}. Many such studies have been conducted within the World Weather Attribution (WWA) initiative (\url{https://www.worldweatherattribution.org}), with the corresponding methodology implemented in the KNMI Climate Explorer \citep{KNMIClimateExplorer}.

Probabilistic attribution studies typically draw on multiple data sources, potentially including several observational products and multiple ensembles of climate-model simulations. Combining these sources is highly valuable because inference based solely on observational records is often subject to substantial uncertainty arising from limited temporal coverage, small effective sample sizes, and the rarity of extreme events. At the same time, synthesizing evidence across heterogeneous data sources poses substantial methodological challenges that have so far been addressed only partially in the attribution literature.

To the best of our knowledge, the approach of \cite{otto24} is the only systematic attempt within the WWA community to formalize such evidence synthesis at the methodological level. Its central idea is to estimate a chosen attribution measure, such as a probability ratio or intensity change, separately for each observational product and climate-model data set and then combine the resulting estimates using tools from meta-analysis. This yields a single synthesized estimate together with an assessment of estimation uncertainty, typically in the form of a confidence interval. However, these procedures, hereafter referred to as the WWA approach, have neither been examined systematically from a statistical perspective nor validated through controlled simulation studies. Our first main contribution is therefore a critical methodological assessment of the WWA approach. Using well-founded statistical arguments and heuristics, we identify several potential shortcomings, explain how they may affect finite-sample performance, and propose corresponding modifications.

Our second main contribution is a fundamentally different synthesis methodology. Rather than combining estimates of a single attribution measure, we synthesize the parameters of the underlying distributional regression models. Once these parameters have been combined, a broad range of attribution measures can be derived from the resulting synthesized model. This parameter-level approach therefore permits simultaneous inference across different event thresholds, exceedance probabilities, and counterfactual climate conditions, providing a substantially richer assessment of the attribution question. Moreover, by construction, it is less prone to infinite estimates, which occur frequently under the original WWA approach and are handled in \cite{otto24} through additional ad hoc modifications. 

Our third main contribution is a comprehensive controlled simulation study comparing the original WWA approach, its modified version, and the proposed parameter-level synthesis method. The simulation results provide empirical support for the concerns raised in our methodological assessment and show that, in the finite-sample settings considered, both the modified WWA procedure and the proposed method perform substantially better than the original WWA approach.

Finally, we illustrate the practical usefulness of the proposed methods in a case study of the heavy precipitation event associated with Storm Boris, which caused severe flooding across parts of Central Europe in September 2024 \citep{Athanase2024}. Both the proposed parameter-level synthesis and the modified WWA approach provide statistically significant evidence of an anthropogenic influence, as represented by changes in global mean surface temperature (GMST), whereas the original WWA approach does not detect a statistically significant signal.

The remainder of the paper is organized as follows. Section~\ref{sec:fundamentals} introduces the main concepts and modeling assumptions underlying probabilistic extreme event attribution. Section~\ref{sec:synthesis-attribution-targets} reviews the WWA approach, which synthesizes evidence at the level of attribution measures, provides a critical methodological assessment, and proposes several modifications. Section~\ref{sec:synthesis-model-parameters} introduces the new synthesis approach operating at the level of distributional regression parameters. Section~\ref{sec:simulation} evaluates all methods in a comprehensive controlled simulation study. Section~\ref{sec:case-study} presents the case study on Storm Boris, and Section~\ref{sec:conclusion} concludes. Additional derivations, simulation results, and case-study analyses are provided in Appendices~\ref{app:model-parameters-to-attribution-targets}–\ref{app:case-study}.
Implementations of all synthesis methods considered in this paper are available in the Python package \href{https://github.com/haufse/synthesis-eea}{\texttt{haufse/synthesis-eea}}. The code used to produce the simulation study and the Storm Boris case study is available in the accompanying GitHub repository \href{https://github.com/haufse/synthesis-eea-paper}{\texttt{haufse/synthesis-eea-paper}}.

\section{Fundamentals of Extreme Event Attribution}
\label{sec:fundamentals}

To formalize probabilistic extreme event attribution (EEA), let $X$ denote a variable of interest, for instance the annual maximum of cumulative daily precipitation at a fixed location. EEA distinguishes between the distribution of $X$ under a factual climate, denoted by $X_{\fact}$ with cumulative distribution function (CDF) $F_{\fact}$, and a counterfactual climate, denoted by $X_{\cntr}$ with CDF $F_{\cntr}$. For example, the factual climate may correspond to present-day conditions, whereas the counterfactual climate represents a pre-industrial world.

Possible key quantities of interest for comparing the distributions of $X_{\fact}$ and $X_{\cntr}$ include the probability ratio
\begin{align}
\label{eq:probability-ratio}
\PR
= \mathrm{PR}(x_0)
= \frac{\mathbb{P}(X_{\fact} > x_0)}{\mathbb{P}(X_{\cntr} > x_0)}
= \frac{1 - F_{\fact}(x_0)}{1 - F_{\cntr}(x_0)},
\end{align}
defined for a high threshold $x_0\in\R$ (often corresponding to an observed extreme event), and the change in intensity, defined at a small exceedance probability level $p_0\in(0,1)$ as
\begin{align}
\label{eq:intensity-change}
\Delta I = \frac{F_{\fact}^{-1}(1-p_0) - F_{\cntr}^{-1}(1-p_0)}{F_{\cntr}^{-1}(1-p_0)}.
\end{align}

Additional modeling assumptions are required to make these expressions tractable. A common approach assumes that the distribution of $X$ depends on the climate only through a low-dimensional proxy for anthropogenic forcing, such as the global mean surface temperature (GMST) or its anomaly relative to a reference climate, e.g., a pre-industrial climate. In fact, GMST has been shown to capture the dominant externally forced signal in many regions \citep{Eyring2021HumanInfluence}. Formally, the assumption is expressed as
\[
F_{\fact} = F( \,\cdot \mid g_{\fact}), 
\qquad 
F_{\cntr} = F( \,\cdot \mid g_{\cntr}),
\]
where $g_{\fact}$ and $g_{\cntr}$ denote the low-dimensional proxy for anthropogenic forcing in the factual and counterfactual climate, respectively. At the level of random variables, this corresponds to
\[
X_{\mathrm{fact}} \overset{d}{=} (X \mid G = g_{\mathrm{fact}}), 
\qquad
X_{\mathrm{cntr}} \overset{d}{=} (X \mid G = g_{\mathrm{cntr}}),
\]
where $(X,G)$ is a generic pair consisting of the variable of interest and the anthropogenic forcing proxy, with $F(\,\cdot \mid g)$ the conditional CDF of $X$ given $G=g$.

Additional structure is obtained by assuming that the conditional distribution of $X$ given $G=g$ belongs to a parametric family. More precisely, there exists a parameter $\theta_g\in\Phi$ with $\Phi\subseteq\mathbb R^D$ such that
\[
    F( \,\cdot\, \mid g) = H_{\theta_g},
    \qquad\text{equivalently}\qquad
    (X \mid G=g) \sim H_{\theta_g}.
\]
Here, $\{H_\theta:\theta\in\Phi\}$ denotes a parametric family of distribution functions, such as the Gaussian, Gamma, or generalized extreme value family; see \cite{philip2020} and Example~\ref{example:GEV} below.
The dependence of $\theta_g$ on $g$ is typically specified
through a known \textit{link function}
\begin{align*}
    f:\Theta\times\mathscr G\to\Phi,
    \qquad
    (\vartheta,g)\mapsto f(\vartheta,g)=\theta_g.
\end{align*}
Here, $\Theta\subseteq\mathbb R^d$, $d\in\mathbb N$, is the parameter space for the model parameter $\vartheta$, while $\mathscr G$ denotes the state space of the forcing proxy, typically $\mathscr G=\mathbb R$.

\begin{example}
\label{example:GEV}
If $X$ is a block maximum variable, the Extremal Types Theorem \citep{Fisher1928,Gnedenko1943} motivates modeling $F(\, \cdot \mid g)$ using the generalized extreme value (GEV) distribution. Recall that the GEV distribution with parameter $\theta = (\gamma, \mu, \sigma) \in \Phi := \R \times \R \times (0,\infty)$ has CDF
\begin{align*}
H_\theta(x)
=
\exp\!\Big(
-\Big[1+\gamma \dfrac{x-\mu}{\sigma}\Big]^{-1/\gamma}
\Big)
\end{align*}
for all $x$ such that $1+\gamma(x-\mu)/\sigma>0$, with $H_\theta(x) = \exp[-\exp\{-(x-\mu)/\sigma\}]$ for $\gamma=0$. Incorporating GMST as a covariate amounts to assuming
$
(X \mid G = g) \sim \mathrm{GEV}(\theta_g),
$
where $\theta_g = (\gamma_g, \mu_g, \sigma_g)$ depends on $g$. Variants of this assumption are widely used in the EEA literature; see, for example, \cite{philip2020,otto23,otto24}.

A common specification for the dependence of $\theta_g$ on $g$ is the \emph{shift model}, frequently used for temperature extremes \citep{philip2020}. It assumes constant shape and scale parameters, $\gamma_g \equiv \gamma$ and $\sigma_g \equiv \sigma$, while the location parameter varies linearly with GMST, $\mu_g = \mu + \alpha g$. This yields a four-parameter nonstationary GEV model with parameter vector $\vartheta = (\gamma, \mu, \sigma, \alpha)$ and link function
\begin{align}\label{eq:linkfunc_shift}
\theta_g = f_{\mathrm{shift}}(\vartheta,g)
=
\begin{pmatrix}
\gamma \\
\mu + \alpha g \\
\sigma
\end{pmatrix}.
\end{align}
For precipitation extremes, a \emph{scale model} is often employed, motivated by the Clausius--Clapeyron relation \citep{philip2020, tradowsky2023}. In this case, both location and scale parameters vary multiplicatively with the GMST anomaly. Two equivalent parametrizations are commonly used:
\begin{align}
\label{eq:linkfunc_scale}
\theta_g = \tilde f_{\mathrm{scale}}(\vartheta,g)
&=
\begin{pmatrix}
\gamma \\
\mu \exp(\alpha g/\mu) \\
\sigma \exp(\alpha g/\mu)
\end{pmatrix},
\qquad
\theta_g = f_{\mathrm{scale}}(\vartheta,g)
=
\begin{pmatrix}
\gamma \\
\mu \exp(\alpha g) \\
\sigma \exp(\alpha g)
\end{pmatrix}.
\end{align}
In both parametrizations, the dispersion ratio $\sigma_g/\mu_g$ remains constant, analogous to the index flood assumption in hydrology \citep{Hanel2009}. While the formulation based on $\tilde f_{\mathrm{scale}}$ is common in the literature, we adopt $f_{\mathrm{scale}}$ in subsequent sections. The two are mathematically equivalent; however, $f_{\mathrm{scale}}$ avoids numerical instability when $\mu$ is close to zero and simplifies maximum likelihood estimation using analytic gradients.
\end{example}

Irrespective of the chosen parametric family and link function, attribution measures such as those in \eqref{eq:probability-ratio} and \eqref{eq:intensity-change} can be expressed as deterministic functions of $\vartheta$, once $g_{\fact}$ and $g_{\cntr}$ are fixed. In the following, we denote a generic attribution measure by $T$ and view it as a function $\vartheta \mapsto T(\vartheta)$ with $\vartheta \in \mathbb{R}^d$, where $d=4$ for the shift and scale GEV models in \eqref{eq:linkfunc_shift} and \eqref{eq:linkfunc_scale}. Estimation of $T$ thus reduces to estimation of the model parameter $\vartheta$; details on the corresponding transformations $\vartheta \mapsto T(\vartheta)$ are given in Appendix~\ref{app:model-parameters-to-attribution-targets}.

In turn, $\vartheta$ can typically be estimated from multiple data sources, including observational records (e.g., weather station data, gridded observations, or reanalysis products) as well as climate model simulations under different forcing scenarios.
In each case, the generic sampling framework is as follows: the pair $(G,X)$ is observed over a time index set $\mathcal{I}$, for example, $\mathcal{I} = \{1900, 1901, \dots, 2026\}$ for annual data, with sample size $n = |\mathcal{I}|$. 
Based on this sample, the parameter $\vartheta$ is estimated, for instance via conditional maximum likelihood, yielding an estimator $\hat{\vartheta}_n$. An estimator of $T$ is then obtained by the plug-in principle,
$
\hat{T}_n = T(\hat{\vartheta}_n).
$
Recall that $T$ is a known function of $\vartheta$, which can typically be evaluated explicitly or approximated to arbitrary accuracy using Monte Carlo methods.
Combined with an assessment of estimation uncertainty, e.g., via an appropriate bootstrap procedure, this yields a statistically sound basis for addressing the attribution question for each individual data set.

The availability of multiple data sources naturally raises the question of how to combine the resulting information about the attribution measure $T$. Such combination procedures are commonly referred to as \emph{synthesis}. In many attribution studies \citep[among others]{Arias2023,VasconcelosJunior2024,Barnes2025,Kimutai2025,Clarke2026}, synthesis is performed directly at the level of the attribution measure: $T$ is estimated separately from each data source, yielding estimators $\hat T_1,\hat T_2,\dots$, which are subsequently combined into an overall estimator $\hat T$. This strategy is implemented, for example, in the KNMI Climate Explorer \citep{KNMIClimateExplorer} and forms the basis of the current WWA synthesis methodology \citep{otto24}. We review these approaches in Section~\ref{sec:synthesis-attribution-targets}.

\section{Synthesis on the Level of Attribution Measures: the WWA approach}
\label{sec:synthesis-attribution-targets}

Attribution-measure-level (AML) synthesis is the prevailing approach in EEA, including in the WWA methodology \citep{philip2020}. A recent formalization of these methods is provided by \cite{otto24}, and the presentation in this section builds largely on their work. While their exposition is primarily aimed at an applied audience and focuses on describing the methods, our objective is to place these approaches within a rigorous statistical framework. In particular, we link the methods to explicit model assumptions, thereby strengthening the justification of their validity and indicating possible avenues for modification. Moreover, in contrast to \cite{otto24}, we systematically distinguish between unknown population parameters and their estimators, thereby helping to prevent potential misinterpretations of the underlying formulas. Whenever a definition is incomplete or missing in \cite{otto24}, we reconstruct their definition based on the accompanying \texttt{R} package \citep{barnesrwwa2024}.

The approach proceeds in three steps, reflecting the nature of the available data sources. First, inference is based solely on observational data sets, which are typically strongly dependent but expected to yield estimates with comparatively small bias. Second, information from multiple climate model simulations is incorporated; these can often be treated as approximately independent, but are more susceptible to systematic biases, as climate models provide only imperfect representations of the true climate system. Third, the observational and model-based sources are combined in a final synthesis step. Each of the three steps is treated in a dedicated subsection.

Throughout, we adopt the general framework introduced in Section~\ref{sec:fundamentals}, with $X$ denoting the variable of interest and $T_0$ the unknown true value of the attribution measure.

\subsection{Observational Synthesis on the Level of Attribution Measures}
\label{subsec:observational-synthesis-attribution-targets}

In practice, multiple observational data products may be used to extract observations from the target variable $X$, typically after appropriate spatial and/or temporal aggregation. 
Formally, we assume that $m=m_{\obs} \in \mathbb{N}$ such data sets are available. Typically, $m$ is small, e.g., 
$m=1$ for \cite{VasconcelosJunior2024}, $m=2$ for \cite{Clarke2026,Barnes2025}, or $m=3$ for \cite{Arias2023}. 
Throughout, we assume that $m\ge 2$, since no synthesis is required when only a single observational data product is available. For each $j \in \{1, \dots, m\}$, we extract a time series $(X_{jt})_{t \in \mathcal{I}_j}$ of the target variable over a sampling period $\mathcal{I}_j$ of length $n_j = |\mathcal{I}_j|$. In addition, GMST anomaly observations relative to a reference period, denoted by $(g_t)_t$, are available throughout all sampling periods.

\begin{figure}
    \centering
    \begin{minipage}[T]{.45\textwidth}
    \includegraphics[width=\linewidth]{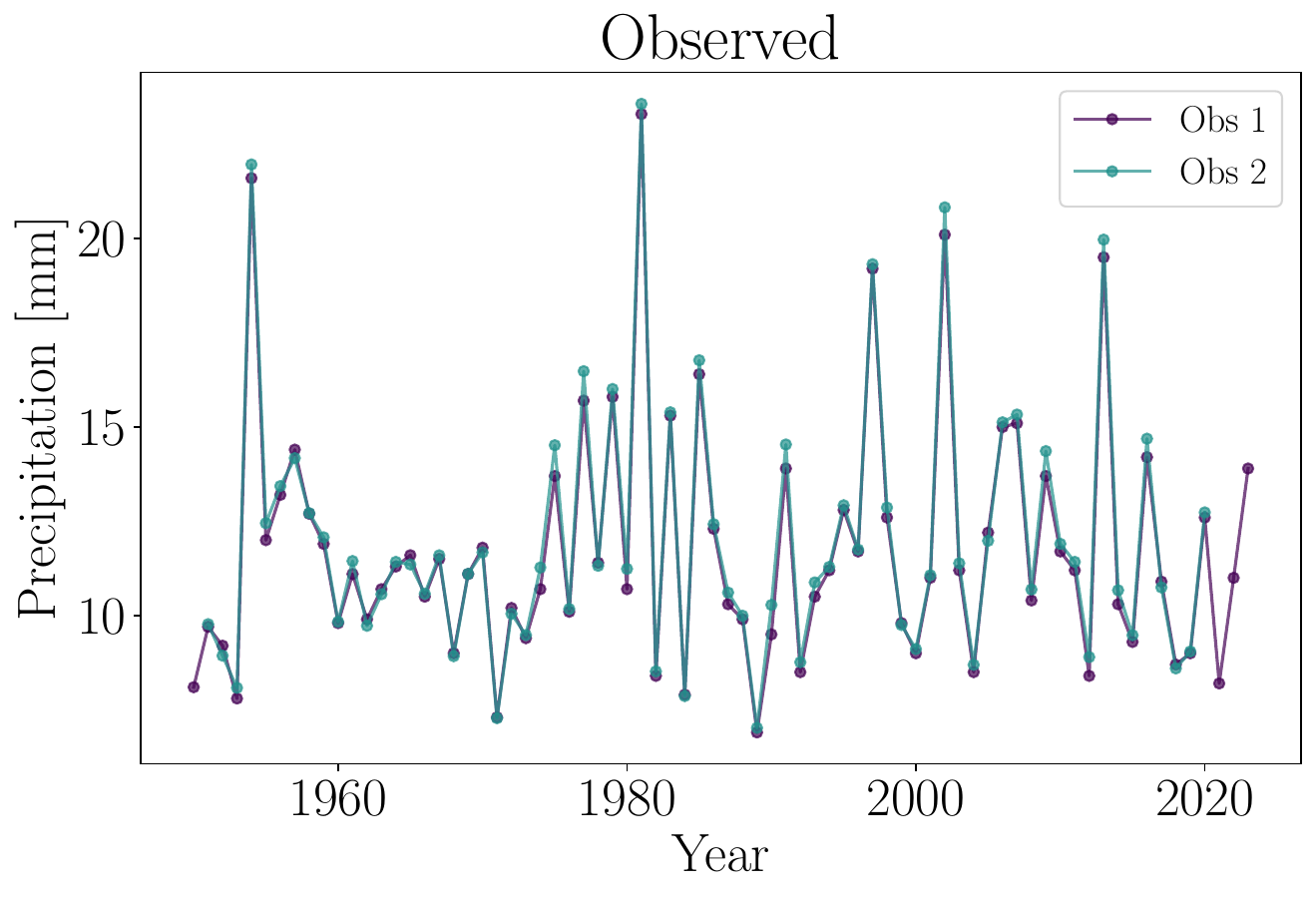}
    \end{minipage}
    \begin{minipage}[T]{.45\textwidth}
    \includegraphics[width=\linewidth]{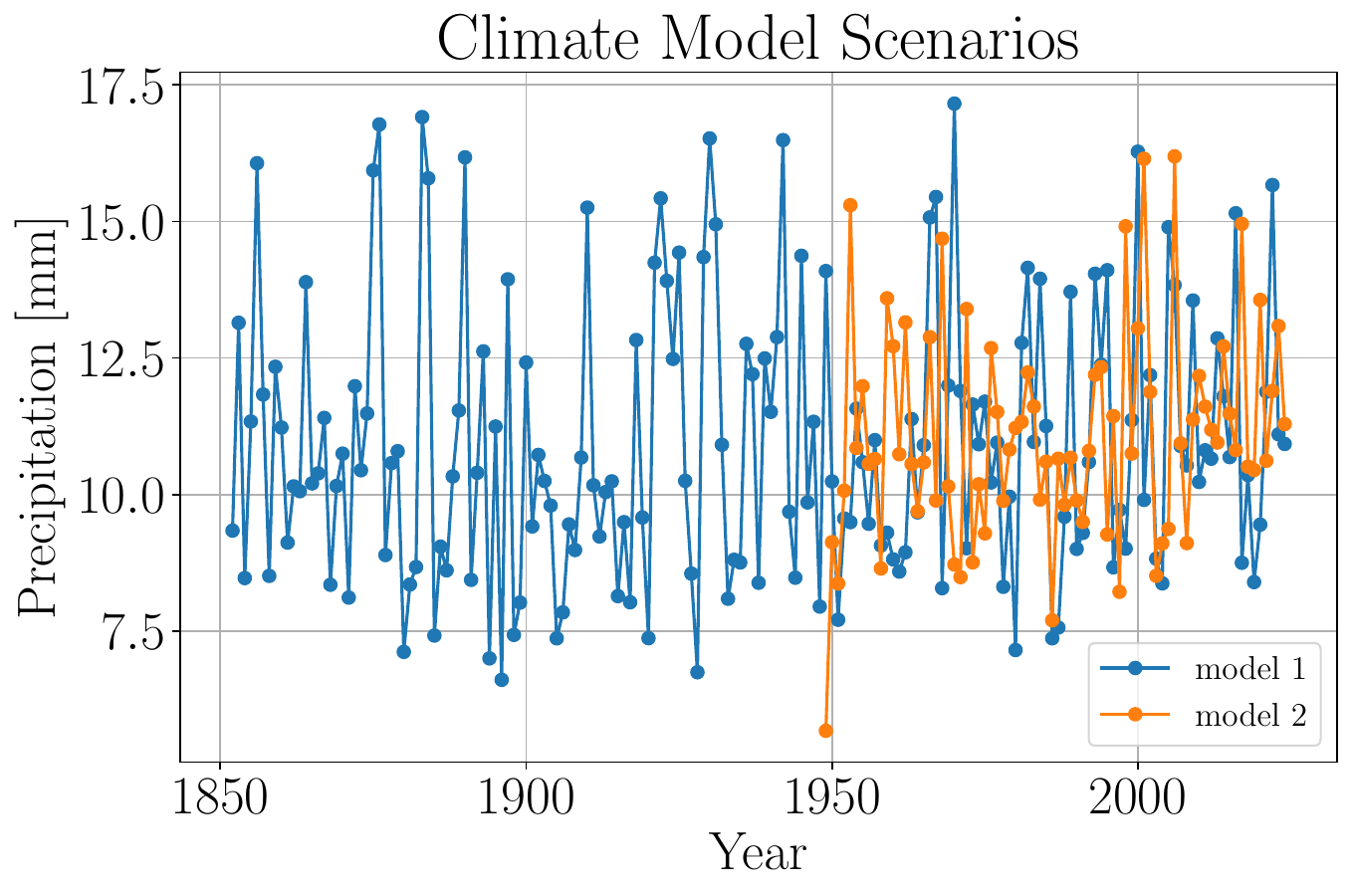}
    \end{minipage}
    \caption{Left: observational data for a precipitation variable analyzed in Section~\ref{sec:case-study}, showing annual maxima 
    of spatially and temporally aggregated precipitation amounts
    from two gridded datasets (HYRAS-DE, 70 years long, 1 km resolution; E-OBS, 74 years long, 0.1$^\circ\times$0.1$^\circ$ resolution). 
    Right: exemplary data for the same variable extracted from two climate models (BCC-CSM2-MR and CLMcom-CCLM4-8-17\_EC-EARTH). While the two time series are strongly correlated on the left, they appear independent on the right. }
    \label{fig:observations-vs-climate-models}
\end{figure}

A key feature of the different data sets is their strong cross-sectional dependence. This dependence arises from the fact that their sampling periods $\mathcal{I}_j$ often overlap. At time points where two data sets are available, the recorded values of $X$ are typically nearly identical.
This phenomenon is illustrated in the left-hand side of Figure~\ref{fig:observations-vs-climate-models}, where the two data sets correspond to a regional composite based on the HYRAS-DE data set \citep{Frick2014} and the E-OBS data set \citep{eobs}, and where the variable $X$ represents annual maxima of spatially averaged (over selected river discharge systems) precipitation totals accumulated over 96 consecutive hours; see Section~\ref{sec:case-study} for further details.

Applying, for instance, the methodology described in Section~\ref{sec:fundamentals} and Appendix~\ref{app:model-parameters-to-attribution-targets}, each data set yields an estimator $\hat T_j = \hat T_j^{\obs}$ of the attribution measure of interest. A natural statistical framework for combining these estimates is based on classical random effects modeling. More specifically, 
we assume that each data set has a data-set-specific ground truth $T_{0j}=T_{0j}^\obs$, which may differ from the true target $T_0$ because of systematic discrepancies.
It is assumed that the data products are constructed in such a way that they are, on average over a hypothetical collection of data products, correct. Mathematically, this requirement can be captured by assuming that the \emph{representation error variables}
\[
\eps_j = \eps_j^\obs = T_{0j}-T_0, \qquad j =1, \dots, m.
\]
form an independent sample from a generic representation error variable $\eps$ with $\Exp[\eps]=0$ and $\tau^2 = \tau_\obs^2  = \Var(\eps)\ge 0$. 
Representation error variables capture the fact that attribution measures derived from the population distributions of observational data products, and possibly from imperfect statistical models fitted to them, need not coincide exactly with the corresponding measure in the true, not perfectly observable climate. Sources of such discrepancies include spatial or temporal aggregation, interpolation and homogenization procedures used in constructing reanalysis products, or misspecification of the distributional regression model used to define the attribution measure. Hence, even in the absence of sampling and estimation variability, the attribution measure associated with a given data product and modeling procedure may systematically deviate from the true value $T_0$. As acknowledged by \cite{otto24}, the assumption that the representation errors are centered at zero may not be fully justified in practice and should therefore be borne in mind when interpreting the results. The independence assumption should likewise be understood as a working model. In particular, observational products may share systematic representation errors because they rely on overlapping observations or similar construction procedures. Such common components cannot be identified from between-product variation and are therefore not captured by the present random-effects formulation.

Next, each estimator $\hat T_j$, constructed solely from the $j$th data product, is assumed to be a reliable estimator for $T_{0j}$ in the sense that
\[
\delta_j = \delta_j^\obs = \hat T_j - T_{0j},  \qquad j =1, \dots, m, 
\]
are random variables that are independent of $(\eps_1, \dots, \eps_m)$ and satisfy $\Exp[\delta_j]=0$ and $\sigma_j^2 = \sigma_{j,\obs}^2 = \Var(\delta_j) >0$. 
Conceptually, the $\delta_j$ represent uncertainty conditional on the data product. They capture internal variability as well as finite-sample and estimation effects, with $\sigma_j^2$ typically driven primarily by the sample size of the $j$th data product. Following \cite{otto24}, we refer to this source of uncertainty as \emph{natural variability}.
In contrast to the representation error variables $\eps_1, \dots, \eps_m$, the $\delta_1, \dots, \delta_m$ are not assumed to be independent; indeed, observations from different data products are typically strongly dependent, and this dependence propagates to the corresponding estimators. We do, however, assume independence between $(\eps_1,\dots,\eps_m)$ and $(\delta_1,\dots,\delta_m)$, which means that the within-product estimation uncertainty is assumed to be unrelated to the population-level discrepancy between a data product and modeling procedure and the true climate.

In summary, we obtain a \emph{random intercept model with heteroskedastic and dependent noise},
\[
\hat T_j - T_0 = \eps_j + \delta_j, \qquad j =1, \dots, m,
\]
where $\bm \eps = (\eps_1, \dots, \eps_m)$ has iid coordinates with mean 0 and variance $\tau^2$ and is independent of $\bm \delta =(\delta_1, \dots, \delta_m)$, a centered vector with possibly dependent coordinates with variance $\sigma_j^2 = \Var(\delta_j)$ that may depend on $j$.

Classical large-sample asymptotics suggest that bootstrap methods applied within the $j$th data product can be used to approximate certain summary characteristics of the distribution of $\delta_j = \hat T_j - T_{0j}$. 
Subsequently, we denote a bootstrap confidence interval of nominal level $1-\alpha$ by $\hat I_j=[\hat I_j^-, \hat I_j^+]$ \citep{Davison1997}.

\begin{figure}
    \centering
    \begin{minipage}[T]{.45\textwidth}
    \includegraphics[width=\linewidth]{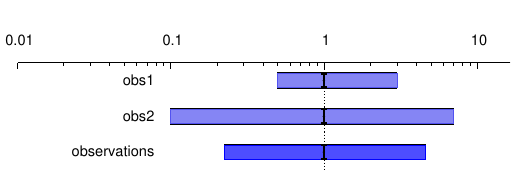}
    \end{minipage}
    \begin{minipage}[T]{.45\textwidth}
    {\includegraphics[width=\linewidth]
    {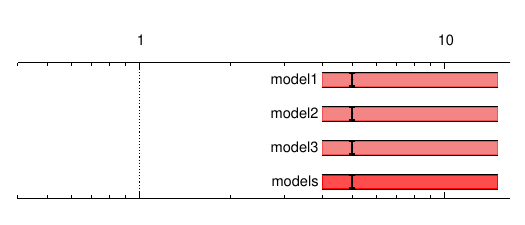}}
    \end{minipage}
    \caption{Synthesis computed in the KNMI Climate Explorer \citep{KNMIClimateExplorer} for mock data sets. Left: observational synthesis with $\hat T_1^\obs=\hat T_2^\obs=1$ and confidence intervals $\hat I_1^\obs=[0.5,3], \hat I_2^\obs=[0.1,7]$. Right: model synthesis with identical point estimates $\hat T_j^{\model}=5$ and confidence intervals $\hat I_j^{\model}=[4,15]$. 
    Despite increasing the number of contributing models, all supporting roughly the same hypothesis about $T$, the synthesized confidence interval does not contract, illustrating a key limitation of the current AML synthesis procedure.
    }
    \label{fig:climate-explorer-synthesis}
\end{figure}

The synthesis approach suggested in \cite{otto24}, visualized in the right hand side of Figure~\ref{fig:climate-explorer-synthesis}, is based on the estimated values $\hat T_j$ and the bootstrap-based confidence intervals $\hat I_j = [\hat I_j^-, \hat I_j^+]$ only. Adapting their notation, introducing several intermediate quantities for clarity, and denoting by $z_\beta$ the $\beta$-quantile of the standard normal distribution, the procedure first defines three estimators of $\sigma_j$,
\begin{align}
\label{eq:variance-estimators-fixed-j}
\hat \sigma_j = \frac{\hat I_j^+ - \hat I_j^-}{2 z_{1-\alpha/2}} ,
\qquad
\hat \sigma_j^{-} = \frac{\hat T_j-\hat I_j^-}{z_{1-\alpha/2}},
\qquad
\hat \sigma_j^{+} = \frac{\hat I_j^+- \hat T_j}{z_{1-\alpha/2}},
\end{align}
(note that the estimators coincide if $\hat I_j$ is a symmetric interval centered at $\hat T_j$),
and then uses
\begin{align} \label{eq:That-sobs}
\hat T_{\sobs}
=
\frac{1}{m} \sum_{j=1}^m \hat T_j,
\qquad 
\hat I_{\sobs} = [\hat I_{\sobs}^- , \hat I_{\sobs}^+]
\end{align}
as a synthesized, observation-based (sobs) estimator for $T_0$ and a synthesized $(1-\alpha)$-confidence interval for $T_0$, respectively, where 
\begin{align}
\label{eq:confidence-intervals-sobs}
\hat I_{\sobs}^\pm
&=
\hat T_{\sobs} \pm z_{1-\alpha/2}
\hat \sigma_{\obs}^{\pm},
\qquad 
(\hat \sigma_{\obs}^\pm)^2 =  (\hat s_{1}^{\pm})^2 + \hat s_2^2.
\end{align}
Here, 
$
\hat s_{1}^\pm =   m^{-1}
\sum_{j=1}^m
\hat \sigma_{j}^\pm$ and 
$
\hat s_2^2 = \widehat{\mathrm{Var}}(\hat T_1, \dots, \hat T_{m})
$ 
denotes
the empirical variance of the sample $\hat T_1, \dots, \hat T_{m}$ \citep[Formula (7)]{otto24}.

The validity of this approach can heuristically be demonstrated under additional normality assumptions. Indeed, if $\eps_j$ and $\delta_j$ (and hence $\hat T_j$) are approximately normal, confidence intervals are typically constructed using a symmetric normal approximation, that is, $\hat I_j^\pm = \hat T_j \pm z_{1-\alpha/2} \hat \sigma_j$, where $\hat\sigma_j^2$ estimates $\sigma_j^2$. The first identity in \eqref{eq:variance-estimators-fixed-j} then follows automatically.
For these choices, we obtain that 
$\hat \sigma_j^\pm = \hat \sigma_j$,
and hence $\hat s_{1}^+ = \hat s_{1}^- = m^{-1}
\sum_{j=1}^m \hat \sigma_{j} =: \hat s_1$,
i.e., $\hat s_1^\pm $ can be interpreted as an estimator for the average standard deviation $m^{-1} \sum_{j=1}^m \sigma_j$. Moreover, $\hat s_2^2$ is an upwardly biased estimator for $\tau^2$; indeed, since $\hat T_j - T_0 = \eps_j +\delta_j$, a straightforward calculation yields
$\E[\hat s_2^2] =\tau^2+(m-1)^{-1} \sum_{j=1}^m\E[(\delta_j-\bar\delta)^2]$,
where $\bar \delta$ denotes the empirical average over the $\delta_j$. 
Overall, $(\hat \sigma_{\obs}^\pm)^2 = (\hat s_{1}^\pm)^2 + \hat s_2^2$ should be considered as an estimator for a quantity that is lower bounded by $t_0^2 := (m^{-1} \sum_{j=1}^m \sigma_j)^2 + \tau^2$. On the other hand, 
$\Var( \hat T_{\sobs} )
=
    \Var( m^{-1} \sum_{j=1}^m \delta_j )  + \tau^2/m
    \le ( m^{-1}\sum_{j=1}^m \sigma_j)^2 +  \tau^2 = t_0^2,
$
where we have used the Cauchy-Schwarz inequality and the fact that $\tau^2/m \le \tau^2$; note that equality in the Cauchy-Schwarz step holds if and only if the variables $\delta_j$ are perfectly positively correlated. Consequently, both $(\hat \sigma_{\obs}^\pm)^2$ should be regarded as overestimating $\Var(\hat T_{\sobs})$. Therefore, $\hat I_{\sobs}$ yields a conservative confidence interval for $T_0$, whose coverage is expected to exceed $1-\alpha$, particularly for larger $m$ or weaker dependence among the $\delta_j$. We illustrate this phenomenon in Section~\ref{sec:simulation}, where we also consider the adapted interval based on replacing $\hat s_2^2$ by $\hat s_2^2/m$ in \eqref{eq:confidence-intervals-sobs}, that is, writing $(\tilde \sigma_{\sobs}^\pm)^2 =(\hat s_{1}^\pm)^2 + \hat s_2^2/m$,
\begin{align}\label{eq:confidence-intervals-sobs-modified}
    \tilde I_\sobs 
    &:= 
    \big[\tilde I_{\sobs}^-, \tilde I_{\sobs}^+\big]
    :=  
    \big[\hat T_{\sobs} - z_{1-\alpha/2}\tilde \sigma_{\sobs}^- ,
    \hat T_{\sobs}+z_{1-\alpha/2}
    \tilde \sigma_{\sobs}^+\big].
\end{align}

\subsection{Model Synthesis on the Level of Attribution Measures}
\label{subsec:model-synthesis-attribution-targets}

The second major line of evidence in EEA consists of estimates derived from climate model simulations. We denote these model-based estimators by $\hat T_1^{\model},\dots, \allowbreak \hat T_{m_\model}^{\model}$, where $m_\model$ denotes the number of climate models. As in the previous section, we write $m=m_\model$ and $\hat T_j =\hat T_j^\model$ for the ease of notation. We also assume that each model has a model-specific ground truth $T_{0j} = T_{0j}^\model$ that may be different from the ground truth measure $T_0$ due to systematic errors in the climate models or the statistical modeling procedure. It is assumed that the $T_{0j}$ are equal to $T_0$ on average over a hypothetical collection of climate models, which we capture mathematically by assuming that the representation error variables
\[
\eps_j := \eps_j^\model = T_{0j} - T_0
\]
form an iid sample from a generic representation error variable $\eps$ with $\E[\eps]=0$ and $\tau^2 = \tau_\model^2 = \Var(\eps)\in(0,\infty)$.
Next, the model-specific estimator $\hat T_j^{\model}$ is assumed to be an  unbiased estimator of the model-specific ground truth  $T_{0j}$. More specifically, the estimation error variables
\[
\delta_j =  \delta_j^\model = \hat T_j^{\model} - T_{0j}, 
\]
are assumed to form an independent sample, also independent from the $\eps_j$, with $\E[\delta_j] = 0$ and $\sigma_j^2 = \sigma_{j, \model}^2 =\Var(\delta_j)\in(0,\infty)$. Contrary to the observational synthesis in the previous section, we assume that the $\delta_j$ are independent. This is in fact a common working assumption in attribution studies, as data sets generated in different climate models may be treated as approximately independent \citep{otto24}. The assumption is visually supported by the apparent spread of block maxima extracted from two climate model runs in of Figure~\ref{fig:observations-vs-climate-models}.

In summary, we obtain the decomposition
\begin{align}
\label{eq:rem}
&\hat T_j = T_0 + \delta_j + \eps_j, \qquad j =1, \dots, m,
\end{align}
where $\bm \eps = (\eps_1, \dots, \eps_m)$ is a vector that has iid coordinates with mean 0 and variance $\tau^2$ that is independent of $\bm \delta =(\delta_1, \dots, \delta_m)$, a vector with independent coordinates with mean 0 and variance $\sigma_j^2 = \Var(\delta_j)$ that may depend on $j$. 

The fact that there is no dependence between the $\delta_j$ allows for more accurate synthesis. Indeed, under the assumption that $\sigma_j^2$ and $\tau^2$ are known, the weighted mean 
\begin{align*}
\hat T_w
=
\Big(\sum_{j=1}^m w_j\Big)^{-1}
\sum_{j=1}^m w_j \hat T_j,
\qquad
w_j = (\sigma_j^2 + \tau^2)^{-1},
\end{align*}
has minimal variance among all convex combinations of the individual estimates $\hat T_j$, namely,
\begin{align} \label{eq:variance-T-w}
\sigma_w^2 = \Var(\hat T_w)= \Big(\sum_{j=1}^m w_j\Big)^{-1}
= 
\Big(\sum_{j=1}^m\frac1{\sigma_j^2+\tau^2}\Big)^{-1}.
\end{align}

In practice, both $\sigma_j^2$ and $\tau^2$ are unknown and need to be estimated. As in the previous section, $\sigma_j^2$ may be estimated using the bootstrap within model $j$; see, for instance, \eqref{eq:variance-estimators-fixed-j} for the estimators suggested by \cite{otto24}. 
For $\tau^2$, several estimators have been proposed in the meta-analysis literature; see  Appendix~\ref{app:tau_estim} for a short survey. Throughout, we follow \citet[Equation (10)]{otto24} and employ an asymmetric version of the Paule-Mandel estimator \citep{paule1982consensus}.

Motivated by the previous derivations, \cite{otto24} suggest to use the synthesized model-based (smod) estimator 
\begin{align} \label{eq:That-smod}
\hat T_{\smod} := \hat T_{\hat w} = 
\Big(\sum_{j=1}^m \hat w_j\Big)^{-1}
\sum_{j=1}^m \hat w_j \hat T_j \quad \text { with } \quad \hat w_j = (\hat \sigma_{j}^2+\hat \tau^2)^{-1},
\end{align}
where $\hat \sigma_{j} = (\hat I_j^+ - \hat I_j^-)/(2z_{1-\alpha/2})$ is as in  \eqref{eq:variance-estimators-fixed-j} and where $\hat \tau^2$ is the aforementioned asymmetric version of the Paule-Mandel estimator. They also suggest the variance estimator
\[
\hat \sigma_{\smod}^2 = \frac{\sum_{j=1}^m \hat w_{j}(\hat \sigma_{j}^2 + \hat \tau^2)}{\sum_{j=1}^m \hat w_{j}} = m \cdot \hat \sigma_{\hat w}^2 ;
\]
see their Equation (5), but in view of \eqref{eq:variance-T-w}, this estimator is likely to substantially exceed its population counterpart.

Based on the same principles that were used in the previous section for deriving \eqref{eq:confidence-intervals-sobs}, \cite{otto24} propose the asymmetric confidence interval 
\begin{align}
\label{eq:confidence-intervals-smod}
\hat I_{\smod}  
:= [\hat I_{\smod}^-, \hat I_{\smod}^+]
:= 
[\hat T_{\smod} - z_{1-\alpha/2}\hat \sigma_{\smod}^-,\hat T_{\smod} + z_{1-\alpha/2}\hat \sigma_{\smod}^+], 
\end{align}
where 
\begin{align}
\label{eq:sigma2-smod}
    (\hat \sigma_{\smod}^\pm)^2 &= \hat \sigma_{\hat w}^2 \sum_{j=1}^m \hat w_{j} \big( (\hat \sigma_{j}^\pm)^2 + \hat \tau^2 \big)
\end{align}
with $\hat \sigma_{j}^\pm$ from \eqref{eq:variance-estimators-fixed-j}.
Although this interval is not stated explicitly in a citable formula in \citet[Section~2.4]{otto24}, it can be reconstructed either from the accompanying discussion or from the implementation provided by \cite{barnesrwwa2024}.
In the case where $\hat \sigma_{j}^\pm = \hat \sigma_{j}$, that is when the model-specific confidence intervals $\hat I_j$ are symmetric around $\hat T_j$, 
we obtain that $\hat \sigma_{\smod}^\pm = \hat \sigma_{\smod}$, and as this expression is likely to substantially exceed $\sigma_{w}$ as explained above, the confidence interval $\hat I_{\smod}$ is expected to be conservative, increasingly so as the number of models grows. This behavior is confirmed in the simulation study in Section~\ref{subsec:simulation-model-synthesis} and is further illustrated in Figure~\ref{fig:climate-explorer-synthesis} using a mock data set with $m=3$ models analyzed via the KNMI Climate Explorer: despite multiple independent sources of evidence supporting the same hypothesis about the attribution measure, the combined interval does not contract. This runs counter to the usual efficiency gains expected when aggregating independent information and reflects the conservativeness of the variance estimator.

In Section~\ref{subsec:simulation-model-synthesis}, we also show that the modified WWA confidence interval
\begin{align}
\label{eq:confidence-intervals-smod-modified}
  \tilde I_{\smod}  
    &:= [\tilde \ell_{\smod}, \tilde u_{\smod}]
    :=[\hat T_{\smod} - z_{1-\alpha/2} \tilde \sigma_{\smod,\mathrm{low}},\hat T_{\smod} + z_{1-\alpha/2}\tilde \sigma_{\smod,\mathrm{up}}]
\end{align}
with $\tilde \sigma_{\smod}^\pm = m^{-1/2} \hat \sigma_{\smod}^\pm$
may provide a substantially more accurate approximation of the nominal level, unless $m$ is very large.

\subsection{Full Synthesis on the Level of Attribution Measures}
\label{subsec:full-synthesis-attribution-targets}

The final step in the WWA synthesis framework consists of combining evidence from observational and climate model data into a single attribution statement. The respective approach described in \cite{otto24} is based on the estimators $\hat T_{\sobs}$ and $\hat T_{\smod}$ from \eqref{eq:That-sobs} and \eqref{eq:That-smod}, respectively, and on the confidence intervals $\hat I_{\sobs} = [\hat \ell_{\sobs}, \hat u_{\sobs} ]$ and $\hat I_{\smod} = [\hat \ell_{\smod}, \hat u_{\smod} ]$ from \eqref{eq:That-sobs} and \eqref{eq:confidence-intervals-smod} only.

More specifically, under the assumption of independence between $\hat T_{\sobs}$ and $\hat T_{\smod}$, the  principles explained in the previous section show that among all estimators of the form 
\[
\hat T_{w_1, w_2} = ( w_1 +w_2)^{-1} (w_1 \hat T_{\sobs} +w_2 \hat T_{\smod})
\]
the one with $w_1 = 1/\sigma^2_{\sobs}$ and $w_2 = 1/\sigma^2_{\smod}$ has lowest possible variance, where $\sigma^2_{\sobs} = \Var(\hat T_{\sobs})$ and $\sigma^2_{\smod} = \Var(\hat T_{\smod})$.  These variances are unknown in practice, and \citet[Formula (11)]{otto24} suggest to use the weights $w_1 = \hat w_{\sobs} $ and $w_2 = \hat w_{\smod} $ instead, where
\begin{align*}
\hat w_{s} 
= 
\big( \hat I_s^+ - \hat I_s^- \big)^{-2}
=
\big[ z_{1-\alpha/2} (\hat \sigma_{s}^- + \hat \sigma_{s}^+)\big]^{-2}, \qquad s \in \{\sobs, \smod\},
\end{align*}
with $\hat \sigma_{s}^\pm$ as defined in \eqref{eq:confidence-intervals-sobs} and \eqref{eq:sigma2-smod}.
We denote the resulting estimator by
\begin{align} \label{eq:That-sfull}
\hat T_{\sfull} = \hat T_{\hat w_{\sobs}, \hat w_{\smod}}.
\end{align}

To relate these weights to conventional inverse-variance weighting, consider the special case in which both confidence intervals are symmetric. Their weights are then proportional to the corresponding inverse estimated variances, with the same proportionality constant $(2z_{1-\alpha/2})^{-2}$. As this common factor has no effect on \eqref{eq:That-sfull}, one could equivalently normalize each interval width by $2z_{1-\alpha/2}$ and redefine $\hat w_{s}  = \hat\sigma_{s}^{-2}$.

The more consequential issue concerns the variance estimates themselves. As explained in the previous two sections, both $\hat\sigma_{\sobs}^2$ and $\hat\sigma_{\smod}^2$ are likely to substantially exceed their population counterparts. Consequently, the relative weights proposed by \cite{otto24} need not approximate the variance-minimizing weights $\sigma_{\sobs}^{-2}$ and $\sigma_{\smod}^{-2}$ well. We illustrate this phenomenon in Section~\ref{sec:simulation}, where we also consider the adapted estimator
\begin{align} \label{eq:That-sfull-adapted}
\tilde T_{\sfull} = \hat T_{\tilde w_{\sobs}, \tilde w_{\smod}}, \qquad
\tilde w_{s}  =  \Big( \frac{\tilde I_{s}^+ - \tilde I_{s}^-}{2z_{1-\alpha/2}} \Big)^{-2}, \qquad s \in \{\sobs, \smod\}.
\end{align}
These weights are based on the modified confidence intervals from \eqref{eq:confidence-intervals-sobs-modified} and \eqref{eq:confidence-intervals-smod-modified}, respectively. Because the normalization by $2z_{1-\alpha/2}$ is common to both weights, it could equivalently be omitted without changing $\tilde T_{\sfull}$. Its inclusion nevertheless makes explicit that the weights correspond to inverse squared uncertainty scales rather than merely to inverse squared interval widths.
The adapted estimator $\tilde T_{\sfull}$ is shown to have substantially smaller mean squared error than the original WWA estimator $\hat T_{\sfull}$; see  Table~\ref{tab:simulation-results-full-synthesis}.

\citet{otto24} additionally propose the synthesized confidence interval
\begin{align}
\label{eq:confidence-intervals-sfull}
\hat I_{\sfull} 
:= [\hat I_{\sfull}^-, \hat I_{\sfull}^+]
:= 
[\hat T_{\sfull} - z_{1-\alpha/2}\hat \sigma_{\sfull}^-,\hat T_{\sfull} + z_{1-\alpha/2}\hat \sigma_{\sfull}^+], 
\end{align}
where
\begin{align}
\label{eq:sigma-sfull}
(\hat \sigma_{\sfull}^\pm)^2
&= \frac{ (\hat \sigma_{\sobs}^\pm)^2  (\hat \sigma_{\sobs}^- + \hat \sigma_{\sobs}^+)^{-2} + (\hat \sigma_{\smod}^\pm)^2  (\hat \sigma_{\smod}^- + \hat \sigma_{\smod}^+)^{-2}}{ (\hat \sigma_{\sobs}^- + \hat \sigma_{\sobs}^+)^{-2} + (\hat \sigma_{\smod}^- + \hat \sigma_{\smod}^+)^{-2} }.
\end{align}
Here, we have corrected an apparent typographical error in equations~(12) and~(13) of \cite{otto24}: the denominator inside the square root should not be squared. The corrected version is also used in the corresponding implementation of \cite{barnesrwwa2024}.

To assess the proposed interval, consider again the special case in which the observational and climate-model confidence intervals are symmetric. In that case,
$
(\hat \sigma_{\sfull}^\pm)^2
=
2(\hat \sigma_{\sobs}^{-2} + \hat \sigma_{\smod}^{-2})^{-1}.
$
By contrast, the variance of the inverse-variance-weighted mean is $(\sigma_{\sobs}^{-2} + \sigma_{\smod}^{-2})^{-1}$.
Thus, even in the symmetric case, the variance estimate proposed by \cite{otto24} is inflated by a factor of $2$. This suggests that the resulting confidence interval is systematically too wide and hence conservative, as illustrated in Section~\ref{sec:simulation}, see Table~\ref{tab:simulation-results-full-synthesis}.

We therefore propose the following modified confidence interval
\begin{align}
\label{eq:confidence-intervals-sfull-adapted}
\tilde I_{\sfull} 
&:= 
[\tilde I_{\sfull}^-, \tilde I_{\sfull}^+]
:=
[\tilde T_{\sfull} - z_{1-\alpha/2}\tilde \sigma_{\sfull}^-,\tilde T_{\sfull} + z_{1-\alpha/2} \tilde \sigma_{\sfull}^+]
\end{align}
that is defined as in \eqref{eq:confidence-intervals-sfull}, but with $\hat T_{\sfull}$ replaced by $\tilde T_{\sfull}$, and with $\hat \sigma_{\sfull}^\pm$ replaced by 
\begin{align}
\label{eq:sigma-sfull-adapted}
(\tilde \sigma_{\sfull}^\pm)^2
&= 
\frac12 \frac{ (\tilde \sigma_{\sobs}^\pm)^2  (\tilde \sigma_{\sobs}^- + \tilde \sigma_{\sobs}^+)^{-2} + (\tilde \sigma_{\smod}^\pm)^2  (\tilde \sigma_{\smod}^- + \tilde \sigma_{\smod}^+)^{-2}}{ (\tilde \sigma_{\sobs}^- + \tilde \sigma_{\sobs}^+)^{-2} + (\tilde \sigma_{\smod}^- + \tilde \sigma_{\smod}^+)^{-2} }.
\end{align}
using $\tilde \sigma_{\sobs}^\pm$ from \eqref{eq:confidence-intervals-sobs-modified} and $\tilde \sigma_{\smod}^\pm$ from \eqref{eq:confidence-intervals-smod-modified}, respectively. The additional factor $1/2$ corrects the inflation identified above.

\section{Synthesis on the Level of Model Parameters}
\label{sec:synthesis-model-parameters}

In this section, we develop an alternative synthesis approach based on combining information at the level of model parameters rather than attribution measures, referred to as parameter-level (PL) synthesis. As in the AML case, we proceed in three steps, beginning with observational data, then turning to climate-model data, and finally combining the two sources of information. The data setting is the same as in Section~\ref{sec:synthesis-attribution-targets}; but different underlying assumptions yield novel synthesis methods.

\subsection{Observational Synthesis on the Level of Model Parameters}
\label{subsec:observational-synthesis-model-parameters}

We begin by recalling the data setting. From each of the $m = m_\obs \in \N$ observational data products, we extract a time series $(X_{jt})_{t \in \mathcal I_j}$ of the target variable over a sampling period $\mathcal I_j$ of length $n_j = |\mathcal I_j|$, where $j \in \{1, \dots, m\}$. In addition, we observe GMST anomalies $(g_t)_{t \in \mathcal I}$ over the combined observation period $\mathcal I = \mathcal I_1 \cup \dots \cup \mathcal I_m$. As before, the restriction $m\ge 2$ reflects that no synthesis is required when only a single observational data product is available.

To account for potential systematic discrepancies between the observational products and the unobservable target process, we introduce representation errors. In contrast to Section~\ref{subsec:observational-synthesis-attribution-targets}, where these discrepancies were modeled at the level of the attribution measures, we now formulate them at the level of the underlying distributional regression parameters. This allows the synthesis to exploit more fully the parametric model structure already used in the preceding WWA approach to obtain the individual attribution estimates $\hat T_j$.

Specifically, we assume that the (unobservable) target variable $X$ satisfies
\(
( X \mid G = g) \sim H_{f(\vartheta, g)},
\)
for some parametric distribution family $\{H_\theta: \theta \in \Phi\}$ (e.g., the GEV from Example~\ref{example:GEV}), some known link function $f:\Theta \times \mathscr G \to \Phi$, and some unknown ground truth parameter $\vartheta \in\Theta \subseteq \R^d$.  
Writing $X_j$ for a generic observation of the target variable in the $j$th data product, we correspondingly assume that
\[
( X_j \mid G = g) \sim H_{f(\vartheta_j, g)},
\]
for some data-product specific ground truth parameter $\vartheta_j=\vartheta_j^\obs$. Thus, in contrast to the AML synthesis, the distributional regression model serves not merely as a working model for estimating the attribution target $T$, but as an explicit component of the synthesis model itself.
The associated \emph{representation error variables}
\[
\eps_j = \eps_j^{\obs} = \vartheta_j - \vartheta, \qquad j =1, \dots, m,
\]
are assumed to form an iid sample with $\Exp[\eps_j] = 0$ and with existing covariance matrix $\Xi := \Xi^\obs := \Cov(\eps_j) \in \R^{d \times d}$. Their motivation is analogous to that of the representation errors from Section~\ref{subsec:observational-synthesis-attribution-targets}.

We acknowledge that the above formulation assumes that the conditional distributions of both the true target variable and the observational data products are adequately represented by the family $\{H_{f(\vartheta,g)}:\vartheta\in\Theta\}$. Potential misspecification of the distributional regression model is therefore not explicitly accounted for. This assumption is stronger than the centeredness condition imposed in Section~\ref{subsec:observational-synthesis-attribution-targets}: at the level of attribution measures, model misspecification may be absorbed into the representation errors, whose centeredness merely requires that the combined effect of misspecification and other sources of discrepancy does not induce a systematic deviation across the hypothetical collection of data products. By contrast, the present PL formulation presumes that the model family itself provides an adequate description of the relevant conditional distributions. In particular, a systematic misspecification shared by all data products would not be captured by the centered PL representation errors. This limitation should be kept in mind when interpreting the resulting synthesis.

For each data product, we furthermore assume the availability of an estimator $\hat \vartheta_j = \hat \vartheta_j^\obs$ that is reliable for $\vartheta_j$ in the sense that the vector $\bm\delta = (\delta_1^\top, \dots, \delta_m^\top)^\top \in \R^{md}$ with entries
\[
\delta_j = \delta_j^{\obs} = \hat \vartheta_j - \vartheta_j, \qquad j =1, \dots, m,
\]
independent of $(\eps_1, \dots, \eps_m)$ with $\Exp[\delta_j] = 0$ and existing covariance matrix $\Sigma \in \R^{md \times md}$ containing block entries $\Sigma_{ij} := \Sigma_{ij}^\obs := \Cov(\delta_i, \delta_j) \in \R^{d \times d}$ for $i,j \in \{1, \dots, m\}$. It follows that the stacked vector $\hat {\bm \vartheta} = (\hat \vartheta_1^\top, \dots, \hat \vartheta_m^\top)^\top \in \R^{md}$ has expectation $(\vartheta^\top, \dots, \vartheta^\top)^\top$ and covariance% matrix
\begin{align}
\label{eq:V}
    V 
    := 
    \Cov(\hat {\bm \vartheta}) 
    = 
    \Sigma + \mathds I_{m \times m} \otimes \Xi = 
    \begin{pmatrix}
        \Sigma_{11} + \Xi & \Sigma_{12} & \dots & \Sigma_{1m}\\
        \Sigma_{21} & \Sigma_{22} + \Xi & \dots & \Sigma_{2m}\\
        \vdots & & \ddots & \vdots \\
        \Sigma_{m1} & \Sigma_{m2} & \dots & \Sigma_{mm} + \Xi
    \end{pmatrix}
    \in \R^{md\times md},
\end{align}
where $\mathds I_{m \times m}$ is the $(m\times m)$-identity matrix, where $\Sigma=(\Sigma_{ij})_{i,j=1}^m \in \R^{md \times md}$ and where $\otimes$ denotes the Kronecker product.
Since we can write $\hat {\bm \vartheta}= \boldsymbol{X}\vartheta + \bm\varepsilon+\bm\delta$ with design matrix $\bm X= \mathds{1}_m \otimes \mathds{I}_{d\times d} \in \R^{md\times d}$ and $\bm\varepsilon = (\eps_1^\top, \dots, \eps_m^\top)^\top$, 
it follows from the classical Gauss-Markov theorem that among all unbiased linear estimators of the form $\hat \vartheta_W = W \hat{\bm \vartheta}$ for some $W \in \R^{d \times md}$, %that are unbiased for $\vartheta$, 
the generalized least squares (GLS) oracle estimator corresponding to
\begin{align} \label{eq:GLS-weights}
    W = \big(\boldsymbol{X}^\top V^{-1}\boldsymbol{X}\big)^{-1}\boldsymbol{X}^\top V^{-1}
\end{align}
has minimal covariance matrix with respect to the Loewner-order for symmetric matrices.

In practice, $\hat \vartheta_W$ is infeasible, since it requires knowledge of $V$ and, consequently, of $\Sigma$ and~$\Xi$. We therefore estimate these quantities as follows. The matrix $\Sigma$ can be estimated by bootstrapping all observational products jointly. First, for each bootstrap iteration $b=1,\dots,B$ (with $B$ large, e.g., $B=1{,}000$), we sample  $n:=|\mathcal I|$ time indices $t_1^{*b}, \dots, t_{n}^{*b}$ from $\mathcal I$ with replacement, resulting in a multiset $\mathcal I^{*b} := \{ t_1^{*b}, \dots, t_{n}^{*b}\}$.
The corresponding observational bootstrap sample is defined as $\{ (g_t, X_{1t},\dots, X_{mt}): t \in \mathcal I^{*b}\}$, where
values of $X_{jt}$ with $t\in \mathcal{I}\setminus \mathcal{I}_j$ are recorded as missing. Next, for every $j$, an estimator $\hat\vartheta_j^{*b}$ for $\vartheta_j$ is computed based on the sample $\{(g_t, X_{jt}): t \in \mathcal I^{*b}\}$ and stacked to $\hat {\bm \vartheta}{}^{*b} = ((\hat \vartheta_1^{*b})^\top, \dots, (\hat \vartheta_m^{*b})^\top)^\top \in \R^{md}$. Finally, the empirical covariance matrix of $\hat {\bm \vartheta}{}^{*1},\dots, \hat {\bm \vartheta}{}^{*B}$ is used to estimate $\Sigma$, namely
\begin{align}\label{eq:hat-Sigma-obs}
    \hat \Sigma = \frac1B \sum_{b=1}^B \big(\hat {\bm \vartheta}{}^{*b}- \overline{\hat {\bm \vartheta}{}^{*1:B}}\big) \big(\hat {\bm \vartheta}{}^{*b}- \overline{\hat {\bm \vartheta}{}^{*1:B}}\big)^\top, \qquad  
    \overline{\hat {\bm \vartheta}{}^{*1:B}}:= \frac1B \sum_{b=1}^B  \hat {\bm \vartheta}{}^{*b}.
\end{align}
Secondly, we need to estimate $\Xi$, for which we use the estimator $\hat \Xi = \hat \Xi(\hat{\bm\vartheta}, \hat\Sigma)$, where
\begin{align}
    \hat \Xi(\hat{\bm\vartheta}, \hat\Sigma)
    &:=
    \frac{1}{m-1}\sum_{j=1}^m \big( \hat \vartheta_j- \overline{\hat \vartheta_{1:m}}\big) \big( \hat \vartheta_j- \overline{\hat \vartheta_{1:m}}\big)^\top 
    - 
    \frac{1}{m} \sum_{j=1}^m \Big(\hat\Sigma_{jj}-\frac{1}{m-1}\sum_{i\ne j} \hat\Sigma_{ij}\Big)\label{eq:xi-hat-obs}, 
\end{align}
where $\overline{\hat \vartheta_{1:m}} = m^{-1}\sum_{j=1}^m \hat \vartheta_j$.
We motivate this estimator as a method-of-moments estimator in Appendix~\ref{subsec:estimating-Xi}. Since $\hat\Xi$ is not guaranteed to be positive semidefinite, we replace it, whenever necessary, by the nearest positive semidefinite matrix obtained by truncating all eigenvalues below a small threshold $\kappa\geq 0$ to $\kappa$; in the applications below, we set $\kappa=10^{-12}$. The same regularization is applied subsequently to all estimated covariance matrices for which positive semidefiniteness or numerical invertibility is not guaranteed.

In view of \eqref{eq:V},~\eqref{eq:hat-Sigma-obs} and~\eqref{eq:xi-hat-obs}, we obtain the plug-in estimator $\hat V:=\hat\Sigma + \mathds I_{m \times m}\otimes\hat \Xi$, which results in the feasible observational synthesis estimator 
\begin{align}\label{eq:hatvartheta_sobs}
    \hat\vartheta_\sobs:=   \big(\boldsymbol{X}^\top \hat V^{-1}\boldsymbol{X}\big)^{-1}\boldsymbol{X}^\top \hat V^{-1} \hat{\bm \vartheta}
\end{align}
with $\bm X= \mathds{1}_m \otimes \mathds{I}_{d\times d} \in \R^{md\times d}$. 
Finally, applying the plug-in principle yields a corresponding estimator of the attribution measure $T$, denoted by $\hat T_{\sobs}$; see Appendix~\ref{app:model-parameters-to-attribution-targets} for details.

The approach used to estimate $\hat\Sigma$ also provides a natural
starting point for uncertainty quantification in the form of a bootstrap sample $\hat\vartheta_\sobs^{*1},\dots,\hat\vartheta_\sobs^{*B}$ of the observational synthesis estimator.
Specifically, we propose to keep $\hat \Sigma$ fixed across bootstrap iterations and define
\[
\hat\vartheta_\sobs^{*b}
=
\big(\bm X^\top(\hat V^{*b})^{-1}\bm X\big)^{-1}
    \bm X^\top(\hat V^{*b})^{-1}
    \hat {\bm \vartheta}{}^{*b}
\]
where 
$
    \hat V^{*b}
    =
    \hat\Sigma+\mathds I_{m\times m}\otimes\hat\Xi^{*b}$ and $
    \hat\Xi^{*b}
    =
    \hat\Xi(\hat {\bm \vartheta}{}^{*b},\hat\Sigma).
$
The method is summarized in Algorithm~\ref{algo:boot_obs}. Keeping $\hat\Sigma$ fixed when computing $\hat\Xi^{*b}$ avoids a computationally demanding double bootstrap. Hence, the procedure should be interpreted as a feasible plug-in bootstrap approximation that accounts for the dominant sources of uncertainty.

\subsection{Model Synthesis on the Level of Model Parameters}
\label{subsec:model-synthesis-model-parameters}

Recall the data setting: from each of the $m=m_\model \in \N$ climate models, we extract GMST-target variable pairs $(g_{jt}, X_{jt})_{t \in \mathcal J_j}$.
Here, the index sets $\mathcal J_j$ are allowed to be multisets, reflecting the fact that several ensemble members may be available for a given climate model, which leads to repeated observations at identical time points. Writing $(G_j,X_j)$  for a generic observation pair, we assume that
\[
(X_{j} \mid G_j = g) \sim  H_{f(\vartheta_j, g)}
\]
for some model-specific ground truth $\vartheta_j=\vartheta_j^{\model}$. Following the constructions from the previous sections, we further assume that
\[
\hat \vartheta_j = \vartheta + \delta_j + \eps_j, \qquad j=1, \dots, m,
\]
where the representation error variables $\eps_j=\vartheta_j-\vartheta$ are iid with mean zero and covariance matrix $\Xi$, while the within-model error variables $\delta_j=\hat \vartheta_j - \vartheta_j$ have mean zero and covariance matrix $\Sigma_j$. As in the model-based synthesis approach on the level of attribution measures from Section~\ref{subsec:model-synthesis-attribution-targets}, we assume that $\eps_1, \dots, \eps_m, \delta_1, \dots, \delta_m$ are mutually independent; we refer to that section for further motivation of this assumption.

The independence assumption between the $\delta_j$ yields a simplification for the GLS oracle estimator $\hat\vartheta_W = W \hat {\bm \vartheta}$ with $W$ from \eqref{eq:GLS-weights}, which simply becomes 
\begin{align}
\label{eq:hatvartheta_W}
\hat \vartheta_{\Sigma,\Xi}
=
\Big(
\sum_{j=1}^m (\Sigma_j+\Xi)^{-1}
\Big)^{-1}
\sum_{j=1}^m
(\Sigma_j+\Xi)^{-1}\hat\vartheta_j,
\end{align}
with respective covariance matrix $\Cov(\hat \vartheta_{\Sigma,\Xi}) = \big[ 
\sum_{j=1}^m (\Sigma_j+\Xi)^{-1}
\big]^{-1}$.
% \[
% \Cov(\hat\vartheta_W) = \Big(
% \sum_{j=1}^m (\Sigma_j+\Xi)^{-1}
% \Big)^{-1}.
% \]
In practice, $\Sigma_j$ and $\Xi$ are unknown and must be estimated from the data. We estimate $\Sigma_j$ using the nonparametric bootstrap applied within the $j$th model: for each bootstrap replicate $b=1, \dots, B$, we sample with replacement $n_j = |\mathcal J_j|$ times from $(g_{jt}, X_{jt})_{t \in \mathcal J_j}$, and calculate the respective estimator $\hat \vartheta_j$ on the bootstrap sample. Denote the estimated value by $\hat \vartheta_j^{*b}$, and let $\hat \Sigma_j^{[\mathrm{prelim}]}$ denote the empirical covariance of $\hat \vartheta_j^{*1}, \dots, \hat \vartheta_j^{*B}$:
\[
\hat \Sigma_j^{[\mathrm{prelim}]} 
= 
\frac1{B-1} \sum_{b=1}^B 
\big (\hat \vartheta_j^{*b} -  \overline{\hat \vartheta{}^{*B}} \big) \big (\hat \vartheta_j^{*b} -  \overline{\hat \vartheta{}^{*B}} \big) ^\top, 
\quad \text{ where} \quad  
\overline{\hat \vartheta{}^{*B}} = \frac1B \sum_{b=1}^B 
\hat \vartheta_j^{*b}.
\]
In principle, one could use the unmodified preliminary estimator $\hat \Sigma_j^{[\mathrm{prelim}]}$. However, this introduces a weighting bias, which can be explained heuristically as follows. The larger an estimate from the $j$th sample is, in a matrix sense, the larger its estimation variance tends to be and, consequently, the smaller its weight in the combined GLS estimator. This mechanism induces a downward bias, which is especially visible in the estimation of the scale parameter.

This issue can be avoided by imposing an additional, natural structural assumption on the covariance matrices. Specifically, suppose that there exists a reference matrix  $\Sigma_{\mathrm{ref}}$ such that $\Sigma_j = |\mathcal J_j|^{-1}\Sigma_{\mathrm{ref}}$. The rationale is that the covariance matrix $\Sigma_j$ is mainly determined by the available sample size $|\mathcal J_j|$ in the $j$th model, while its shape is approximately shared across models. Under this assumption, the rescaled preliminary estimators $|\mathcal J_j|\hat \Sigma_j^{\mathrm{prelim}}$ all target the same reference matrix, suggesting the estimator 
\begin{align*}
    \hat \Sigma_j := \frac{1}{|\mathcal J_j|} \hat \Sigma_{\mathrm{ref}}, 
    \qquad 
    \hat\Sigma_{\mathrm{ref}} 
:=
\frac{1}{m}\sum_{j=1}^m
|\mathcal J_j| \hat \Sigma_j^{\mathrm{prelim}}.
\end{align*}
This construction decouples the estimated covariance shape from the random magnitude of the preliminary estimates and therefore avoids the mechanism by which large estimates in the $j$th sample would automatically receive smaller weights in the combined GLS estimator.

Finally, for estimating $\Xi$, we apply the multivariate DerSimonian-Laird estimator from \eqref{eq:multivariate-DerSimonian-Laird}, using $(\hat \vartheta_j, \hat \Sigma_j)$ instead of $(X_j,  \Sigma_j)$. 
The final estimator, obtained by replacing $(\Sigma_j, \Xi)$ in \eqref{eq:hatvartheta_W} by $(\hat \Sigma_j, \hat \Xi)$, is 
\begin{align}
\label{eq:hatvartheta_smod}
\hat \vartheta_{\smod} := \hat \vartheta_{\hat \Sigma, \hat \Xi}
=
\Big(
\sum_{j=1}^m (\hat \Sigma_j+\hat \Xi)^{-1}
\Big)^{-1}
\sum_{j=1}^m
(\hat \Sigma_j+ \hat \Xi)^{-1}\hat\vartheta_j.
\end{align}
Applying the plug-in principle then yields a corresponding estimator of the attribution measure~$T$, denoted by $\hat T_{\smod}$; see Appendix~\ref{app:model-parameters-to-attribution-targets} for details.

The uncertainty of $\hat \vartheta_{\smod}$ and $\hat T_{\smod}$ is assessed using hierarchical bootstrap methods; see Section~3.8 of \cite{Davison1997} for background. We describe two such bootstrap schemes below. Both generate bootstrap replicates $\hat \vartheta_{\smod}^{*1}, \dots , \hat \vartheta_{\smod}^{*B}$, which may, for instance, be used to estimate the covariance matrix of $\hat \vartheta_{\smod}$. In addition, they induce bootstrap replicates
$\hat T_{\smod}^{*1}, \dots , \hat T_{\smod}^{*B}$, which, together with the original estimate $\hat T_{\smod}$, can be used to construct bootstrap confidence intervals \citep[Section 5.3.1]{Davison1997}.

\smallskip
\noindent
\emph{Strategy 1: Group bootstrap.} 
For each bootstrap replicate $b=1, \dots, B$ (with $B$ large), we draw, with replacement, $m$ pairs from
$(\hat \vartheta_1, \hat \Sigma_1), \dots, (\hat \vartheta_m, \hat \Sigma_m)$. Let $(\hat \vartheta_1^{*b}, \hat \Sigma_1^{*b}), \dots,\allowbreak (\hat \vartheta_m^{*b}, \hat \Sigma_m^{*b})$ denote the resulting bootstrap sample.
Based on this sample, we compute the multivariate DerSimonian--Laird estimator \eqref{eq:multivariate-DerSimonian-Laird}, denoted by $\hat\Xi^{*b}$. Finally, we obtain $\hat\vartheta_{\smod}^{*b}$ from \eqref{eq:hatvartheta_W} by replacing $(\vartheta_j,\Sigma_j,\Xi)$ with $(\hat\vartheta_j^{*b},\hat\Sigma_j^{*b},\hat\Xi^{*b})$, which in turn yields $\hat T_{\smod}^{*b}$. The procedure is summarized in Algorithm~\ref{algo:boot_model_group}.

Strictly speaking, this approach involves a computational simplification. Since the covariance estimators $\hat\Sigma_j$ are themselves obtained via bootstrap procedures, a fully nested bootstrap would require recomputing each $\hat\Sigma_j$ within every outer bootstrap replicate based on resampled raw data. To avoid this substantial computational burden, we instead treat the estimated covariance matrices $\hat\Sigma_j$ as fixed throughout the outer bootstrap procedure described above.

\smallskip
\noindent
\emph{Strategy 2: Hybrid group parametric bootstrap.} 
For each bootstrap replicate $b=1, \dots, B$, we draw $m$ triples with replacement from 
$
(\hat\vartheta_1,\hat\Sigma_1, (g_{1t})_{t \in \mathcal J_1}), \dots, (\hat\vartheta_m,\hat\Sigma_m, (g_{mt})_{t \in \mathcal J_m}).
$
Denote the resulting sample by  $(\hat \vartheta_1^{*b}, \hat \Sigma_1^{*b},(g_{1t}^{*b})_{t \in \mathcal J_1^{*b}}), \dots, (\hat \vartheta_m^{*b}, \hat \Sigma_m^{*b},(g_{mt}^{*b})_{t \in \mathcal J_m^{*b}})$. For each $j \in \{1, \dots, m\}$, simulate new observations from the target variable using the parametric model:
$X_{jt}^{*b} \sim H_{f(\hat \vartheta_j^{*b}, g_{jt}^{*b})}$ for $t \in \mathcal J_j^{*b}$.
Write $\hat \vartheta_{j}^{**b}$ for the re-estimate of the parameter vector based on the sample $(g_{jt}^{*b}, X_{jt}^{*b})_{t \in \mathcal J_j^{*b}}$. We then re-estimate $\Xi$ using the multivariate DerSimonian--Laird estimator from \eqref{eq:multivariate-DerSimonian-Laird}, with $(X_j,  \Sigma_j)$ replaced by $(\hat \vartheta_j^{**b}, \hat \Sigma_j^{*b})$ and denote the resulting estimate by $\hat \Xi^{*b}$. Finally, we compute  $\hat\vartheta_{\smod}^{*b}$ as in \eqref{eq:hatvartheta_W} with $(\vartheta_j, \Sigma_j, \Xi)$  replaced by $(\hat\vartheta_j^{**b},\hat\Sigma_j^{*b},\hat\Xi^{*b})$, and let $\hat T_{\smod}^{*b} = T(\hat\vartheta^{*b}_{\smod})$. The procedure is summarized in Algorithm~\ref{algo:boot_model_hybrid}.

\subsection{Full Synthesis on the Level of Model Parameters}
\label{subsec:full-synthesis-model-parameters}

It remains to combine the observational and model syntheses from the previous two sections. For the estimation step, we proceed analogously to the WWA synthesis framework described in Section~\ref{subsec:full-synthesis-attribution-targets}; the uncertainty quantification, however, will be handled differently.

Assuming that $\hat\vartheta_{\sobs}$ and $\hat\vartheta_{\smod}$ are unbiased, independent, and have covariance matrices $\Sigma_{\sobs}$ and $\Sigma_{\smod}$, respectively, the GLS estimator for $\vartheta$ is given by the precision-weighted average
$
(\Sigma_{\sobs}^{-1}+ \Sigma_{\smod}^{-1})^{-1} (\Sigma_{\sobs}^{-1} \hat \vartheta_{\sobs} + \Sigma_{\smod}^{-1} \hat\vartheta_{\smod})$. This motivates the following fully synthesized estimator:
\begin{align}
\label{eq:hatvartheta_sfull}
\hat \vartheta_{\sfull} 
= 
\big(\hat\Sigma_{\sobs}^{-1}+\hat\Sigma_{\smod}^{-1}\big)^{-1}
\big(\hat\Sigma_{\sobs}^{-1}
\hat\vartheta_{\sobs}
+\hat\Sigma_{\smod}^{-1}
\hat\vartheta_{\smod}\big),
\end{align}
where, motivated by \eqref{eq:hatvartheta_sobs} and \eqref{eq:hatvartheta_smod}, 
\begin{align*}
    \hat\Sigma_{\sobs}&= \big(\boldsymbol{X}^\top \hat V^{-1}\boldsymbol{X}\big)^{-1}, \qquad 
    \hat\Sigma_{\smod}=\Big(
\sum_{j=1}^m (\hat \Sigma_j+\hat \Xi)^{-1}
\Big)^{-1}.
\end{align*}

The uncertainty of $\hat T_{\sfull}$ is assessed using the following bootstrap procedure: first, run both the observational synthesis bootstrap from Section~\ref{subsec:observational-synthesis-model-parameters} (Algorithm~\ref{algo:boot_obs}) and the model synthesis bootstrap from Section~\ref{subsec:model-synthesis-model-parameters} (Algorithm~\ref{algo:boot_model_group} or~\ref{algo:boot_model_hybrid}). This results in bootstrap estimates $\hat\vartheta_{\sobs}^{*1}, \dots, \hat\vartheta_{\sobs}^{*B}$ and   $\hat V^{*1}, \dots ,\hat V^{*B}$ for observational synthesis, as well as $\hat\vartheta_{\smod}^{*1}, \dots, \hat\vartheta_{\smod}^{*B}$ and $(\hat\Xi^{*b}, \hat\Sigma_{1}^{*b} ,\dots, \hat\Sigma_{m}^{*b})_{b=1}^B $ for model synthesis. The fully synthesized $b$-th bootstrap replicate is then
\[
\hat \vartheta_{\sfull}^{*b}
= 
\big((\hat\Sigma_{\sobs}^{*b})^{-1}+(\hat\Sigma_{\smod}^{*b})^{-1}\big)^{-1}
\big((\hat\Sigma_{\sobs}^{*b})^{-1}
\hat\vartheta_{\sobs}^{*b}
+(\hat\Sigma_{\smod}^{*b})^{-1}
\hat\vartheta_{\smod}^{*b}\big),
\]
where
$\hat \Sigma_{\sobs}^{*b} = (\bm X^\top (\hat V^{*b})^{-1} \bm X)^{-1}$ and 
$\hat\Sigma_{\smod}^{*b} = \{ \sum_{j=1}^m (\hat\Sigma_j^{*b}+\hat\Xi^{*b})^{-1}\}^{-1}.$
Bootstrap replicates for the estimated attribution measure are finally $\hat T_{\sfull}^{*b} = T(\hat\vartheta^{*b}_{\sfull})$. The method is summarized in Algorithm~\ref{algo:boot_full}.

\section{Monte Carlo Simulation Study}
\label{sec:simulation}

The finite-sample performance of the synthesis approaches introduced in the previous sections is assessed through a controlled Monte Carlo simulation study. We consider both systematically varied input parameters, such as sample sizes and ensemble counts, and a design calibrated to the case study in Section~\ref{sec:case-study}. For brevity, the main text reports only the latter design, which is described in Section~\ref{subsec:setup}; results for the systematically varied settings are provided in Appendix~\ref{app:additional-simulation-results}.
We further focus on the results for the full synthesis, presented in Section~\ref{subsec:results}. Detailed results for observational and model-based synthesis are deferred to Appendices~\ref{subsec:simulation-observational-synthesis} and~\ref{subsec:simulation-model-synthesis}, respectively.

\subsection{Setup}
\label{subsec:setup}
The general setup follows Section~\ref{sec:fundamentals}, particularly Example~\ref{example:GEV}, with the true target variable $X$ representing annual maxima of cumulative daily precipitation amounts. 
%For additional implementation details, we refer to Appendix~\ref{sec:simulation-details}.

\subsubsection{Ground truth and attribution measure}  \label{subsec:setup:groundtruth}
 We assume that, conditionally on the GMST value $G=g$, the distribution of $X$ is GEV with parameters linked to $g$ through the scale-link function in \eqref{eq:linkfunc_scale}, that is,
\begin{align}
\label{eq:simulation-ground-truth-gev}
(X \mid G = g)  \sim \mathrm{GEV}\big(f_{\mathrm{scale}}(\vartheta_0, g)\big).
\end{align}
The true parameter vector $\vartheta_0=(\gamma_0,\mu_0,\sigma_0,\alpha_0)$ is set to $\vartheta_0=(0.1, 10, 2, 0.1)$, which is close to the estimated parameter in the case study in Section~\ref{sec:case-study}. As attribution measure, we consider the logarithmic  probability ratio $\log \PR= \log \PR(x_0)$ 
at the event threshold $x_0=28.1$, corresponding to the extreme event analyzed in the case study. The factual GMST value is fixed at $g_{\fact}=\SI{1.13}{\celsius}$, the observed four-year running average GMST anomaly in 2024 compared to the average GMST over the period 1951--1980, while the counterfactual value is taken as 
$g_{\cntr} = g_{\fact}-\SI{1.3}{\celsius}$ as in the case study.
The resulting attribution measure is
\[
\log\PR = \log\PR(\vartheta_0, x_0,g_{\fact},g_{\cntr}) = \log \frac{1-H_{f_\mathrm{scale}(\vartheta_0, g_{\fact})}(x_0)}{1-H_{f_\mathrm{scale}(\vartheta_0, g_{\cntr})}(x_0)}
=0.9439.
\]

\subsubsection{Observational data} \label{subsec:setup:obs}
As discussed previously, the observational data consist of a single GMST trajectory $(G_t)_{t \in \mathcal I}$ together with observations $(X_{jt})_{t \in \mathcal I_j}$ of the target variable extracted from $m_\obs$ observational data products. Throughout, we fix $m_\obs=2$ and consider the sampling periods
$\mathcal I= \mathcal I_1 =\{1951,\dots,2025\}$ and $\mathcal I_2=\{1976,\dots,2025\}$,
so that the corresponding sample sizes are $n_1=75$ and $n_2=50$. Extension of this setup illustrating the effect of increasing sample sizes are discussed in Appendix~\ref{subsec:simulation-observational-synthesis}.

The observational data are then generated as follows. First, for each \(t\in\mathcal I\), we set $G_t=g_{\obs,t}$ where $g_{\obs,t}$ denotes the observed GMST in year $t$. Next, for each $j\in\{1,2\}$, we generate data-set specific ground-truth parameters according to $\vartheta_j \sim \mathcal N_4(\vartheta_0,\Xi_{\obs})$ with $\Xi_{\obs}$ equal to the estimated matrix $\hat \Xi_{\obs}$ from the case study; see \eqref{eq:case-study-xis} in the Appendix. 
%Recall that the $\vartheta_j$ reflect small systematic discrepancies between observational products.
Next, we generate independent pairs $(U_{1t},U_{2t})_{t\in\mathcal I}$ from a Gaussian copula with correlation parameter \(\rho=0.98\). Finally, using inverse transform sampling, we define, for $t\in\mathcal I_j$,
$
X_{jt}
=
H^{-1}_{f_{\mathrm{scale}}(\vartheta_j,G_t)}(U_{jt}),
$
where $H^{-1}_{\theta}$ denotes the quantile function of the GEV distribution with parameter $\theta$. Overall, this yields the simulated observations
\[
(G_t)_{t\in\mathcal I},
\qquad
(X_{1t})_{t\in\mathcal I_1},
\qquad
(X_{2t})_{t\in\mathcal I_2}.
\]

Because the random parameters \(\vartheta_j\) are tightly concentrated around \(\vartheta_0\), the conditional distribution of \(X_{jt}\) given \(G_t=g\) remains close to the ground-truth GEV model with parameter \(f_{\mathrm{scale}}(\vartheta_0,g)\) from \eqref{eq:simulation-ground-truth-gev}. At the same time, the large copula correlation \(\rho=0.98\) induces strong dependence between \(X_{1t}\) and \(X_{2t}\). In this way, the simulation setup reproduces two characteristic features of real observational data products: small systematic differences in the underlying distributions and substantial cross-sectional dependence between the observations of the target variable; see also   Figure~\ref{fig:obs-sumlated-vs-obs} in the appendix for a visual comparison.

\subsubsection{Model data} \label{subsec:setup:model}
The model data consist of bivariate samples $(G_{jt}, \allowbreak X_{jt})_{t\in\mathcal J_j}$ obtained from $m_\model$ climate models. 
We fix $m_\model=10$ as in the case study.
The index sets $\mathcal J_j$ are allowed to be multisets, which reflects the practical situation in which several ensemble members are available for a given climate model.
Throughout, we denote by $e_j$ the number of ensemble members available for model $j$, and write $\mathcal J_{j,a}$  for the observation period associated with the $a$th ensemble member of that model; we then have $\mathcal J_j=\mathcal J_{j,1} \uplus \dots \uplus \mathcal J_{j,e_j}$ as a multiset union.  Consistent with typical climate-model data sets, we assume that all ensemble members of a given model share the same observation period, and sample the pairs of ensemble counts and observation periods $(e_j,\mathcal J_{j,1})$ independently from the respective empirical distribution observed in the case study (Figure~\ref{fig:forestnfit}).

Next, to introduce variability in the GMST trajectories across models and ensembles, we construct synthetic GMST series by perturbing the observed GMST trajectory with discretized Brownian motion noise. Specifically, if $\mathcal J_{j,1} \subset \{1880, \dots, 2025\}$, we set $G_{jt}=g_{\obs, t} +\sum_{\{t'\in\mathcal{I}:~t'\le t\}}\eps_{jt'}$, where the variables $\eps_{jt}\sim \mathcal N(0,1/750)$ are independent across both $j$ and $t$. Otherwise, if $\mathcal J_{j,1}$ extends beyond this period, we  instead linearly interpolate the observed GMST trajectory over 1880--2025, evaluate the interpolant on an equally spaced grid with spacing $145/|\mathcal J_{j,1}|$, and then apply the same perturbation scheme.
Consequently, all models share the same large-scale warming trend while still exhibiting moderate ensemble-specific deviations from the observed GMST trajectory. A visual comparison of the simulated GMST curves with those extracted from climate model data is provided in the Appendix, Figure~\ref{fig:gmst-simlated-vs-obs}.

To reflect structural differences between climate models, we further generate model-specific ground-truth parameter vectors independently according to $\vartheta_j \sim \mathcal N_4(\vartheta_0,\Xi^{\model})$, where the covariance matrix $\Xi^{\model}$ is chosen as the empirical inter-model covariance estimated in the case study; see \eqref{eq:case-study-xis} in the Appendix. Conditional on $\vartheta_j$ and $G_{jt}$, we then generate $X_{jt}\sim \mathrm{GEV}(f_{\mathrm{scale}}(\vartheta_j,G_{jt}))$. Overall, this yields simulated observations
\[
(G_{jt},X_{jt})_{t\in\mathcal J_j}, 
\quad \text{where} \quad
\mathcal J_j=\mathcal J_{j,1} \uplus \dots \uplus \mathcal J_{j,e_j}
\quad\text{and}\quad 
j \in \{1, \dots, m_\model\}.
\]

\subsection{Alignment between the simulation design and the statistical model assumptions}

We briefly examine how well the simulation design aligns with the statistical model assumptions underlying the two synthesis approaches, beginning with the representation variables $\eps_j$ governing between-product heterogeneity. For PL synthesis, these variables coincide with the parameter effects specified in the simulation design: $\eps_j:=\vartheta_j-\vartheta_0$ is centered by construction. For AML synthesis, the natural source-specific target is $T_{0j}=T(\vartheta_j)$. Since $T$ is nonlinear, $\eps_j:=T_{0j}-T_0$ is generally not centered, with mean $\Exp[T(\vartheta_j)]-T_0$.

To assess the practical importance of this asymmetry, we computed $\Exp[T(\vartheta_j)]$ under the data-generating processes above. The resulting values are $0.9308$ for the observational component and $0.9727$ for the climate-model component, compared with $T_0=T(\vartheta_0)=0.9439$. The corresponding squared deviations, approximately $1.7\times10^{-4}$ and $8.3\times10^{-4}$, are considerably smaller than the squared bias of order $10^{-2}$ observed for the full AML synthesis below. Hence, although the simulation design is structurally better aligned with PL synthesis, the lack of exact centeredness under AML is unlikely to explain more than a small part of the observed bias difference, and the comparison between AML and PL remains reasonably fair.

For both observational and model synthesis, the individual estimators are obtained by fitting the same distributional GEV models used to generate the data. This concerns the construction of the estimators rather than an explicit assumption on the within-product variables $\delta_j$ in either synthesis approach. In both cases, these variables arise from estimation error, with approximate unbiasedness motivated by standard large-sample theory and dependence determined by the simulation design. There is therefore no comparable structural asymmetry.

\subsection{Methods}
\label{subsec:methods}
For each observational or model-based data set, we fit the distributional GEV regression model by conditional maximum likelihood. Confidence intervals are constructed using the basic bootstrap interval \citep[Section~2.4]{Davison1997} at the $95\%$ nominal level.
Estimated probability ratios and their upper confidence bounds may occasionally be infinite. For the model-based WWA procedures, we replace such values by finite surrogates following \citet{otto24}; see Section~\ref{subsec:infties}. For all other procedures, infinite values are retained and reported.
We then apply and compare the following six full-synthesis approaches:
\begin{compactitem}

\item \textbf{WWA}: the WWA estimator from \eqref{eq:That-sfull} and confidence interval from \eqref{eq:confidence-intervals-sfull}.

\item \textbf{mWWA}: the modified WWA estimator from \eqref{eq:That-sfull-adapted} and its associated confidence interval from \eqref{eq:confidence-intervals-sfull-adapted}.

\item \textbf{WWA(2)}: a variant of WWA  based on replacing $\widehat{\PR}{}^{(1)}$ from \eqref{eq:pr-as-a-function-of-x0-estimator} by $\widehat{\PR}{}^{(2)}$ from \eqref{eq:pr-as-a-function-of-p0-estimator} for each individual model estimate, with 
$
\hat p_0 = H_{f_{\mathrm{scale}}(\hat \vartheta_\sobs, g_{\obs,2024})}(x_0)
$
based on $\hat \vartheta_\sobs$ from~\eqref{eq:hatvartheta_sobs}.

\item \textbf{mWWA(2)}: same as in the previous item, but using mWWA for synthesis.

\item \textbf{gPar}: the plug-in estimator of $\log \PR$ from \eqref{eq:pr-as-a-function-of-x0-estimator} that is based on $\hat \vartheta_{\sfull}$ from \eqref{eq:hatvartheta_sfull}, together with basic bootstrap confidence intervals based on the group bootstrap method as an input to Algorithm~\ref{algo:boot_full}.
\item \textbf{hPar}: the same procedure as gPar, but with the group bootstrap replaced by the hybrid bootstrap.
\end{compactitem}

\subsection{Results}
\label{subsec:results}

The results from $N=1{,}000$ independent simulation runs are summarized in Table~\ref{tab:simulation-results-full-synthesis}; the performance metrics are described in Appendix~\ref{subsec:performance}. Infinite values occur only for the individual observational data sets and the four observational WWA synthesis procedures, since infinities arising in the model-based WWA procedures are replaced by finite surrogates as described in Section~\ref{subsec:infties}. For all four WWA-based procedures, the observational synthesis yields an infinite upper confidence bound in the same $39.2\%$ of the runs, causing the observational component to receive effectively zero weight in the inverse-variance-weighted full synthesis. By construction, gPar and hPar do not encounter this problem.

\begin{table}[htbp]
\centering
\caption{
Performance Metrics for the \textbf{full synthesis}. Variance, squared bias and MSE are reported in units of $10^{-3}$.
} \label{tab:simulation-results-full-synthesis}
\begin{tabular}{l|rrr|rrr}
\toprule
Method & Var. & Bias$^2$ & MSE & Coverage (\%)& Length & Score \\
\midrule
gPar
& \multirow{2}{*}{30.6}
& \multirow{2}{*}{\textbf{1.06}}
& \multirow{2}{*}{31.7}
& 87.9
& \textbf{0.64}
& 0.96
\\
hPar
& %31.7
& %0.83
& %32.5
& \textbf{94.3}
& 0.84
& 0.95
\\
\midrule
WWA
& 36.4
& 14.6
& 51.0
& 100
& 3.09
& 3.09
\\
mWWA
& 23.3
& 9.95
& 33.3
& 92.0
& 0.75
& 0.95
\\
\midrule
WWA(2)
& 28.3
& 14.8
& 43.1
& 99.9
& 2.71
& 2.71
\\
mWWA(2)
& \textbf{18.4}
& 10.6
& \textbf{29.1}
& 91.4
& \textbf{0.64}
& \textbf{0.84}
\\
\bottomrule
\end{tabular}
\end{table}

The PL estimator underlying gPar and hPar has by far the smallest squared bias, at $1.06$, compared with values between $9.95$ and $14.8$ for the four WWA-based estimators. Its variance, at $30.6$, is somewhat larger than those of mWWA and mWWA(2), which attain variances of $23.3$ and $18.4$, respectively. Consequently, the PL estimator has an MSE of $31.7$: this is substantially smaller than the MSEs of WWA and WWA(2), at $51.0$ and $43.1$, respectively, and close to that of mWWA, at $33.3$, but larger than the minimum MSE of $29.1$ achieved by mWWA(2). Thus, the modified WWA procedures benefit from lower variance, whereas the principal advantage of the PL estimator lies in its markedly reduced bias.

Regarding the performance of the confidence interval, the results reveal a similar trade-off between calibration and efficiency. The original WWA and WWA(2) intervals are strongly conservative, both attaining coverage of (close to9 $100\%$, but are also considerably longer than all competing intervals, with lengths of $3.09$ and $2.71$, respectively. Their interval scores are correspondingly poor. The modified procedures yield substantially shorter intervals and coverage much closer to the nominal level: mWWA has a coverage of $92\%$, length $0.75$, and score $0.95$, whereas mWWA(2) has similar coverage at $91.4\%$, the shortest interval length of $0.64$, and the lowest score of $0.84$.

Among the PL procedures, gPar also produces intervals of length $0.64$, but its coverage of $87.9\%$ indicates noticeable undercoverage. The hPar construction improves coverage to the nominal level of $94.3\%$, at the cost of increasing the interval length to $0.84$. The corresponding interval scores, $0.96$ for gPar and $0.95$ for hPar, remain substantially smaller than those of WWA and WWA(2), although they are slightly larger than those obtained by the modified WWA procedures. Overall, the PL approach provides the clearest reduction in squared bias, with hPar achieving nominal coverage. By contrast, mWWA(2) attains the smallest variance and MSE and the most favorable interval score, but at the cost of a substantially larger bias.

\section{Case Study}
\label{sec:case-study}

We illustrate the proposed PL synthesis framework using a heavy precipitation event associated with Storm Boris, which led to severe flooding across parts of Central Europe in September 2024. The event was characterized by persistent and spatially extensive rainfall, resulting in major river flooding, widespread infrastructure damage, and at least 27 reported fatalities across several affected countries, including Austria, Czechia, Poland, and Romania.
The spatial context of the event is summarized in Figure~\ref{fig:case_boris} of the appendix.

The target variable $X$ is constructed separately for each data product as follows. First, we compute the spatial average of precipitation over the study region illustrated in Figure~\ref{fig:case_boris} of the appendix. We then calculate, for each day of the year, the cumulative spatially averaged precipitation over the four-day period starting at 06:00 UTC on the given day. Finally, we extract the annual maximum of these daily four-day accumulations, a generic version of which is denoted by~$X$.

\begin{figure}[thb]
    \centering
    \includegraphics[width=.72\textwidth]{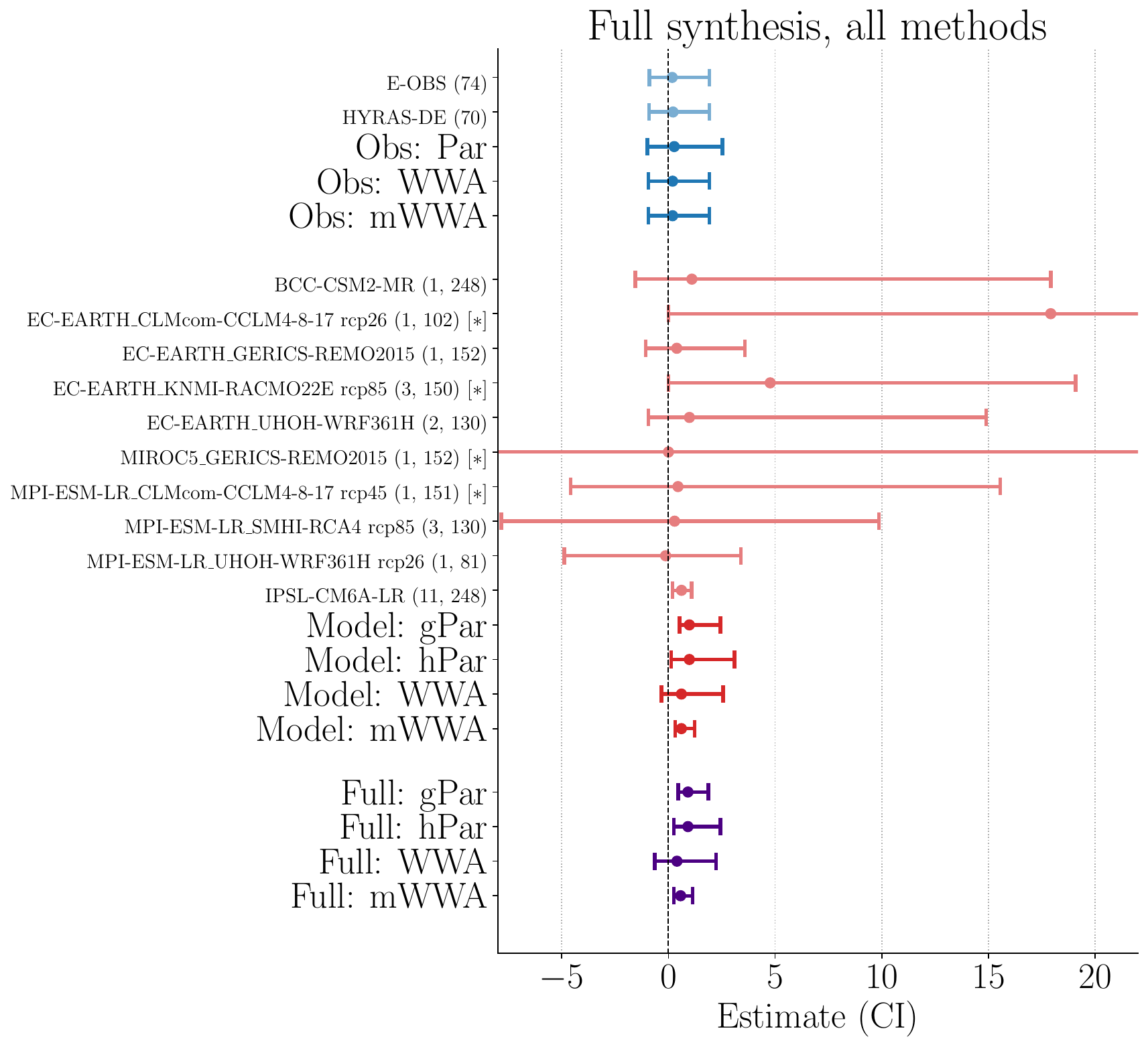}
    \caption{Forest plot of log-probability ratio estimates obtained from individual observational and climate model datasets, together with synthesized estimates based on different synthesis strategies. The numbers in the tick labels indicate the observation length for observational products and tuples of ensemble size and ensemble length for climate models. An asterisk [$*$] indicates that an infinite estimate or upper bound has been replaced according to the description in Appendix~\ref{subsec:infties}.}
    \label{fig:forestnfit}
\end{figure}

This approach is applied to two observational data products, E-OBS and a regional composite based on HYRAS-DE, SPARTACUS, RainGRS and SRA4d, covering 74 and 70 years, respectively, as well as to the ten climate-model 
data sets listed in Figure~\ref{fig:forestnfit}. The climate-model data sets comprise up to 11 ensemble members; two selected ensemble time series are shown in Figure~\ref{fig:observations-vs-climate-models}. As a covariate, we use the global mean surface temperature (GMST) anomaly relative to the 1951–1980 reference period. Figure~\ref{fig:gmst-simlated-vs-obs} (Appendix) compares the observed GMST time series with the corresponding time series from the ten climate models, where the model-specific GMST is averaged across all available ensemble members.

As the attribution measure, we consider the logarithmic probability ratio $\log \PR(x_0,g_{\fact},\allowbreak g_{\cntr})$, where $x_0=28.1$ denotes the observed precipitation amount during Storm Boris in the HYRAS-DE based data set, where $g_{\fact}=\SI{1.13}{\celsius}$ is the observed GMST anomaly in 2024, and where $g_{\cntr}=g_{\fact}-\SI{1.3}{\celsius}$ is the GMST roughly corresponding to the pre-industrial climate.
For a generic pair $(G,X)$, we use the same GEV model as in Section~\ref{sec:simulation}, namely, $(X\mid G=g)
\sim
\mathrm{GEV}(f_{\mathrm{scale}}(\vartheta,g))$
and then apply the methods described in Section~\ref{subsec:methods}.

The resulting estimates and confidence intervals are summarized in Figure~\ref{fig:forestnfit}. For all individual data sets, the $\log\PR$ estimates and confidence intervals shown there are based on the plain plug-in estimator from \eqref{eq:pr-as-a-function-of-x0-estimator}. Figure~\ref{fig:forestnfit2} (Appendix) presents the corresponding results obtained by instead using the alternative estimator in \eqref{eq:pr-as-a-function-of-p0-estimator} for each individual data set. Remarkably, the individual model confidence intervals are much narrower for the latter approach, but the obtained synthesis results are rather similar.

We observe for the observational products that the PL, WWA, and modified WWA syntheses yield similar positive estimates, but all corresponding confidence intervals include zero. In the model synthesis, both PL approaches and the modified WWA approach produce positive estimates with confidence intervals excluding zero, whereas the classical WWA interval still overlaps zero. The hierarchical-bootstrap parameter synthesis yields a wider interval than the group-bootstrap parameter synthesis, 
which aligns with the simulation findings from Section~\ref{sec:simulation}. While being wider, preceding results indicate a more accurate approximation of the nominal level.
By contrast, the modified WWA approach produces the most concentrated estimate, while the classical WWA approach remains comparatively uncertain.

Combining the observational and climate-model evidence in the full synthesis preserves this overall pattern while generally stabilizing the estimates. Under the assumed GMST-based attribution model, the gPar and hPar procedures, as well as the modified WWA approach, provide statistically significant evidence that an event at least as extreme as Storm Boris is more likely under the factual than under the counterfactual climate. By contrast, the classical WWA synthesis does not yield a statistically significant result.

\begin{figure}
    \centering
    \includegraphics[width=0.89\linewidth]{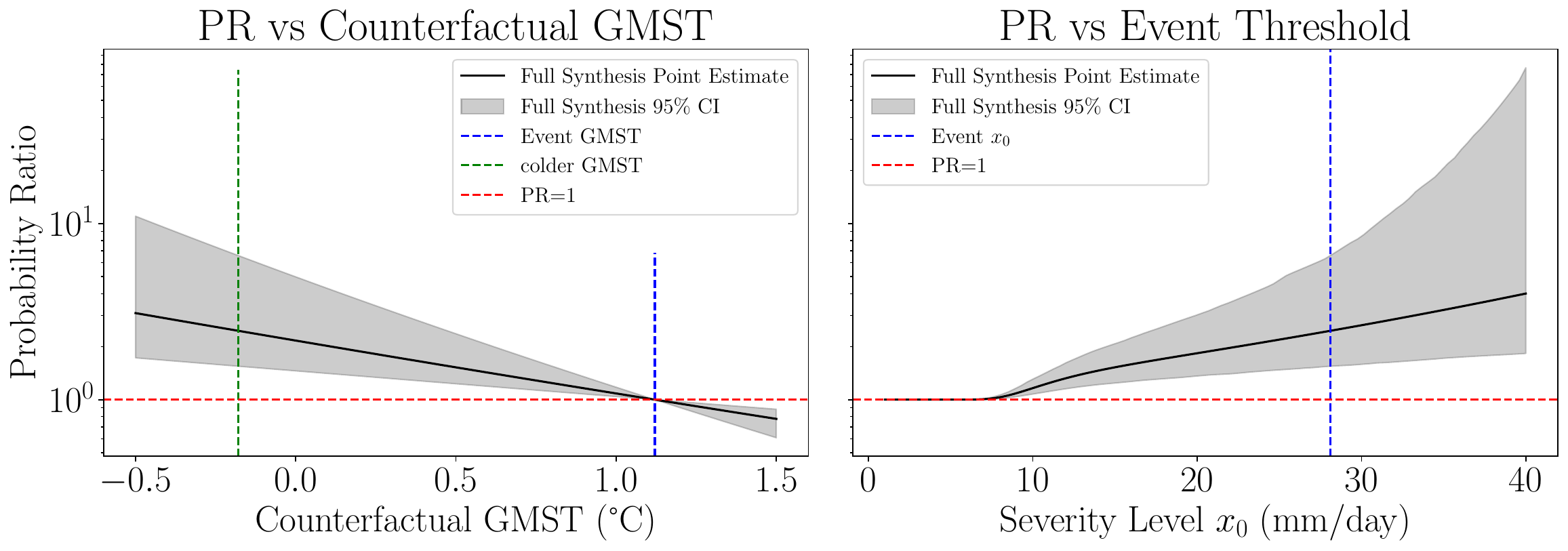}
    \caption{Estimated probability ratio $\widehat\PR(x_0,g_\fact,g_\cntr)$ as a function of the counterfactual GMST $g_\cntr$, with $x_0=28.1$ and $g_\fact=\SI{1.13}{\celsius}$ fixed (left), and as a function of the event threshold $x_0$, with $g_\fact=\SI{1.13}{\celsius}$ and $g_\cntr=g_\fact-\SI{1.3}{\celsius}$ fixed (right). }
    \label{fig:TrendsWithGmstAndX0}
\end{figure}

Figure~\ref{fig:TrendsWithGmstAndX0} provides a further summary of the full PL synthesis. The left panel shows the estimated probability ratio $\widehat\PR(x_0,g_\fact,g_\cntr)$ as a function of the counterfactual GMST $g_\cntr$, with $x_0=28.1$ and $g_\fact=\SI{1.13}{\celsius}$ held fixed. For counterfactual climates colder than the factual climate, the event is estimated to be less likely, whereas the reverse holds for warmer counterfactual conditions. The uncertainty increases as $g_\cntr$ moves farther away from the factual GMST.
The right panel shows the estimated probability ratio $\log\widehat\PR(x_0,g_\fact,g_\cntr)$ as a function of the threshold $x_0$, with $g_\fact=\SI{1.13}{\celsius}$ and $g_\cntr=g_\fact-\SI{1.3}{\celsius}$ held fixed. The estimated anthropogenic influence increases with the severity of the precipitation event, accompanied by a corresponding increase in uncertainty.

Importantly, the curves in Figure~\ref{fig:TrendsWithGmstAndX0} can be obtained only through PL synthesis. Indeed, by combining the estimated GEV parameters, the probability ratio can be evaluated jointly over a continuum of counterfactual GMST values and event thresholds. By contrast, a synthesis conducted separately for a fixed attribution measure, such as the WWA approaches, would yield estimates only at the specific values considered in that analysis. The curves should nevertheless be interpreted only for GMST values reasonably close to the range covered by the observational and climate-model data, since values outside this range require extrapolation and may therefore be unreliable.

\section{Conclusion}
\label{sec:conclusion}

This paper has examined statistical approaches for synthesizing evidence in probabilistic EEA. We identified several weaknesses of existing AML procedures, proposed targeted modifications, and introduced a PL framework that synthesizes the underlying distributional regression models before attribution measures are derived. The latter yields a common distributional model and thereby supports coherent inference across thresholds, exceedance probabilities, and counterfactual climate conditions.

In the simulation study, the original synthesis procedure exhibited substantial bias and overly conservative confidence intervals. Both the modified AML procedures and PL synthesis performed markedly better, although neither approach was uniformly superior across all criteria. The Storm Boris case study further showed that the choice of synthesis method can materially affect the resulting attribution statement.

Both approaches rely on substantive assumptions. AML synthesis requires source-specific attribution effects to be meaningfully centered around a common target, whereas PL synthesis additionally depends on an adequate common model, a scientifically meaningful parameter target, and a suitable parametrization. In both cases, between-source heterogeneity is difficult to estimate from few data products, and independence remains a working assumption.

The simulation design is more closely aligned with PL synthesis because heterogeneity is introduced at the parameter level within the fitted model family. Future work should therefore examine model misspecification, alternative parametrizations, shared biases, dependent climate models, and more comprehensive uncertainty procedures.

%%%%%%%%%%%%%%%%%%%%%%%%%%%%%%%%%%%%%%%%%%%%%%
%% Appendix---Please move all appendices to %%
%% a Supplementary file.                    %%
%%%%%%%%%%%%%%%%%%%%%%%%%%%%%%%%%%%%%%%%%%%%%%
%% Support information, if any,             %%
%% should be provided in the                %%
%% Acknowledgements section.                %%
%%%%%%%%%%%%%%%%%%%%%%%%%%%%%%%%%%%%%%%%%%%%%%
\begin{acks}[Acknowledgments]
We acknowledge the World Climate Research Programme's Working Group on Regional Climate, and the Working Group on Coupled Modelling, former coordinating body of CORDEX and responsible panel for CMIP5. We also thank the climate modelling groups listed in Figure~\ref{fig:forestnfit} for producing and making available their model output. Finally, we acknowledge the Earth System Grid Federation infrastructure, an international effort led by the U.S.\ Department of Energy's Program for Climate Model Diagnosis and Intercomparison, the European Network for Earth System Modelling and other partners in the Global Organisation for Earth System Science Portals (GO-ESSP).

% During the preparation of this manuscript, the authors used ChatGPT, developed by OpenAI, to improve the language and clarity of the text and to assist with the presentation of figures. The authors reviewed and edited all AI-assisted content and take full responsibility for the final manuscript.
% \end{itemize}

\end{acks}
%%%%%%%%%%%%%%%%%%%%%%%%%%%%%%%%%%%%%%%%%%%%%%
%% Funding information, if any,             %%
%% should be provided in the                %%
%% funding section.                         %%
%%%%%%%%%%%%%%%%%%%%%%%%%%%%%%%%%%%%%%%%%%%%%%
\begin{funding}
All three authors have been supported by the integrated project ``Climate Change and Extreme Events – ClimXtreme Module B Statistics Phase II'' (AB and EH: project B3.3, grant number 01LP2323L; JS: project B1.2, grant number 01LP2323B) funded by the German Federal Ministry of Research, Technology and Space (BMFTR). 
EH is grateful for support from the Studienstiftung des deutschen Volkes. This work used resources of the Deutsches Klimarechenzentrum (DKRZ) granted by its Scientific Steering Committee (WLA) under project ID bb1152.\\
\end{funding}

% \textbf{Ethical Approval.} Not applicable.\\

% \textbf{Availability of supporting data. } Code for the simulation study and case study are available in this Github repository \href{https://github.com/haufse/synthesis-eea-paper}{\texttt{haufse/synthesis-eea-paper}}. Functions and algorithms accompanying the paper are provided as a \texttt{python} package in \linebreak\href{https://github.com/haufse/synthesis-eea}{\texttt{haufse/synthesis-eea}} [\cite{haufse_synthesis_eea}]. The datasets used in Section~\ref{sec:case-study} are available at the \href{https://esgf-metagrid.cloud.dkrz.de}{DKRZ Metagrid}  for CMIP6 models, at \href{https://esgf-data.dwd.de/}{Earth System Grid Federation} of the DWD for EURO-CORDEX models, at \href{https://opendata.dwd.de/climate\_environment/CDC/}{DWD Opendata} for HYRAS and at \href{https://surfobs.climate.copernicus.eu/dataaccess/access\_eobs.php}{Copernicus} for E-OBS.\\

% \textbf{Competing interests.} The authors declare that they have no conflict of interest.\\

\textbf{Authors' contributions.} \hspace{1em}
EH developed the methodological details, implemented the proposed procedures, conducted the simulation study and case-study analysis, and prepared the initial draft of the manuscript. AB conceived the study and its overall methodological direction, provided continuous supervision and feedback, and contributed to revising and improving the manuscript. JS advised the other authors throughout the project on applied aspects of extreme-event attribution and contributed expertise on the case study and the underlying data sets. All authors contributed to revising the manuscript, approved the final version, and agree to be accountable for the work.
During the preparation of this manuscript, the authors used ChatGPT, developed by OpenAI, to improve the language and clarity of the text and to assist with the presentation of figures. The authors reviewed and edited all AI-assisted content and take full responsibility for the final manuscript.

\bibliographystyle{imsart-nameyear} % Style BST file
%\bibliography{bibliography}       % Bibliography file (usually '*.bib')
\putbib[bibliography]
\end{bibunit}
% Supplement citation unit: citations from this point are kept separate.
\begin{bibunit}[imsart-nameyear]
%% or include bibliography directly:
% \begin{thebibliography}{}
% \bibitem[\protect\citeauthoryear{???}{???}]{b1}
% \end{thebibliography}

\newpage

\appendix
\thispagestyle{empty}
\numberwithin{equation}{section}

\counterwithin{figure}{section}
\counterwithin{table}{section}

\begin{center}
	
	{\bfseries SUPPLEMENT TO THE PAPER:  \\  ``\MakeUppercase{Evidence Synthesis in Probabilistic Extreme Event Attribution: From Attribution Measures to Model Parameters'' }}
	\vspace{.5cm}
	
	{\textsc{By Erik Haufs, Axel Bücher and Jonas Schröter}}
	
	\vspace{.28cm}
	
	{\textit{Ruhr-Universität Bochum and Deutscher Wetterdienst}}
	
	\vspace{.28cm}

	\begin{center}
		\begin{minipage}{.6\textwidth}
			{\small \hspace{.5cm}
					Appendix~\ref{app:model-parameters-to-attribution-targets} derives explicit expressions for the attribution measures under the proposed parametric framework, while Appendix~\ref{app:tau_estim} reviews the random-effects models underlying the synthesis procedures. Appendix~\ref{sec:simulation-details} provides additional details on the simulation study, and Appendix~\ref{app:additional-simulation-results} presents further simulation results. Supplementary results for the Storm Boris case study are given in Appendix~\ref{app:case-study}. Finally, Appendix~\ref{app:algo} collects the bootstrap algorithms used throughout the paper.
				 }
		\end{minipage}
	\end{center}

\end{center}

\vspace{.5cm}

\begingroup

% Hide entries tagged as belonging to the main document
\etocsettagdepth{main}{none}

% Permit entries through subsection level in the appendix
\etocsettagdepth{Supplementary}{subsection}

% Hide section entries while retaining subsection entries
% \etocsetlevel{section}{3}
% \etocsetlevel{subsection}{2}

\setcounter{tocdepth}{2}
\tableofcontents

\endgroup

\etocdepthtag.toc{Supplementary}

\appendix

\section{From Model Parameters to Attribution Measures}
\label{app:model-parameters-to-attribution-targets}

Under parametric model assumptions, attribution measures such as the probability ratio in \eqref{eq:probability-ratio} or the intensity change in \eqref{eq:intensity-change} can be expressed explicitly as functions of the model parameters, naturally leading to corresponding plug-in estimators. Different representations of the same measure may thereby give rise to different estimators. In this section, we discuss two such approaches for estimating probability ratios.

We start by recalling the probability ratio at threshold level $x_0$
\begin{align*}
\PR
= 
\PR(x_0)
= 
\frac{1 - F_{\fact}(x_0)}{1 - F_{\cntr}(x_0)}
\end{align*}
from \eqref{eq:probability-ratio}. The assumptions imposed in Section~\ref{sec:fundamentals} imply that
\begin{align} \label{eq:gev-fact-cntr}
F_{\fact}(x) = H_{f(\vartheta, g_{\fact})}(x), 
\qquad
F_{\cntr}(x) = H_{f(\vartheta, g_{\cntr})}(x),
\end{align}
such that $\PR$ becomes an explicit function of $x_0$, $\vartheta$, $g_{\fact}$, and $g_{\cntr}$, namely
\begin{align*} 
%\label{eq:pr-as-a-function-of-x0}
\PR = \frac{1-H_{f(\vartheta, g_{\fact})}(x_0)}{1-H_{f(\vartheta, g_{\cntr})}(x_0)}.
\end{align*}
Given an estimator $\hat \vartheta$, together with associated values of $(g_{\fact}, g_{\cntr})$ (for instance, from the observed GMST trajectory or the GMST trajectory obtained from a specific climate model) and a threshold $x_0$, the previous equation yields the plug-in estimator
\begin{align}
\label{eq:pr-as-a-function-of-x0-estimator}
\widehat{\PR}{}^{(1)} 
= 
\widehat{\PR}{}^{(1)}(\hat \vartheta, x_0, g_{\fact}, g_{\cntr})
= 
\frac{1-H_{f(\hat \vartheta, g_{\fact})}(x_0)}{1-H_{f(\hat \vartheta, g_{\cntr})}(x_0)}.
\end{align}

Typically, the threshold $x_0$ is chosen based on observational data from the factual climate. 
Climate-model data intended to approximate the factual climate, however, may systematically deviate from it (for instance through a model-specific factual GMST $g_{\model,\fact}$ differing from the observed $g_{\obs,\fact}$) so that the same threshold $x_0$ may correspond to an unrealistic event in some models. To obtain more stable and comparable probability ratios across models, one may therefore allow for model-specific thresholds through an approach commonly referred to as \emph{quantile mapping} \citep{Cannon2015}.

Specifically, note that we can write
\[
\PR = \frac{1-p_0}{1-F_{\cntr}(F_{\fact}^{{-1}}(p_0))}, \qquad \text{where } p_0 = F_{\fact}(x_0),
\]
whence $\PR=\PR(p_0)$ can also be interpreted as a function of the non-exceedance probability $p_0$ rather than a function of the threshold $x_0$. The assumption in \eqref{eq:gev-fact-cntr} yields the representation
\begin{align*} 
%\label{eq:pr-as-a-function-of-p0}
\PR = \frac{1-p_0}{1-H_{f(\vartheta, g_{\cntr})}(H^{-1}_{f(\vartheta, g_{\fact})}(p_0))},
\end{align*}
which, given an estimator $\hat \vartheta$ together with associated values of $(g_{\fact}, g_{\cntr})$  and a non-exceedance probability $p$ approximating $p_0$, yields the alternative plug-in estimator
\begin{align}
\label{eq:pr-as-a-function-of-p0-estimator}
\widehat{\PR}{}^{(2)} &= \widehat{\PR}{}^{(2)}(\hat \vartheta, p, g_{\fact}, g_{\cntr})
=
\frac{1-p}{1-H_{f(\hat \vartheta, g_{\cntr})}(H^{-1}_{f(\hat \vartheta, g_{\fact})}(p))}.
\end{align}

The estimator $\widehat{\PR}{}^{(2)}$ is typically evaluated at $p = \hat p_0= H_{f(\hat \vartheta_{\obs}, g_{\obs,\fact})}(x_0)$, where $\hat \vartheta_{\obs}$ is estimated from a reference data product and where $g_{\obs,\fact}$ denotes the observed factual GMST. The quantity
\[
x_{\model} 
= 
H^{-1}_{f(\hat \vartheta, g_{\fact})}(H_{f(\hat \vartheta_{\obs}, g_{\obs,\fact})}(x_0))
\]
appearing in the denominator of $\widehat{\PR}{}^{(2)}$ may then be interpreted as a model-specific threshold associated with the original threshold $x_0$. The mapping $x \mapsto H^{-1}_{f(\hat \vartheta, g_{\fact})}(H_{f(\hat \vartheta_{\obs}, g_{\obs,\fact})}(x))$ is commonly referred to as a \emph{quantile map}.

\section{Meta-analyses and random-effects models}
\label{app:tau_estim}

Meta-analysis originates in the work of \cite{glass1976} and \cite{cochran1954combination} and is widely used to synthesize results from multiple studies, particularly in medical statistics. In this appendix, we review univariate and multivariate random-effects models (REMs) and collect the specific variance-component estimators and variance formulas underlying Sections~\ref{subsec:model-synthesis-attribution-targets} and~\ref{subsec:model-synthesis-model-parameters}. The goal is not to provide a comprehensive survey, but to justify the particular choices made in the main text.

\subsection{Univariate random-effects models}

We recall the univariate random-effects model that forms the theoretical basis of the considerations in Section~\ref{subsec:model-synthesis-attribution-targets}. 
For each study $j=1,\dots,m$, 
let $X_j$ denote a univariate quantity of interest (typically resulting from aggregating data collected for the $j$th study) and assume that
\begin{align*}
X_j = \mu + \delta_j + \varepsilon_j,
\end{align*}
where $\mu$ denotes the overall target parameter, and $\delta_1, \dots, \delta_m,\varepsilon_1, \dots, \varepsilon_m$  are independent random errors with
\begin{align*}
\E[\delta_j]=\E[\varepsilon_j]=0,
\qquad
\Var(\varepsilon_j)=\tau^2,
\qquad
\Var(\delta_j)=\sigma_j^2.
\end{align*}
Here, $\tau^2$ measures inter-study variability, while $\sigma_j^2$ captures within-study uncertainty. The variances $\sigma_j^2$ are typically assumed known or easily estimable, whereas $\tau^2$ must be inferred from the data.

Given an estimator $\hat\tau$ of $\tau$, the corresponding precision-weighted estimator of $\mu$ is
\begin{align}
\hat\mu(\hat\tau) \label{eq:muhat}
=
\Big(\sum_{j=1}^m \frac{1}{\sigma_j^2+\hat\tau^2}\Big)^{-1}
\sum_{j=1}^m \frac{X_j}{\sigma_j^2+\hat\tau^2}.
\end{align}

A wide range of estimators for $\tau^2$ has been proposed. Two of the most prominent are the Paule--Mandel estimator \citep{paule1982consensus} and the DerSimonian--Laird estimator \citep{dersimonian1986meta}. We focus on these two approaches, as the latter admits a natural multivariate extension (Appendix~\ref{app:multi_rem}) that is required in our PL synthesis.

\smallskip \noindent \emph{Paule--Mandel estimator.}
The Paule--Mandel (PM) estimator is a method-of-moments estimator based on the statistic
\begin{equation}
\label{eq:qstat}
Q(t)
=
\sum_{j=1}^m
\frac{(X_j-\hat\mu(t))^2}{\sigma_j^2+t^2},
\end{equation}
with $\hat\mu(t)$  as defined in \eqref{eq:muhat}. A direct calculation yields $\E[Q(\tau)]=m-1$. Moreover, if $X_j\sim\mathcal{N}(\mu,\sigma_j^2+\tau^2)$, then $Q(\tau)$ is $\chi_{m-1}^2$ distributed with $m-1$ degrees of freedom; see \cite{dersimonian1986meta,hardy1998detecting}.
The PM estimator $\hat\tau_{\mathrm{PM}}$ is defined as the unique solution to $Q(t)=m-1$ if $Q(0)\ge m-1$, and $\hat\tau_{\mathrm{PM}}=0$ otherwise. Uniqueness follows from the fact that $Q(t)$ is a continuous and strictly decreasing function of $t$ \citep{Knapp2006,vanAert2018}. Finally, we note that  \cite{otto24} use a slightly modified version of $Q(t)$ that is based on estimating $\hat \sigma_j^2$ from confidence intervals only; see their Equation (10).

\smallskip \noindent \emph{DerSimonian--Laird estimator.}
Let $Q=Q(0)$ denote the statistic in \eqref{eq:qstat} evaluated at $t=0$.
Its expectation satisfies \citep{Kacker2004}
\begin{align*}
\E[Q]
=
m-1 + \Big(S_1-\frac{S_2}{S_1}\Big)\tau^2,
\qquad
S_r := \sum_{j=1}^m \sigma_j^{-2r}.
\end{align*}
Solving this expression for $\tau^2$ yields the DerSimonian--Laird (DL) estimator
\begin{align*}
\hat\tau^2_{\mathrm{DL}}
=
\max\Big\{0,\,
\frac{Q-(m-1)}{S_1-S_2/S_1}
\Big\}.
\end{align*}
The DL estimator has two properties that are particularly attractive in the present context: it is non-iterative (computationally light) and admits natural extensions to multivariate REMs, as illustrated in Appendix~\ref{app:multi_rem} below.

\smallskip \noindent \emph{Likelihood-based estimators.} 
Beyond moment estimators such as DL and PM, likelihood-based estimators of $\tau^2$ are frequently used, in particular restricted maximum likelihood (REML); see \citep{veroniki2016methods} for an overview. REML is appealing because it is derived from a likelihood principle under the Gaussian REM and often exhibits reduced small-sample bias relative to DL. However, in practice it has two drawbacks in the present application: (i) it is iterative (higher computational cost in large simulation studies), and (ii) in multivariate and bootstrap-driven settings it requires additional modeling assumptions to specify and maximize the likelihood reliably. Since our main objective is to build a synthesis procedure that generalizes seamlessly to the multivariate PL framework and interacts transparently with hierarchical bootstrap inference, we employ DL-type estimators as a computationally robust default. 

\subsection{Multivariate random-effects models}\label{app:multi_rem}

Multivariate random-effects models were introduced by \cite{Raudenbush1988}. Overviews and methodological developments are provided in \citep{jackson2011multivariate,vanHouwelingen2002}. In particular, \citep{Chen2012} propose a multivariate extension of the DerSimonian--Laird estimator, which we adopt here.

In analogy to the univariate case, let
\begin{align*}
X_j = \mu + \delta_j + \varepsilon_j,
\qquad j=1,\dots,m,
\end{align*}
where $X_j,\mu,\delta_j,\varepsilon_j\in\R^p$.
We assume
\begin{align*}
\E[\delta_j]=\E[\varepsilon_j]=0,
\qquad
\Cov(\varepsilon_j)=\Xi,
\qquad
\Cov(\delta_j)=\Sigma_j,
\end{align*}
with $\Sigma_j$ known and $\mu,\Xi$ unknown. Additionally assuming that $\Xi + \Sigma_j$ is invertible for each $j$, we obtain the precision-weighted estimator of $\mu$ 
\begin{align}
\label{eq:hat-mu-multivariate}
\hat\mu(\Xi)
=
\Big(\sum_{j=1}^m (\Sigma_j+\Xi)^{-1}\Big)^{-1}
\sum_{j=1}^m (\Sigma_j+\Xi)^{-1}X_j.
\end{align}

The between-study covariance $\Xi$ encodes both marginal heterogeneity of individual components (diagonal entries) and cross-component co-variation (off-diagonal entries). In the present synthesis context, $X_j$ represents a parameter estimate (e.g., a nonstationary GEV parameter vector), and off-diagonal elements of $\Xi$ quantify the extent to which models that differ in one parameter component (e.g., location) systematically differ in another (e.g., scale or trend). Accounting for such correlations is important whenever inference is propagated through nonlinear maps $T(\cdot)$, as cross-covariances affect the uncertainty of derived estimators for attribution measures.

\smallskip \noindent \emph{Multivariate DerSimonian--Laird estimator.}
Following \cite{Chen2012}, define the auxiliary matrices
\begin{align*}
\Psi
=
\Big(\sum_{j=1}^m \Sigma_j^{-1}\Big)^{-1},
\qquad
\Phi
=
\Psi^{-1}
-
\sum_{j=1}^m \Sigma_j^{-1}\Psi\Sigma_j^{-1},
\end{align*}
and
\begin{align*}
A
=
\sum_{j=1}^m
\Sigma_j^{-1}
\big(X_j-\hat\mu(\mathbf{0})\big)
\big(X_j-\hat\mu(\mathbf{0})\big)^\top
-
(m-1)\mathds{I}_{p\times p},
\end{align*}
with $\hat \mu(\mathbf 0)$ from \eqref{eq:hat-mu-multivariate}.
The multivariate DerSimonian--Laird estimator of the between-study covariance matrix is then defined as
\begin{align} \label{eq:multivariate-DerSimonian-Laird}
\hat{\Xi}
=
\frac{1}{2}
\Big(
\Phi^{-1}A + A^\top\Phi^{-1}
\Big).
\end{align}

\subsection{Method-of-moments estimator for $\Xi$ under dependence}
\label{subsec:estimating-Xi}
In the setting of Section~\ref{subsec:observational-synthesis-model-parameters}, the $\delta_i$ and $\delta_j$ are cross-correlated, that is, $\Sigma_{ij} = \Cov(\delta_i, \delta_j) \neq 0$ for $i\neq j$. Thus, standard tools as the multivariate DerSimonian--Laird estimator above are not directly applicable here. Instead, define 
\begin{align}
\label{eq:estimator-S}
    S = \frac{1}{m-1}\sum_{j=1}^m \Big(X_j- \overline{X_{1:m}}\Big)\Big(X_j- \overline{X_{1:m}}\Big)^\top, \qquad \overline{X_{1:m}}=\frac{1}{m}\sum_{j=1}^m X_j.
\end{align}
Using the identities 
\[
S
=
\frac{1}{2m(m-1)}
\sum_{i,j=1}^m
(X_i-X_j)
(X_i-X_j)^\top
\]
and
\[
\Cov(X_i, X_j) = \Xi \cdot \bm 1(i=j) + \Sigma_{ij},
\] 
we obtain that
\begin{align*}
\mathbb E[S]
=
\frac{1}{2m(m-1)}
\sum_{i,j=1}^m
\Cov(X_i-X_j)
&=
\frac{1}{2m(m-1)}
\sum_{i \ne j}
\Big\{
2\Xi+\Sigma_{ii}+\Sigma_{jj}-\Sigma_{ij}-\Sigma_{ji}
\Big\}
\\
&=
\Xi
+
\frac1m
\sum_{j=1}^m
\Big(
\Sigma_{jj}
-
\frac{1}{m-1}
\sum_{i \ne j}
\Sigma_{ij}
\Big).
\end{align*}
Solving for $\Xi$ and replacing $\Exp[S]$ by its unbiased estimator $S$ yields the matching moment estimator
\begin{align*}
    \hat \Xi(\bm X,(\Sigma_{ij})_{i,j=1}^m)
    &:= 
    S
    - 
    \frac{1}{m} \sum_{j=1}^m \Big(\Sigma_{jj}-\frac{1}{m-1}\sum_{i\ne j} \Sigma_{ij}\Big),
\end{align*}
where $\bm X=(X_1^\top, \dots, X_m^\top)^\top$; 
see also \eqref{eq:xi-hat-obs}.

\section{Additional Details on the Simulation Study}
\label{sec:simulation-details}

This section provides additional details on the simulation study, including a description of the performance metrics.

\subsection{Visual assessment of the simulation design}
\begin{figure}[thp]
    \centering
    \includegraphics[width=0.48\linewidth]{figs_arxiv/obs_pr_vs_year_without_mean.pdf}
    \includegraphics[width=0.48\linewidth]{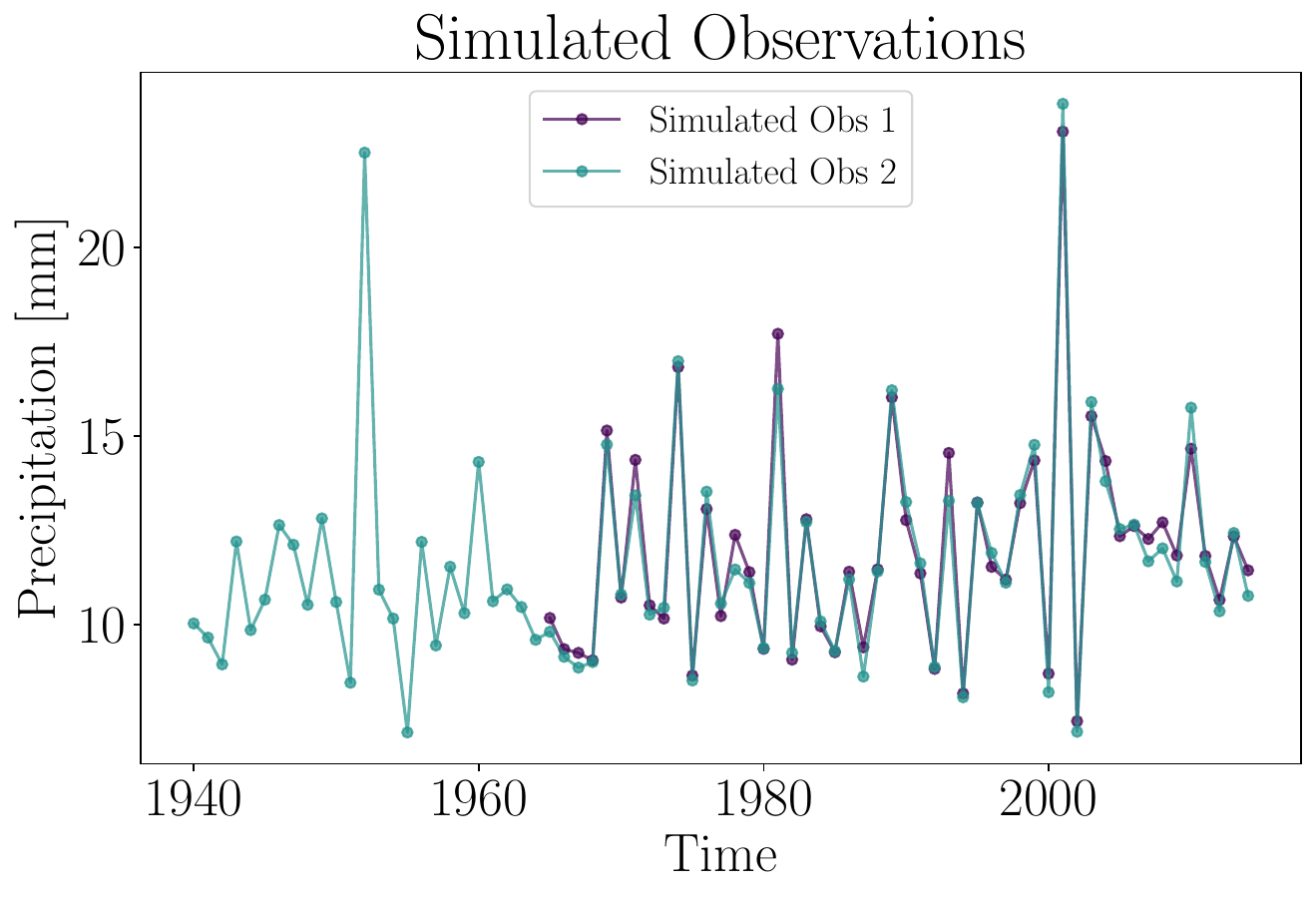}
    \caption{Observed and simulated annual maxima of cumulative daily precipitation. Left: annual maxima extracted from two observational data products (HYRAS-DE and E-OBS). Right: simulated time series designed to reproduce similar characteristics.
    }
    \label{fig:obs-sumlated-vs-obs}
\end{figure}

\begin{figure}[thp]
    \centering
    \includegraphics[width=0.45\linewidth]{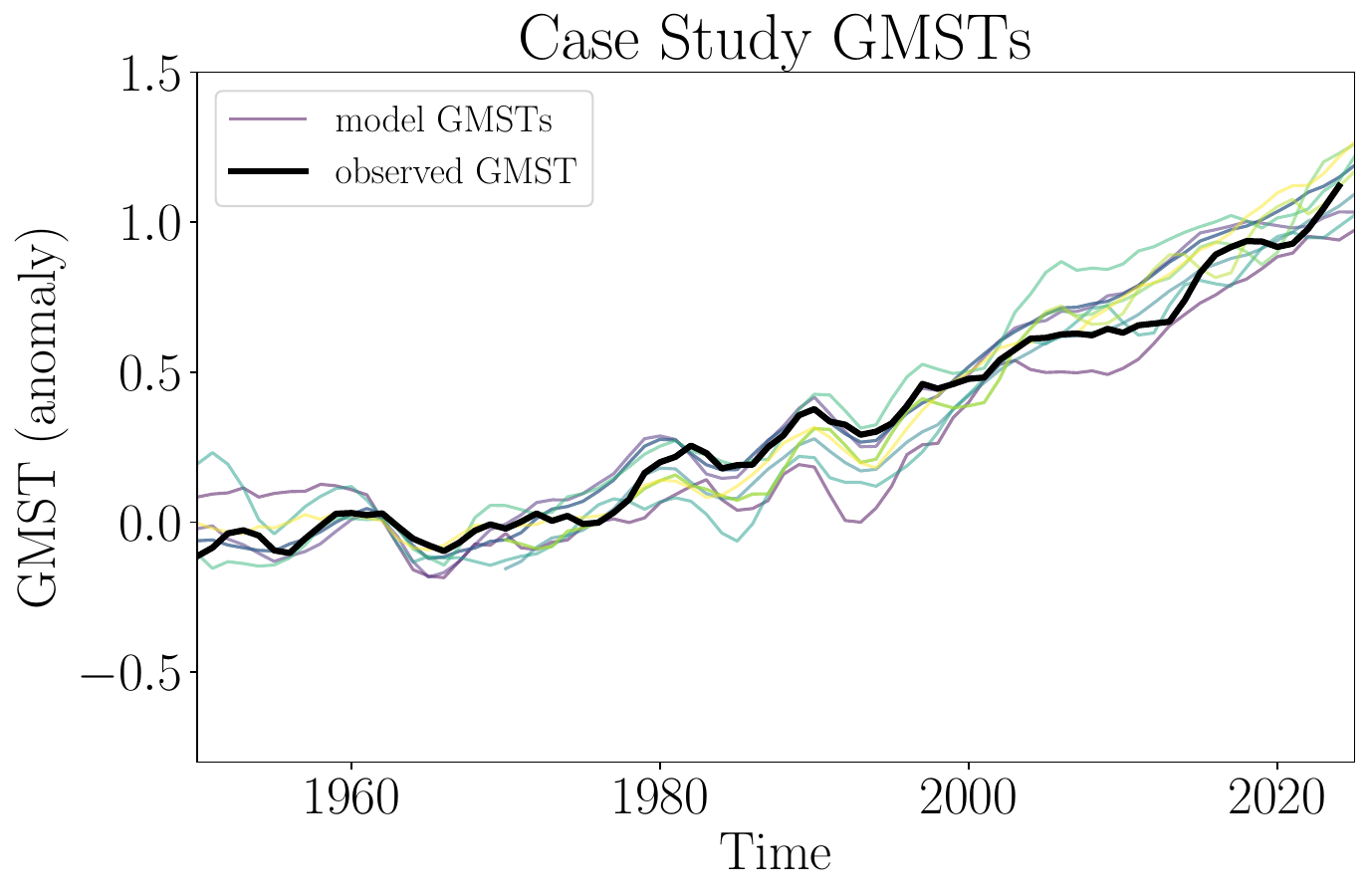}
    \includegraphics[width=0.45\linewidth]{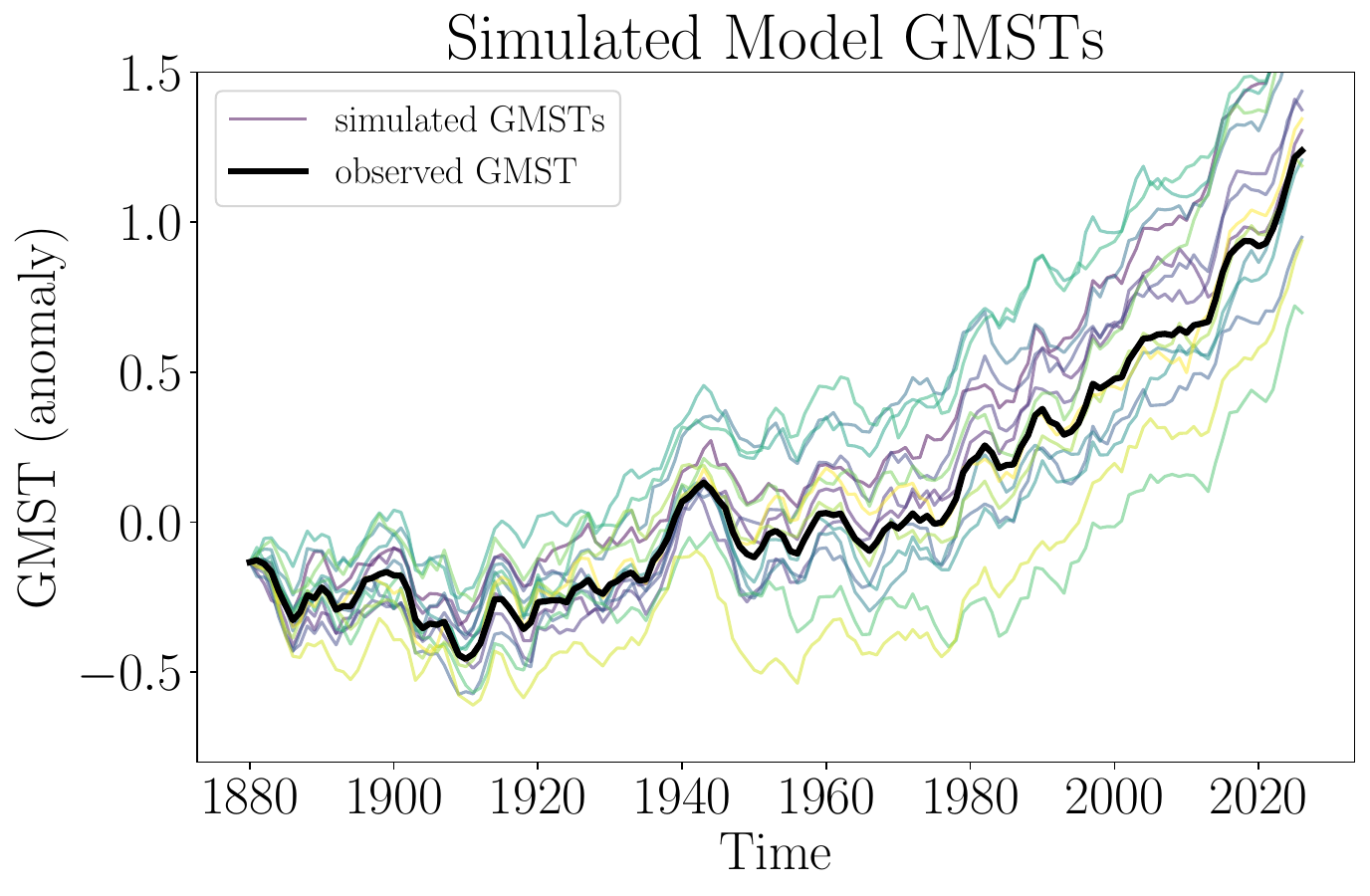}
    \caption{Simulated and climate-model-based GMST curves. Left: GMST extracted from climate model data used in the case study in Section~\ref{sec:case-study}. Right: Simulated GMST curves designed to reproduce similar characteristics. }
    %\eh{update: die Modelle EC-EARTH (KNMI / UHOH / Gerics) haben dieselbe GMST}
    \label{fig:gmst-simlated-vs-obs}
\end{figure}

Figures~\ref{fig:obs-sumlated-vs-obs} and~\ref{fig:gmst-simlated-vs-obs} compare simulated and observed time series for the block maxima and model GMST trajectories, respectively. In both cases, the simulations reproduce the main temporal patterns and overall variability of the observed series. The close visual resemblance provides an informal check that the simulation design generates data with realistic temporal behavior.

\subsection{Dataset specific estimation}\label{subsec:datasetspecific}
Each data set, observational or model-based, involves a distributional regression model for the GEV family that can, in principle, be estimated using conditional maximum likelihood. 
More precisely, for a generic sample $(g_t,x_t)_{t \in \mathcal J}$, one may define
\[
\hat \vartheta \in \argmax_{\vartheta \in \Theta} \sum_{t \in \mathcal J} \log h_{f_{\mathrm{scale}}(\vartheta, g_t)}(x_t),
\]
where $h_\theta$ denotes the density of the GEV family and where $f_{\mathrm{scale}}$ is from \eqref{eq:linkfunc_scale}. 
Owing to its well-established efficiency and general optimality properties under correct model specification, this estimator is employed throughout the main paper.

For the more comprehensive simulation designs described in Section~\ref{subsec:simulation-observational-synthesis} and \ref{subsec:simulation-model-synthesis}, however, repeated numerical maximization for every simulated data set would constitute a computational bottleneck. We therefore use a substantially faster estimation procedure based on probability-weighted moments (PWM) after eliminating the temporal trend. 
More specifically, observe that $(X \mid G=g) \sim H_{f_{\mathrm{scale}}(\vartheta, g)}$ implies that $Z:= \exp(-\alpha g) X$ satisfies $(Z \mid G=g) \sim H_{(\mu, \sigma, \gamma)}$. Hence, the conditional distribution of $Z$ given $G=g$ no longer depends on $\alpha$. Taking logarithms yields the mean-regression equation 
\[
\log X = \nu + \alpha g + \eps, 
\]
where $\nu := \mathbb E[\log Z]$ and $\varepsilon := \log Z-\nu$ has mean zero. For a given sample $(g_t,x_t)_{t \in \mathcal J}$, we therefore estimate $\alpha$ by ordinary least squares; the resulting estimator is denoted by $\hat\alpha^{\mathrm{OLS}}$. We then construct the pseudo-sample $(\hat z_t)_{t \in \mathcal J}$ by $\hat z_t := \exp(-\hat \alpha^{\mathrm{OLS}} g_t) x_t$, which may be regarded as an approximate sample from the GEV distribution with parameter $(\mu,\sigma,\gamma)$. Finally, we fit the GEV family to this pseudo-sample using the PWM estimator of \cite{Hosking1985}. 

As illustrated in Appendix~\ref{app:conditional-mle-vs-pwm}, the two-stage PWM estimator exhibits similar finite-sample behavior to the conditional ML estimator, while requiring only about $20$--$25\%$ of the computation time needed for the ML estimator. While the conditional ML estimator remains the default procedure in the main paper, we use the PWM estimator for its numeric properties in the computationally intensive parts of the simulation study, Appendices~\ref{subsec:simulation-observational-synthesis} and~\ref{subsec:simulation-model-synthesis}.

Finally, once the GEV distribution has been fitted, (logarithmic) probability ratios are estimated using the plain plug-in estimator from \eqref{eq:pr-as-a-function-of-x0-estimator}, unless stated otherwise. Dataset specific confidence intervals are then always constructed using the basic bootstrap confidence interval \citep[Section 2.4]{Davison1997}.

\subsection{Handling Infinite Estimates and Confidence Bounds} \label{subsec:infties}
Probability-ratio estimates and their upper confidence bounds may occasionally equal $+\infty$. As illustrated in the following sections, this occurs only rarely for point estimates, both at the individual data-set level and at the synthesized levels, but considerably more often for upper confidence bounds when the individual sample sizes or the number of available data sets are small.

One possible response is to report a one-sided confidence interval whenever the upper endpoint is infinite. For the model-based approaches, however, \cite{otto24} propose a modification inspired by the six-sigma rule that replaces infinite values by finite surrogates. Specifically, they define the truncation threshold
\[
\tilde T^{\model}
:=
\max\big\{\hat I_j^{+,\model}:j\in[m_{\model}],\ \hat I_j^{+,\model}<\infty\big\},
\]
which is finite for every real and simulated model data set considered in this paper. Whenever a model-specific log-probability-ratio estimate
$\hat T_j^{\model} = \log \widehat{\PR}{}_j^{\model}$ is infinite, it is replaced by $\tilde T^{\model}$. Moreover, whenever an upper confidence bound $\hat I_j^{+,\model}$ is infinite, it is replaced by $\tilde I_j^{+,\model}:=\hat T_j^{\model}+3(\hat T_j^{\model}-\hat I_j^{-,\model})$. 

We apply this modification throughout the simulation study and case study. It affects only the WWA approaches for model synthesis introduced in Section~\ref{subsec:model-synthesis-attribution-targets}; in all other settings, infinite values are reported explicitly.

\subsection{Performance Metrics}
\label{subsec:performance}

The performance of the synthesis approaches are evaluated using four performance metrics for the synthesized estimators and four corresponding metrics for the associated confidence intervals. All metrics are computed empirically based on $N=1,000$ independent simulation runs.

For the synthesized estimators of $\log\PR$, we report the proportion of infinite estimates. Across all applications considered in the paper and supplement, such estimates occurred only for the observational synthesis approaches; recall that potential infinities in the WWA model synthesis are removed using the construction described in the Section~\ref{subsec:infties}. Even in these cases, all full-synthesis approaches remain well defined and yield finite estimates. For the WWA full-synthesis estimator, this is because the observational component receives zero weight. We further report the empirical bias, variance, and mean squared error (MSE), conditioning on finiteness whenever instructive or necessary.

For the corresponding confidence intervals, we analogously report the proportion of infinite upper endpoints, which were likewise observed only for the observational synthesis approaches. Even in such cases, the parametric full-synthesis intervals remain well defined and have finite upper endpoints. The same holds for the WWA full-synthesis intervals, which are then defined by taking the appropriate limits in \eqref{eq:sigma-sfull} and \eqref{eq:sigma-sfull-adapted}. We further report the average interval width, empirical coverage probability, and average interval score \citep{Gneiting2007}, conditioning on a finite upper endpoint whenever instructive or necessary.

Recall that the interval score provides a joint assessment of interval sharpness and calibration. Writing $I=[I^-, I^+]$ for a confidence interval for the attribution measure $T_0$ with nominal coverage level $1-\alpha$, the interval score is defined by
\begin{align*}
\mathrm{IS}_\alpha(I;T_0)
=
(I^+-I^-)
+
\frac{2}{\alpha}(I^--T_0)\,\mathds{1}\{T_0<I^-\}
+
\frac{2}{\alpha}(T_0-I^+)\,\mathds{1}\{T_0>I^+\}. 
\end{align*}
Smaller values of $\mathrm{IS}_\alpha$ indicate better overall performance. The first term rewards shorter intervals, whereas the second and third terms penalize undercoverage, with penalties increasing linearly in the distance between $T_0$ and the interval whenever $T_0\notin I$. Hence, the interval score favors confidence intervals that are simultaneously narrow and well-calibrated.

\section{Additional simulation results}
\label{app:additional-simulation-results}

This section supplements Section~\ref{sec:simulation} with separate analyses of observational synthesis (Section~\ref{subsec:simulation-observational-synthesis}) and model synthesis (Section~\ref{subsec:simulation-model-synthesis}). It further compares conditional ML and two-stage PWM estimation (Section~\ref{app:conditional-mle-vs-pwm}) and examines infinite estimates in individual model data sets (Section~\ref{subsec:infinite-individual}).

\subsection{Additional simulation results for observational synthesis}
\label{subsec:simulation-observational-synthesis}

The simulation results in Section~\ref{sec:simulation} concern only the full synthesis. Here, we report additional results for the observational component alone.

To investigate the effect of sample size on performance, we embed the original simulation design in a broader setup. Specifically, we fix $m_\obs=2$ and consider sample-size pairs $(n_1,n_2)$ satisfying $n_1=1.5 \cdot n_2$, ranging from $(75,50)$ to $(300,200)$. For $n_1\le145$, the corresponding samples are obtained by extending the observational window backwards in time, using the available GMST values from 1880 to 2025. Since the GMST record is only available from 1880 onwards, larger sample sizes cannot be generated by extending the period further into the past. For $n_1>145$, we therefore use a linearly interpolated version of the GMST trajectory on the period 1880–2025 and evaluate it on a finer grid with spacing $145/n_1$. All remaining aspects of the simulation design are as specified in Section~\ref{subsec:setup:obs}.

\begin{figure}[thb!]
    \centering
    \includegraphics[width=0.99\linewidth]{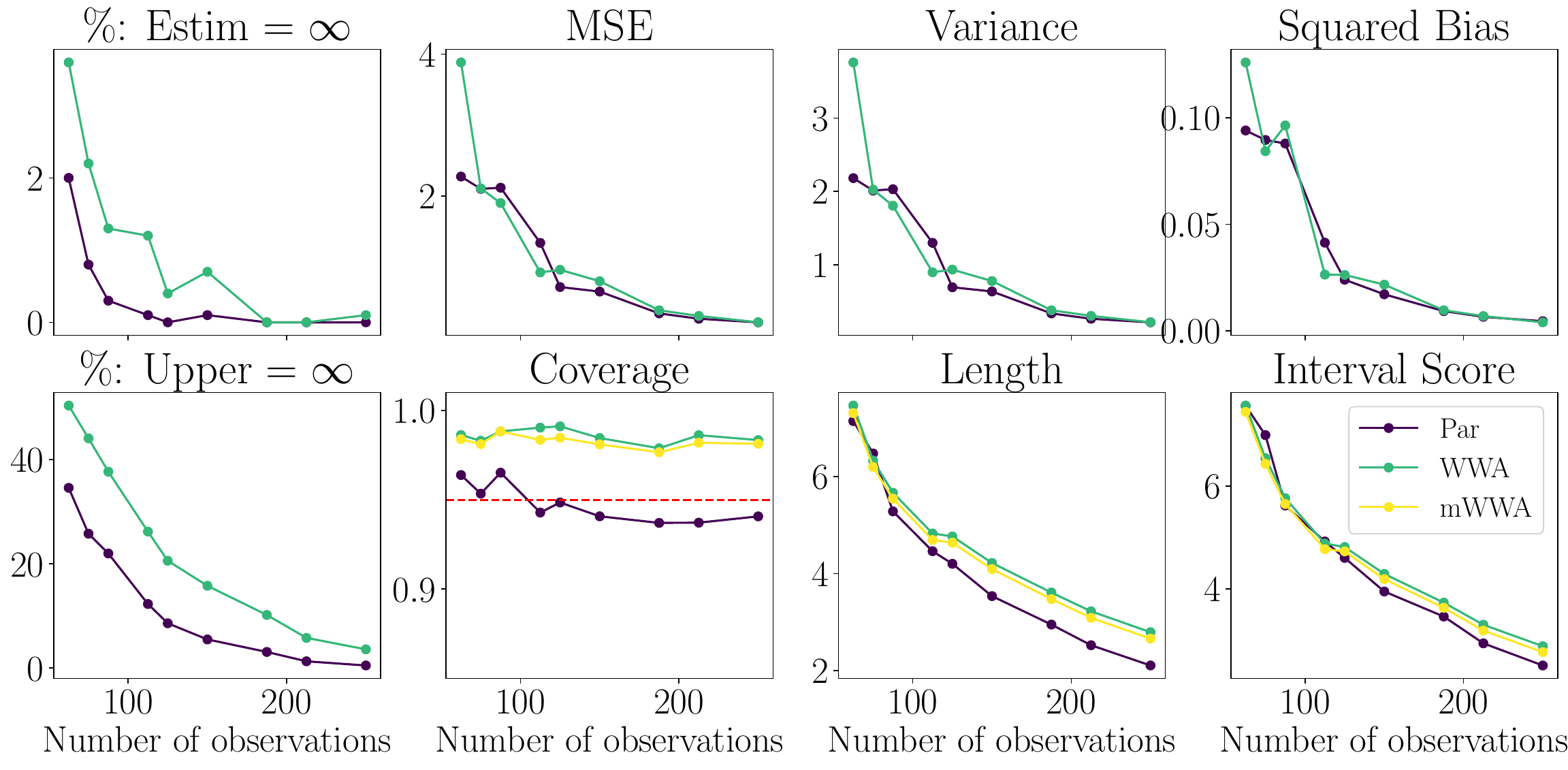}
    \caption{Performance Metrics for the \textbf{observational synthesis}. Top row: performance of synthesized point estimate. Bottom row: performance of synthesized confidence interval.}
    \label{fig:results-observational-synthesis}
\end{figure}

We compare the three observational synthesis approaches corresponding to the observational components of the full synthesis methods described in Section~\ref{subsec:methods}:

\smallskip
\begin{compactitem}
    \item \textbf{WWA}: the WWA estimator and confidence interval from \eqref{eq:That-sobs}.
    \item \textbf{mWWA}: the modified WWA confidence interval from \eqref{eq:confidence-intervals-sobs-modified}.
    \item \textbf{Par}: the plug-in estimator of $\log \PR$ from \eqref{eq:pr-as-a-function-of-x0-estimator} that is based on the synthesized PL estimator $\hat \vartheta_{\sobs}$ from \eqref{eq:hatvartheta_sobs}, as well as respective basic bootstrap confidence intervals based on the bootstrap method described in Algorithm~\ref{algo:boot_obs}.
\end{compactitem}
\smallskip \noindent
Throughout, the two-stage PWM estimator is used instead of the conditional maximum-likelihood estimator to reduce the computational burden (Section~\ref{subsec:datasetspecific}). 
The results are summarized in Figure~\ref{fig:results-observational-synthesis}. With respect to the proportion of infinite point estimates and upper confidence bounds, the PL approach performs substantially better than the WWA approach, although the proportion remains considerable for small sample sizes. Conditional on finiteness, the two approaches exhibit broadly similar point-estimation performance, with only minor differences across sample sizes. More pronounced differences arise for the confidence intervals. As expected, the plain WWA approach is conservative, and the modified version reduces this conservativeness only slightly. By contrast, the PL approach achieves coverage closer to the nominal level while producing shorter confidence intervals. These advantages are also reflected in the interval score, for which the PL approach performs best. Overall, the PL synthesis is therefore preferable, although the improvements are moderate and most apparent for larger sample sizes.

\subsection{Additional simulation results for model synthesis}
\label{subsec:simulation-model-synthesis}

The simulation results in Section~\ref{sec:simulation} concern only the full synthesis. Here, we report additional results for the model component alone, considering both the original simulation design from Section~\ref{sec:simulation} and an extended design for investigating the effects of ensemble size and the number of ensemble members.

\subsubsection{The simulation design from Section~\ref*{sec:simulation}}
The model-component results corresponding to the full synthesis analysis in Section~\ref{subsec:results} are reported in Table~\ref{tab:simulation-results-model-synthesis-empirical-ensemble-sizes}.

We observe that the PL synthesis estimator has a somewhat larger variance than the WWA estimator, but a substantially smaller squared bias, resulting in a slightly lower MSE. Compared with WWA, WWA(2) further reduces the variance, leading to an MSE that is nearly identical to that of PL synthesis despite its larger squared bias.
The performance of the confidence intervals depends on the calibration method. The gPar intervals are the shortest overall, with an average length of $0.61$, but undercover, attaining a coverage probability of only $89.8\%$. By contrast, the hPar intervals achieve coverage close to the nominal level ($97.5\%$), while remaining considerably shorter than the WWA intervals and yielding a substantially smaller interval score: $0.77$ versus $2.44$ in average length and $0.81$ versus $2.45$ in interval score. The modified WWA intervals are slightly shorter than the hPar intervals, but they also exhibit lower coverage ($93.2\%$ versus $97.5\%$), which results in a smaller interval score for hPar than mWWA ($0.81$ versus $0.94$). Relative to mWWA, mWWA(2) further reduces the average interval length from $0.77$ to $0.61$, matching the length of the gPar intervals. This gain comes at the expense of a small reduction in coverage (from $93.2\%$ to $91.6\%$), but the interval score nevertheless improves from $0.94$ to $0.82$. Despite this improvement, hPar still attains the best overall interval score while providing coverage much closer to the nominal level.
Overall, PL synthesis substantially reduces squared bias while retaining competitive point-estimation accuracy. Among the shorter interval procedures, hPar provides the most reliable coverage.

\begin{table}[htbp]
\centering
\caption{
Performance Metrics for the \textbf{model synthesis}. Variance, squared bias and MSE are reported in units of $10^{-3}$.
}
\label{tab:simulation-results-model-synthesis-empirical-ensemble-sizes}
\begin{tabular}{l|rrr|rrr}
\toprule
Method & Var. & Bias$^2$ & MSE & Coverage (\%) & Length & Score \\
\midrule
gPar
  & \multirow{2}{*}{26.5}  
  & \multirow{2}{*}{\textbf{0.72}} 
  & \multirow{2}{*}{\textbf{27.2}} 
  & 89.8 & \textbf{0.61} & 0.89 \\
hPar
  &  &  &
  & \textbf{97.5} & 0.77 & \textbf{0.81} \\
\midrule
WWA
  & \multirow{2}{*}{{22.7}}
  & \multirow{2}{*}{8.77} 
  & \multirow{2}{*}{31.4} 
  & 99.9 & 2.44 & 2.45\\
mWWA
  &  &  & 
  & 93.2 & 0.77 & 0.94 \\
\midrule
WWA(2)
  & \multirow{2}{*}{\textbf{18.1}}
  & \multirow{2}{*}{9.67}
  & \multirow{2}{*}{27.7}
  & 99.9 & 1.94 & 1.95\\
mWWA(2)
  &  &  & 
  & 91.6 & \textbf{0.61} & 0.82 \\
\bottomrule
\end{tabular}
\end{table}

\subsubsection{An extended simulation design}
\label{subsec:simulation-model-synthesis-extended}
Complementing the previous analyses, we investigate performance trends as the number of models $m_\model$ and the number of ensemble members $e_j$ vary. We consider $m_\model\in \{5,10,\ldots,120\}$. For each value of $m_\model$, we set $e_1= \cdots= e_{m_\model}\in \{1,\ldots,10\}$ and let all ensemble members cover the common observation period $\mathcal J_{j,a}=\{1951,\ldots,2025\}$. Thus, the total sample size for model $j$ is $n_j=75e_j$. All remaining aspects of the simulation design are as specified in the second and third paragraph of Section~\ref{subsec:setup:model}.

Because the focus here is on model synthesis alone, we do not consider WWA(2) and mWWA(2), which also depend on the observational synthesis component. We therefore compare the following four synthesis approaches:
\smallskip
\begin{compactitem}
    \item \textbf{WWA}: the WWA estimator from \eqref{eq:That-smod} and the WWA confidence interval from \eqref{eq:confidence-intervals-smod}.
    \item \textbf{mWWA}: the modified WWA confidence interval from \eqref{eq:confidence-intervals-smod-modified}.
    \item \textbf{gPar}: the plug-in estimator of $\log \PR$ from \eqref{eq:pr-as-a-function-of-x0-estimator} that is based on the synthesized PL estimator $\hat \vartheta_{\smod}$ from \eqref{eq:hatvartheta_smod}, together with corresponding basic bootstrap confidence intervals based on the bootstrap method described in Algorithm~\ref{algo:boot_model_group}.
    \item \textbf{hPar}: same as in the previous item, but with the basic bootstrap confidence intervals based on the bootstrap method described in Algorithm~\ref{algo:boot_model_hybrid}.
\end{compactitem}
\smallskip \noindent
Throughout, the two-stage PWM estimator is used instead of the conditional maximum-likelihood estimator to reduce the computational burden.(Section~\ref{subsec:datasetspecific}). Moreover, both WWA approaches use the procedure described in Section~\ref{subsec:infties} to handle infinite values.
The need for this modification is illustrated in Appendix~\ref{subsec:infinite-individual}: the proportion of infinite upper confidence bounds across the individual model data sets exceeds $25\%$ for small ensemble sizes. After applying the modification, no infinite point estimates or upper confidence bounds were observed for any of the synthesis methods, and hence conditioning on finiteness is unnecessary in this section.

\begin{figure}[t!]
    \centering

    \captionsetup[subfigure]{
    labelformat=simple,
    labelsep=colon,
    justification=raggedright,
    singlelinecheck=false
    }
    \renewcommand{\thesubfigure}{Panel~(\Alph{subfigure})}

    \begin{subfigure}[t]{\linewidth}
        \caption{Fixed number of models, $m_{\model}=60$:}

        \vspace{.1cm}
        
        \label{fig:results-model-synthesis-fixed-models}
        \centering
        \includegraphics[width=0.8\linewidth]{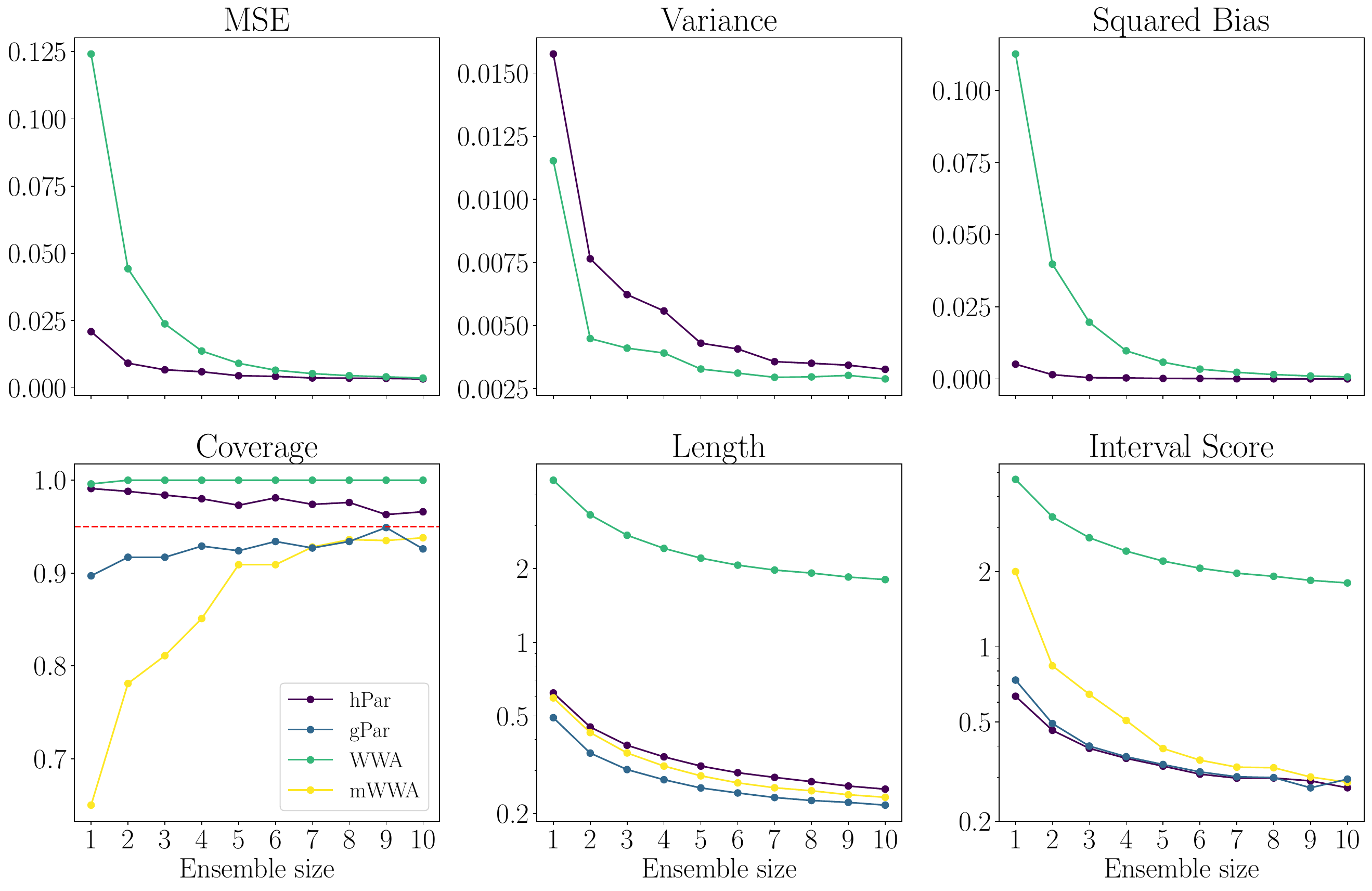}
    \end{subfigure}

    \vspace{1em}

    \begin{subfigure}[t]{\linewidth}
        \caption{Fixed ensemble size of $5$:}
        
        \vspace{.1cm}
        
        \label{fig:results-model-synthesis-fixed-ensemble-size}
        \centering
        \includegraphics[width=0.8\linewidth]{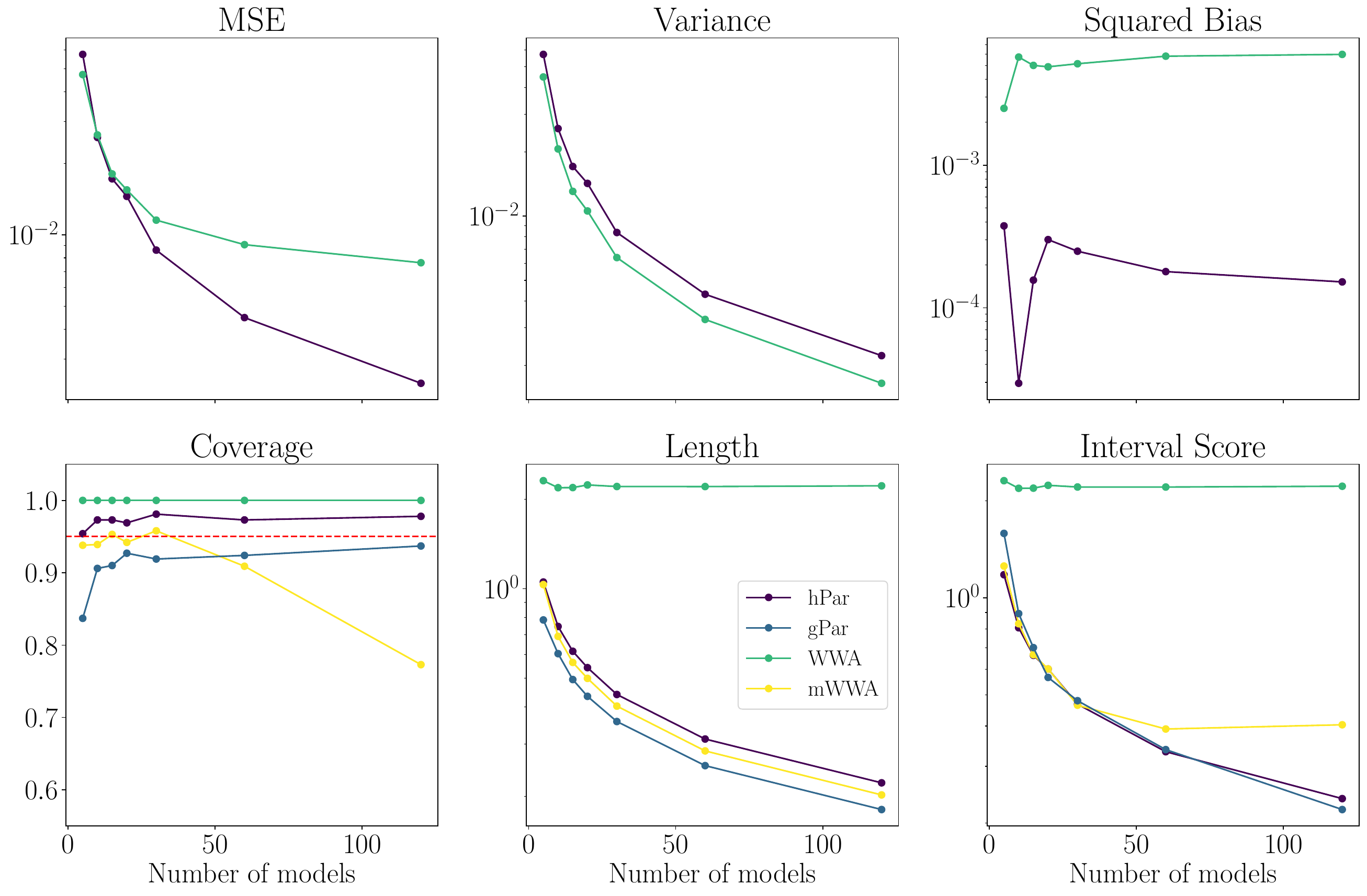}
    \end{subfigure}

    \caption{Performance of the \textbf{model-synthesis} approaches under two simulation settings. Panel~(A) shows the performance as a function of the ensemble size for a fixed number of models, $m_{\model}=60$, whereas Panel~(B) shows the performance as a function of the number of models for a fixed ensemble size of $5$. In each panel, the top row summarizes the performance of the synthesized point estimators, and the bottom row summarizes the performance of the associated confidence intervals.}
    \label{fig:results-model-synthesis-all}
\end{figure}

\begin{figure}[thb]
    \centering
    \includegraphics[width=0.85\linewidth]{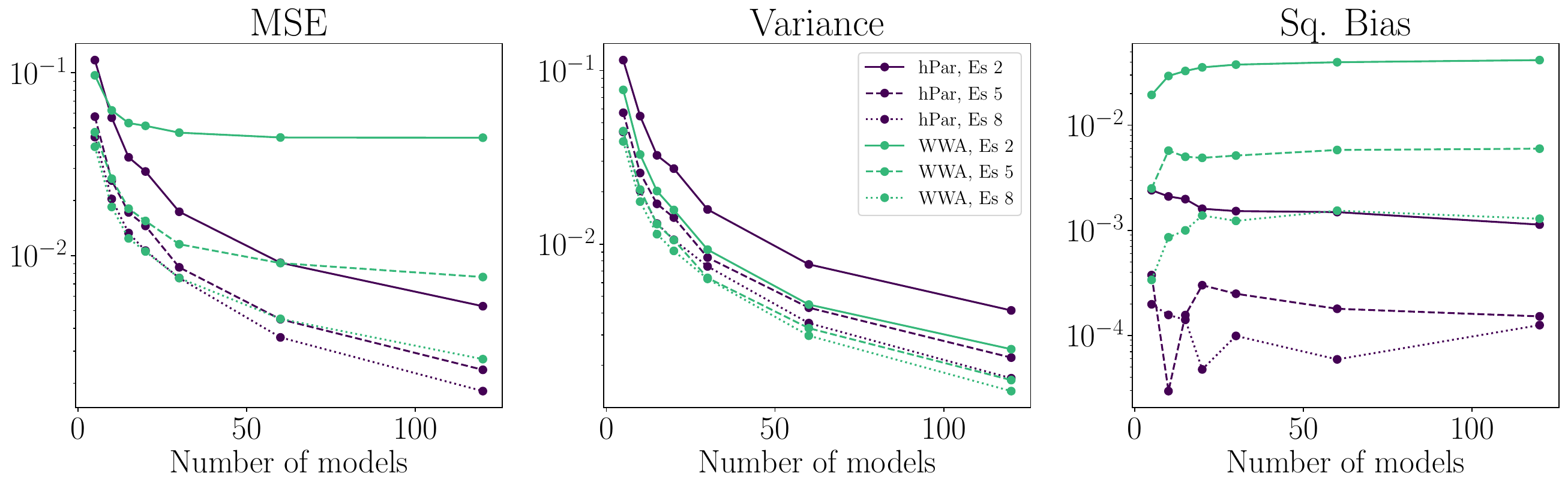}
    \caption{Performance metrics for \textbf{model-synthesis} estimators as a function of the number of models for three different ensemble sizes.}
    \label{fig:results-model-synthesis-estimation-multiple-ensembles}
\end{figure}

\begin{figure}[thb]
    \centering
    \includegraphics[width=0.99\linewidth]{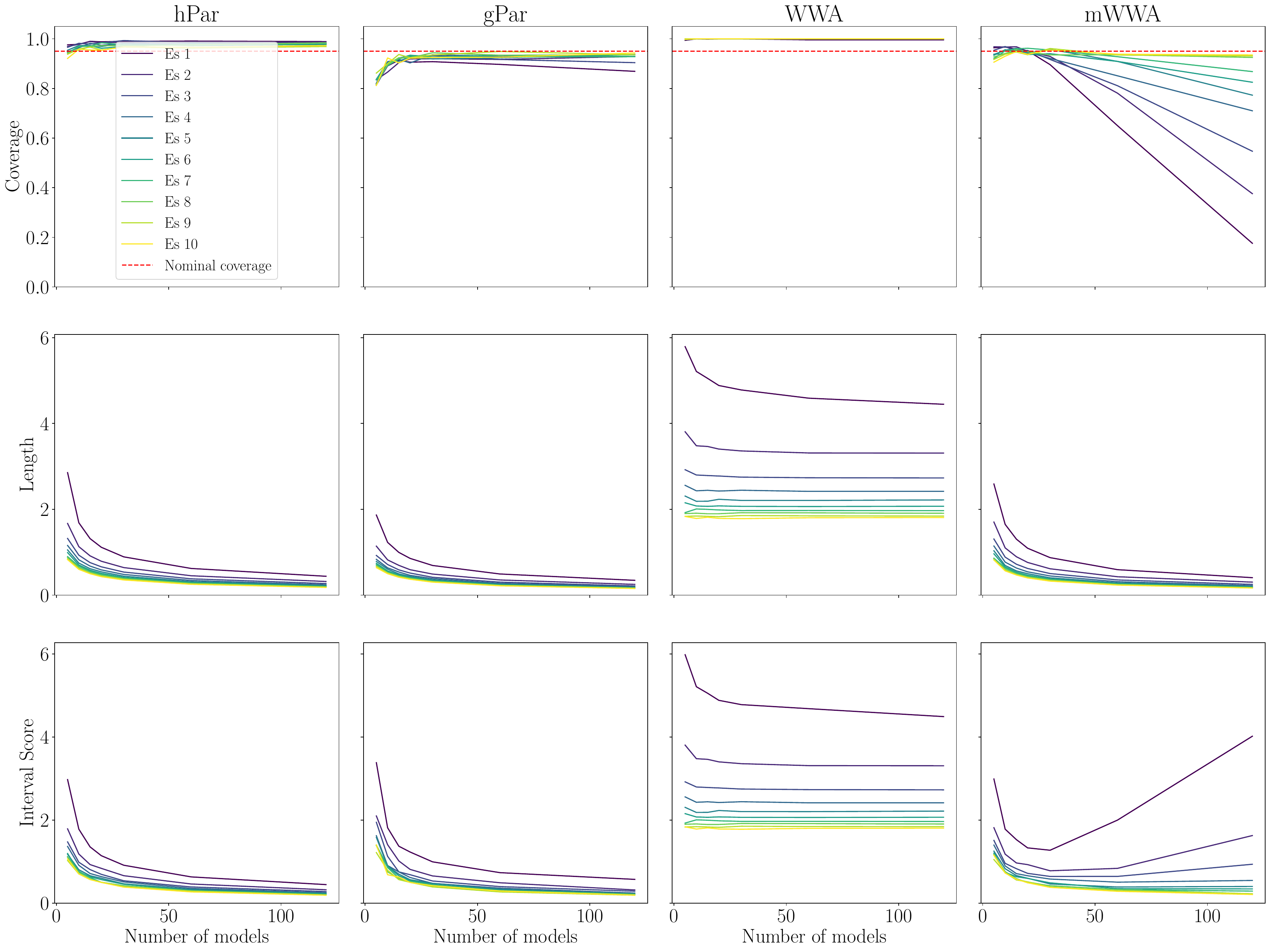}
    \caption{Performance metrics for \textbf{model-synthesis} confidence intervals as a function of the number of models for ten different ensemble sizes.}
    \label{fig:results-model-synthesis-intervals-multiple-ensembles}
\end{figure}

The results are presented from several complementary perspectives in Figures~\ref{fig:results-model-synthesis-all}, \ref{fig:results-model-synthesis-estimation-multiple-ensembles}, and~\ref{fig:results-model-synthesis-intervals-multiple-ensembles}. Figure~\ref{fig:results-model-synthesis-all} summarizes the performance of both the point estimators and the confidence intervals. Its upper panel shows performance as a function of ensemble size for a fixed number of models, $m_\model=60$, whereas the lower panel shows performance as a function of the number of models for a fixed ensemble size of~$e=5$. Figure~\ref{fig:results-model-synthesis-estimation-multiple-ensembles} focuses on the point estimators and displays their performance as a function of the number of climate models for ensemble sizes of $2$, $5$, and~$8$. Finally, Figure~\ref{fig:results-model-synthesis-intervals-multiple-ensembles} illustrates the joint effects of ensemble size and the number of models on the confidence intervals.

Overall, the simulation results favor the PL synthesis approach. Across Figures~\ref{fig:results-model-synthesis-all} and~\ref{fig:results-model-synthesis-estimation-multiple-ensembles}, the PL estimator exhibits substantially smaller squared bias than the AML estimator, although the latter achieves slightly smaller variance in some settings. The MSE decreases for both approaches as either the ensemble size or the number of models increases, but PL provides the more favorable and stable bias--variance trade-off.

The differences are particularly pronounced for the confidence intervals. In Figure~\ref{fig:results-model-synthesis-all}, the original WWA interval is highly conservative, with coverage close to one, but at the cost of substantially greater interval length and a much larger interval score. The modified WWA interval is considerably shorter, but may undercover. By contrast, the PL bootstrap intervals attain coverage close to the nominal level while maintaining comparatively small interval lengths and interval scores. Figure~\ref{fig:results-model-synthesis-intervals-multiple-ensembles} leads to similar conclusions. The two PL methods perform well across the range of ensemble sizes, with slight undercoverage for gPar and slight overcoverage for hPar. The WWA intervals remain very wide throughout and consequently attain empirical coverage of $100\%$. The mWWA intervals are substantially shorter, but tend to undercover when the number of climate models is large and the ensemble size is small.

\subsection{Conditional ML Estimation vs.\ two-stage PWM estimation}
\label{app:conditional-mle-vs-pwm}

\begin{figure}[!htp]
    \centering
    \includegraphics[width=0.85\linewidth]{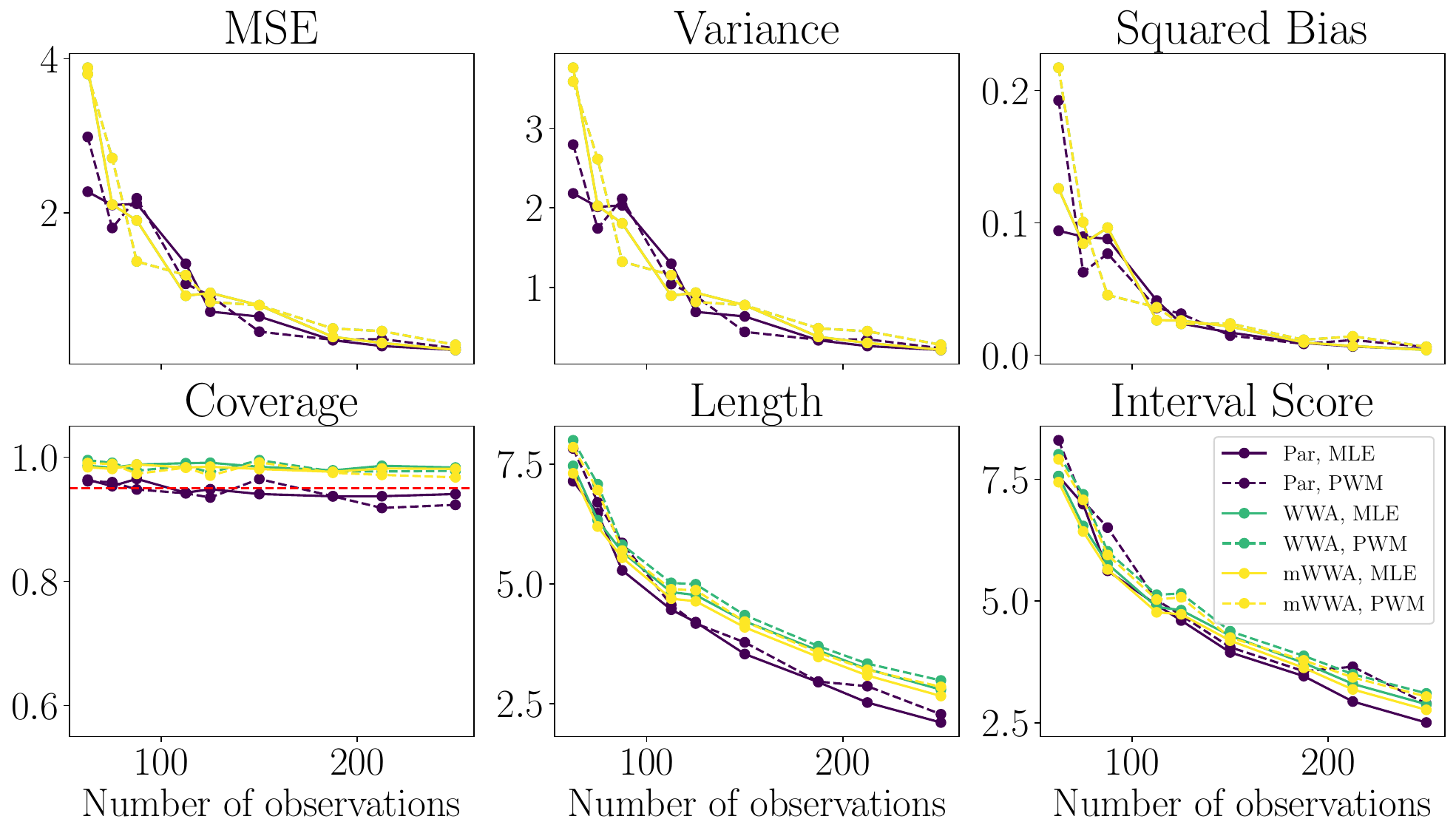}
    \caption{Observational synthesis based on either two-stage PWM or conditional ML estimation. See also Figure~\ref{fig:results-observational-synthesis}.}
    \label{fig:conditional-mle-vs-pwm}
\end{figure}

The results reported in Sections~\ref{subsec:simulation-observational-synthesis} and \ref{subsec:simulation-model-synthesis-extended} are all based on the two-stage PWM estimator rather than the conditional ML estimator. 
This choice is motivated by the comparison in Figure~\ref{fig:conditional-mle-vs-pwm} (which is an extended version of Figure~\ref{fig:results-observational-synthesis}): for the setup of observational data sets described in Section~\ref{subsec:simulation-observational-synthesis}, the respective estimators and confidence intervals exhibit very similar finite-sample behavior, with only minor differences in the resulting estimates. Since the two-stage PWM estimator is substantially more efficient computationally, requiring only about $20$--$25\%$ of the computation time needed for conditional ML, we use the PWM-based procedure throughout Sections~\ref{subsec:simulation-observational-synthesis} and \ref{subsec:simulation-model-synthesis-extended}.

\subsection{Infinite estimates for individual model data sets}
\label{subsec:infinite-individual}

\begin{figure}[thb]
    \centering
    \includegraphics[width=0.7\linewidth]{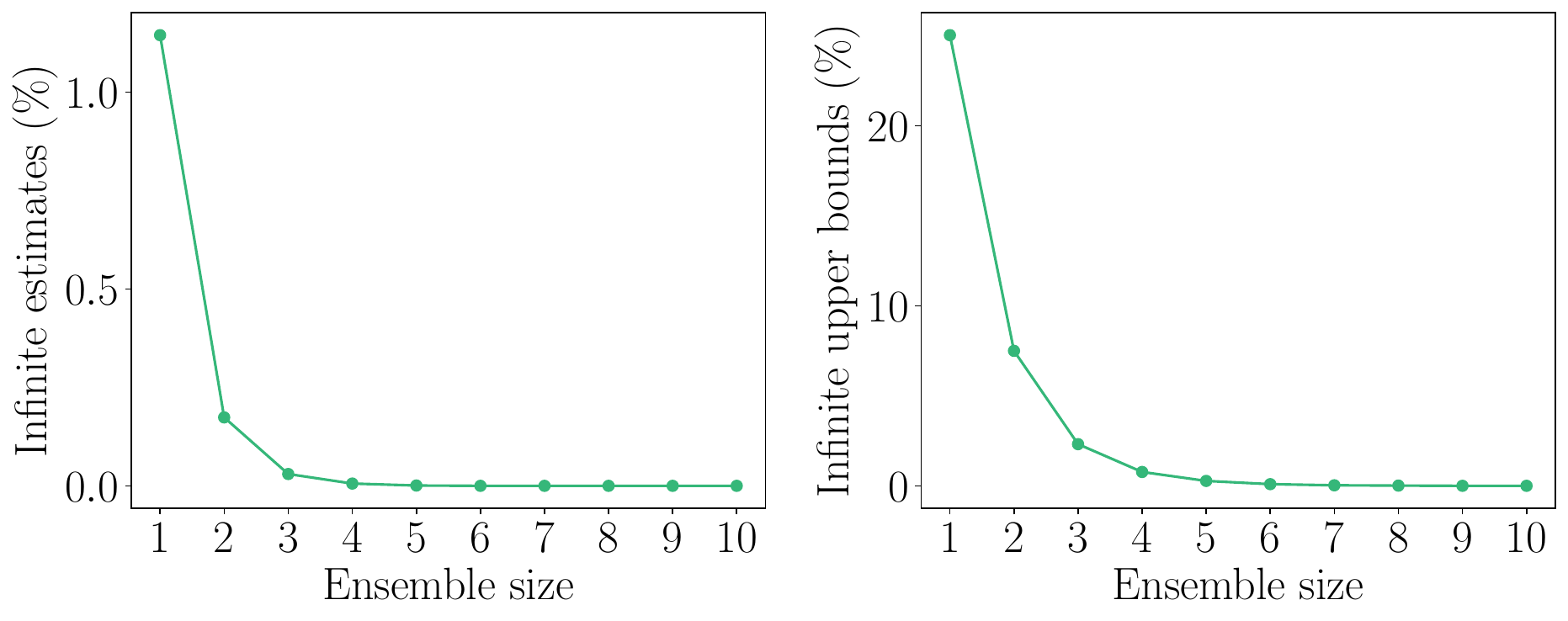}
    \caption{Frequency of infinite probability-ratio estimates (left) and upper confidence bounds (right) for individual model data across simulation runs. }
    \label{fig:inftys2}
\end{figure}

Figure~\ref{fig:inftys2} shows the proportions of infinite point estimates and infinite upper confidence bounds for the individual model data sets. Both proportions decrease sharply with ensemble size. For an ensemble size of one, approximately \(1\%\) of the point estimates and \(25\%\) of the upper bounds are infinite. From an ensemble size of four onward, infinite point estimates are virtually absent, while the proportion of infinite upper bounds is below \(1\%\) and continues to approach zero. Thus, infinite values arise primarily for upper confidence bounds based on very small ensembles. The corresponding proportions are largely stable across different numbers of available models and are therefore not shown separately.

\section{Additional results for the case study}
\label{app:case-study}

\begin{figure}[thb]
    \centering
        \includegraphics[width=.75\linewidth]{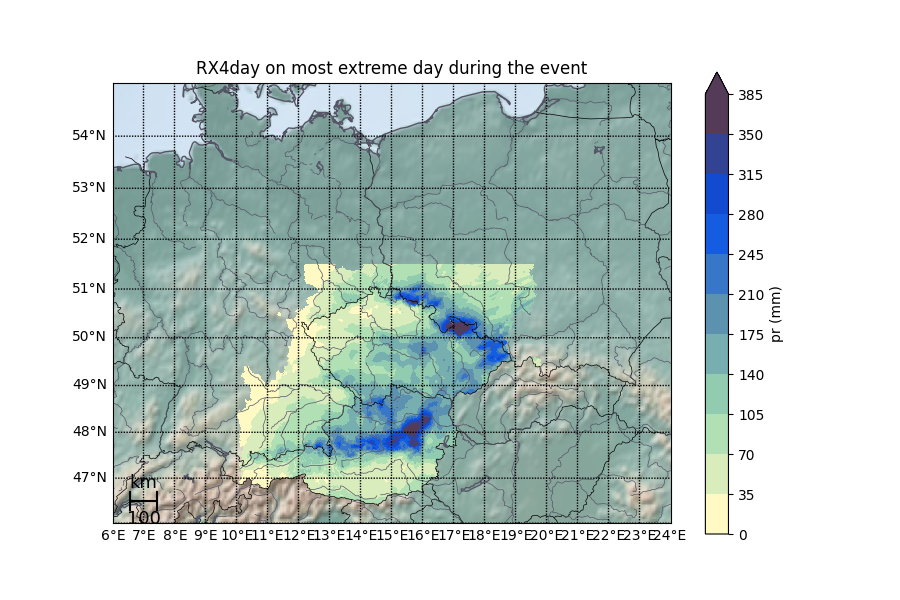}
    
    \vspace{-.7cm}
    \caption{
    Spatial context of the Storm Boris precipitation event in September 2024. The figure shows cumulative precipitation over four consecutive days, beginning on 12 September 2024 at 06:00 UTC.
    }
    \label{fig:case_boris}
\end{figure}

Figure~\ref{fig:case_boris} illustrates the spatial extent and precipitation pattern of Storm Boris.

\begin{figure}[thb]
    \centering
    \includegraphics[width=.72\textwidth]{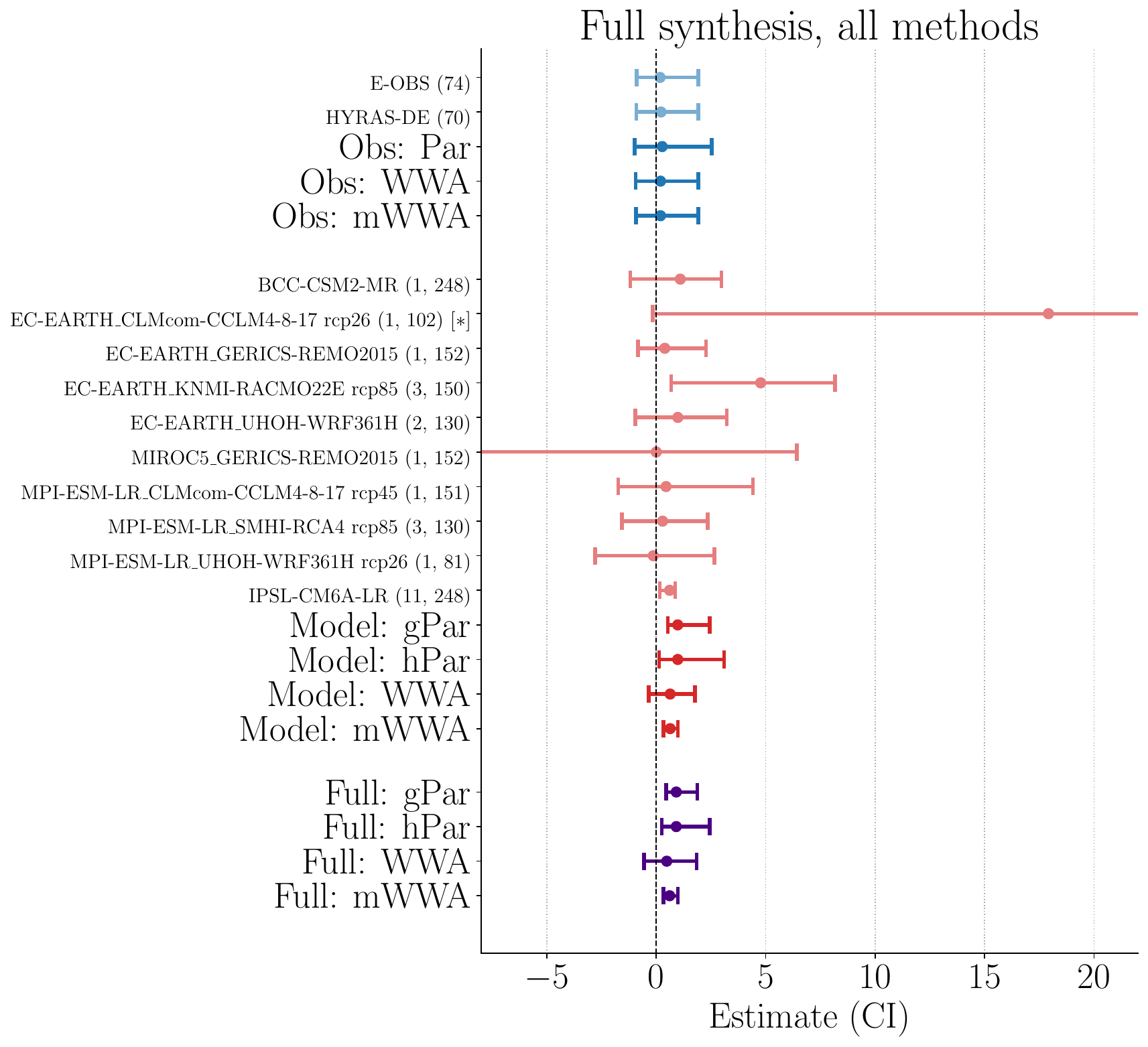}
    \caption{Same as Figure~\ref{fig:forestnfit}, but with the plug-in estimator~\eqref{eq:pr-as-a-function-of-x0-estimator} replaced by~\eqref{eq:pr-as-a-function-of-p0-estimator}. The modifications only affect the (modified) WWA synthesis, PL syntheses remain unchanged. 
    }
    \label{fig:forestnfit2}
\end{figure}

As noted in Section~\ref{sec:case-study}, the results in Figure~\ref{fig:forestnfit} are based on applying the direct probability-ratio estimator from \eqref{eq:pr-as-a-function-of-x0-estimator} to each individual model data set. Figure~\ref{fig:forestnfit2} presents the corresponding results obtained by instead using the alternative estimator from \eqref{eq:pr-as-a-function-of-p0-estimator}. For this purpose, we define
\[
\hat p_0 = H_{f_{\mathrm{scale}}(\hat \vartheta_\sobs, g_{\fact})}(x_0)
=
H_{f_{\mathrm{scale}}(\hat \vartheta_\sobs, 1.13)}(28.1)
\]
where $\hat\vartheta_{\sobs}$ denotes the synthesized observational parameter estimator from \eqref{eq:hatvartheta_sobs}.  This modification affects only the individual model estimates and, consequently, the WWA model- and full-synthesis results.

The results in Figure~\ref{fig:forestnfit2} are qualitatively similar to those in Figure~\ref{fig:forestnfit}. Although the individual confidence intervals are substantially narrower under the alternative estimator, the synthesized WWA intervals change only moderately, particularly with respect to their lower endpoints. A detailed comparison is provided in Table~\ref{tab:casestudy}.

\begin{table}[thb]
\caption{Synthesis estimates and confidence intervals for the case study on Storm Boris. WWA methods based on individual intervals estimated using \eqref{eq:pr-as-a-function-of-x0-estimator} are denoted by WWA and mWWA, whereas those using \eqref{eq:pr-as-a-function-of-p0-estimator} are denoted by WWA(2) and mWWA(2).}
\label{tab:casestudy}
\centering
\begin{tabular}{l|cccccc}
\toprule
\multirow{2}{*}{\textbf{Method}} &
\multicolumn{2}{c}{\textbf{Observed}} &
\multicolumn{2}{c}{\textbf{Model}} &
\multicolumn{2}{c}{\textbf{Full}} \\
\cmidrule(lr){2-3}
\cmidrule(lr){4-5}
\cmidrule(lr){6-7}
&
\textbf{Est.} & \textbf{CI} &
\textbf{Est.} & \textbf{CI} &
\textbf{Est.} & \textbf{CI} \\
\midrule

gPar
& \multirow{2}{*}{0.27}
& \multirow{2}{*}{$(-0.99,\ 2.53)$}
& \multirow{2}{*}{0.98}
& $(0.52,\ 2.44)$
& \multirow{2}{*}{0.91}
& $(0.45,\ 1.86)$ \\

hPar
&
& 
&
& $(0.13,\ 3.10)$
& 
& $(0.25,\ 2.44)$\\

\midrule

WWA
& \multirow{4}{*}{0.19}
& \multirow{4}{*}{$(-0.93,\ 1.92)$}
& \multirow{2}{*}{0.61}
& $(-0.33,\ 2.56)$
& 0.40
& $(-0.64,\ 2.24)$ \\

mWWA
&
&
&
& $(0.31,\ 1.22)$
& 0.57
& $(0.25,\ 1.13)$ \\

% \midrule

WWA(2)
& %\multirow{2}{*}{0.19}
& %\multirow{2}{*}{$(-0.93,\ 1.92)$}
& \multirow{2}{*}{0.63}
& $(-0.35,\ 1.76)$
& 0.48
& $(-0.56,\ 1.85)$ \\

mWWA(2)
&
&
&
& $(0.32,\ 0.99)$
& 0.61
& $(0.33,\ 0.98)$ \\

\bottomrule
\end{tabular}
\end{table}

Finally, we report estimates of the representation-error covariance matrices $\Xi^{\obs}$ and $\Xi^{\model}$ introduced in Sections~\ref{subsec:observational-synthesis-model-parameters} and \ref{subsec:model-synthesis-model-parameters}, respectively. For $\Xi^{\obs}$, we use the estimator $\hat\Xi^{\obs}$ from \eqref{eq:xi-hat-obs}; for $\Xi^{\model}$, we use the multivariate DerSimonian–Laird estimator from \eqref{eq:multivariate-DerSimonian-Laird}, as employed in the construction of $\hat\vartheta_{\smod}$ in \eqref{eq:hatvartheta_smod}. The resulting estimates, which are used to define the data-generating processes in the simulation study, are given by
{\small\begin{align}
\label{eq:case-study-xis}
\hat \Xi^{\obs} \approx 10^{-5}
\begin{pmatrix}
31 & 140 & 24 & 47 \\
140 & 650 & 110 & 220 \\
24 & 110 & 19 & 37 \\
47 & 220 & 37 & 71
\end{pmatrix},
\quad 
\hat \Xi^{\model} \approx 10^{-5}
\begin{pmatrix}
61 & 1900  & -100  & -37 \\
1900 & 72000 & -3900 & -1400 \\
-100 & -3900 & 220   & 75 \\
-37 & -1400 & 75 & 37
\end{pmatrix}.
\end{align}}Both matrices are strongly dominated by their leading eigenvalue and are therefore close to rank-one in terms of their spectral structure. In the observational case, the second largest eigenvalue is several orders of magnitude smaller than the leading one, while in the model-based case it is about four orders of magnitude smaller. At the same time, the overall scale of $\hat\Xi^{\model}$ is roughly two orders of magnitude larger than that of $\hat\Xi^{\obs}$, as reflected by their leading eigenvalues, which are approximately $7.21\cdot 10^{-1}$ and $7.75\cdot 10^{-3}$, respectively.

\newpage

\section{Algorithms}\label{app:algo}
In this section, we collect the algorithms described in the main paper.

\begin{algorithm}[H]
\caption{Observational Synthesis Bootstrap for Model Parameters}
\label{algo:boot_obs}
\begin{algorithmic}[1]
\State Extend the time series $(X_{jt})_{t \in \mathcal I_j}$ to a time series
$(X_{jt})_{t \in \mathcal I}$ by filling missing values with ``NaN''
(not-a-number) for $t \in \mathcal I \setminus \mathcal I_j$.
\Require $(g_t, X_{1t}, \dots, X_{mt})_{t \in \mathcal I}$ with $m \ge 2$

\State \textbf{Step 1: Estimating $\bm \Sigma$}
\For{$b = 1, \dots, B$}
    \State Draw $|\mathcal I|$ time indices from $\mathcal I$ with replacement, and collect them in a multiset $\mathcal I^{*b}$.
    \For{$j = 1, \dots, m$}
        \State Compute $\hat \vartheta_j^{*b}$ from $(g_t, X_{jt})_{t\in \mathcal I^{*b}}.$
    \EndFor
    \State Stack $\hat {\bm \vartheta}{}^{*b}=(\hat \vartheta_1^{*b},\dots,\hat \vartheta_m^{*b}) \in \R^{md}.$
\EndFor
\State Put $\hat \Sigma=\widehat{\Cov}(\hat {\bm \vartheta}{}^{*1},\dots,\hat {\bm \vartheta}{}^{*B})\in \mathbb R^{md \times md}.$

\State \textbf{Step 2: Joint bootstrap}
\For{$b = 1, \dots, B$}
    \State Define $\hat \Xi^{*b}:=\hat \Xi(\hat {\bm \vartheta}{}^{*b}, \hat\Sigma)$.
    \State Estimate $V$ via $\hat V^{*b}=\hat\Sigma + \mathds I_{m \times m}\otimes\hat \Xi^{*b}$.
    \State Put $\hat \vartheta^{*b}_\sobs=  (\bm X^\top (\hat V^{*b})^{-1} \bm X)^{-1}
\bm X^\top (\hat V^{*b})^{-1}\hat {\bm \vartheta}{}^{*b}$, where $\bm X= \mathds{1}_m \otimes \mathds{I}_{d\times d} \in \R^{md\times d}$.
\EndFor
\State \Return bootstrap sample $ \hat\vartheta^{*1}_\sobs, \dots, \hat\vartheta^{*B}_\sobs. $
\State \Return (optional)  $\hat V^{*1}, \dots, \hat V^{*B}$.
\end{algorithmic}
\end{algorithm}

\begin{algorithm}
\caption{Model Synthesis for Model Parameters: the Group Bootstrap}
\label{algo:boot_model_group}
\begin{algorithmic}[1]
\Require Model-specific estimates $(\hat\vartheta_1,\hat\Sigma_1),\dots,(\hat\vartheta_{m},\hat\Sigma_m)$; number of bootstrap replications $B$.
\For{$b=1,\dots,B$}
    \For{$j=1,\dots,m$}
        \State (a) Randomly select $(\hat\vartheta_j^{*b}, \hat\Sigma_j^{*b})$ from $(\hat\vartheta_1,\hat\Sigma_1),\dots,(\hat\vartheta_{m},\hat\Sigma_m)$.
    \EndFor
    \State (b) Compute $\hat{\Xi}^{*b}$ as in \eqref{eq:multivariate-DerSimonian-Laird} with $(X_j,  \Sigma_j)$ replaced by $(\hat \vartheta_j^{*b}, \hat \Sigma_j^{*b})$.
    \State (c) Compute $\hat\vartheta_{\smod}^{*b}$ as in \eqref{eq:hatvartheta_smod} with $(\hat \vartheta_j, \hat \Sigma_j, \hat\Xi)$  replaced by $(\hat\vartheta_j^{*b}, \hat\Sigma_j^{*b}, \hat\Xi^{*b})$.
\EndFor
\State \Return bootstrap sample $\hat \vartheta_{\smod}^{*1}, \dots , \hat \vartheta_{\smod}^{*B}$. 
\State \Return (optional)  $(\hat\Xi^{*b}, \hat\Sigma_1^{*b}, \dots, \hat\Sigma_m^{*b})$ for $b=1, \dots,B$.
\end{algorithmic}
\end{algorithm}

\begin{algorithm}
\caption{Model Synthesis: Bootstrap Strategy 2 (Hybrid Group-Parametric Bootstrap)}
\label{algo:boot_model_hybrid}
\begin{algorithmic}[1]
\Require 
Model-specific estimates $(\hat\vartheta_1,\hat\Sigma_1),\dots,(\hat\vartheta_{m},\hat\Sigma_m)$; corresponding GMST data $(g_{jt})_{t\in\mathcal J_j}$ for $j\in [m]$; number of bootstrap replications $B$.
\For{$b=1,\dots,B$}
    \For{$j=1,\dots,m$}
        \State (a) Randomly select $(\hat\vartheta_j^{*b}, \hat\Sigma_j^{*b}, (g_{jt}^{*b})_{t \in \mathcal J_j^{*b}})$ from %\\ \hspace{7.4cm} 
        \Statex \hspace{7.4cm} 
        $(\hat\vartheta_1,\hat\Sigma_1, (g_{1t})_{t \in \mathcal J_1}), \dots, (\hat\vartheta_m,\hat\Sigma_m, (g_{mt})_{t \in \mathcal J_m})$.
        \For{$t\in\mathcal J_j^{*b}$}
            \State (b) Sample
            $
            X_{jt}^{*b}\sim
            \GEV\bigl(f(\hat\vartheta_j^{*b},g_{jt}^{*b})\bigr).
            $
        \EndFor
        \State (c) Re-fit $\hat\vartheta_j^{**b}$ from $(g_{jt}^{*b},X_{jt}^{*b})_{t\in\mathcal J_j^{*b}}$.
    \EndFor
    \State (d) Compute $\hat{\Xi}^{*b}$ as in \eqref{eq:multivariate-DerSimonian-Laird} with $(X_j,  \Sigma_j)$ replaced by $(\hat \vartheta_j^{**b}, \hat \Sigma_j^{*b})$ 
    \State (e) Compute $\hat\vartheta_{\smod}^{*b}$ as in \eqref{eq:hatvartheta_smod} with $(\hat \vartheta_j, \hat \Sigma_j, \hat \Xi)$  replaced by $(\hat\vartheta_j^{**b},\hat\Sigma_j^{*b},\hat\Xi^{*b})$.
\EndFor
\State \Return bootstrap sample $\hat \vartheta_{\smod}^{*1}, \dots , \hat \vartheta_{\smod}^{*B}$. 
\State \Return (optional)  $(\hat\Xi^{*b}, \hat\Sigma_1^{*b}, \dots, \hat\Sigma_m^{*b})$ for $b=1, \dots,B$.
\end{algorithmic}
\end{algorithm}

\begin{algorithm}
\caption{Bootstrap for Full Synthesis}
\label{algo:boot_full}
\begin{algorithmic}[1]
\Require $(\hat\vartheta_{\sobs}^{*b}, \hat V^{*b})_{b=1}^B$ from Algorithm~\ref{algo:boot_obs} and $(\hat\vartheta_{\smod}^{*b},\hat\Xi^{*b}, \hat\Sigma_{1}^{*b} ,\dots, \hat\Sigma_m^{*b})_{b=1}^B$ from Algorithm~\ref{algo:boot_model_group} or~\ref{algo:boot_model_hybrid}; number of bootstrap replications $B$.
\For{$b=1,\dots,B$}
    \State Compute $\hat\Sigma_{\sobs}^{*b} = (\bm X^\top (\hat V^{*b})^{-1} \bm X)^{-1}$ and $\hat\Sigma_{\smod}^{*b} = \big(\sum_{j=1}^m \big(\hat\Sigma_j^{*b}+\hat\Xi^{*b}\big)^{-1}\big)^{-1}$.
    \State Put $\hat \vartheta_{\sfull}^{*b}
= 
\big( (\hat\Sigma_{\sobs}^{*b})^{-1}+(\hat\Sigma_{\smod}^{*b})^{-1}\big)^{-1}
\big((\hat\Sigma_{\sobs}^{*b})^{-1}
\hat\vartheta_{\sobs}^{*b}
+(\hat\Sigma_{\smod}^{*b})^{-1}
\hat\vartheta_{\smod}^{*b}\big)$
\EndFor
\State \Return bootstrap samples $\hat\vartheta_{\sfull}^{*1}, \dots, \hat\vartheta_{\sfull}^{*B}$.
\end{algorithmic}
\end{algorithm}

\newpage

% Supplement-only bibliography.
\putbib[bibliography]
\end{bibunit}
\end{document}